\documentclass{aa}

\usepackage{txfonts}
\usepackage{mathtools}
\usepackage{amsmath}	% Advanced maths commands
\usepackage{rotating}
\usepackage{placeins}
\usepackage{xcolor}
\usepackage[breaklinks=true]{hyperref}
\hypersetup{
    colorlinks = true,
    citecolor  = blue,
    linkcolor  = blue,
    urlcolor   = blue
}
\newcommand{\Msun}{M$_{\odot}$}

\newcommand{\Ms}{M$_{\star}$}

\newcommand{\alphaCO}{$\alpha_{\rm{CO}}$}

\begin{document}

%%%%%%%%%%%%%%%%%%%%%%%%%%%%%%%%%%%%%%%%
% if you use custom commands in your title,
% ensure to check your title when submitting!
%%%%%%%%%%%%%%%%%%%%%%%%%%%%%%%%%%%%%%%%
%   \title{Metal-THINGS: spatially resolved analysis of the Dust-to-Gas ratio, the CO-to-H$_2$ conversion factor and oxygen metallicity for low mass galaxies}

\title{Metal-THINGS: Spatially Resolved Dust–to-Gas and CO–to-H$_2$ scaling relations in Low-Mass Galaxies}
%and metallicity?

% Titulo muy largo, que tal algo como:
%The Metal-THINGS Survey: Resolved Dust-to-Gas ratios, CO-to-H$_2$ conversion, and metallicity in Low-Mass Galaxies

   %\subtitle{Subtitle} [Metal-THINGS: scaling relationships]

%%%%%%%%%%%%%%%%%%%%%%%%%%%%%%%%%%%%%%%%
% Please separate each author with the \and command
%
% Please do not include ORCIDs next to author names.
% Only ORCIDs authenticated by individual authors in EDPS
% editorial system will be taken into account.
% ORCIDs included here will be removed.
%%%%%%%%%%%%%%%%%%%%%%%%%%%%%%%%%%%%%%%%

   \author{L. E. Garduño\inst{1,2}\fnmsep\thanks{Corresponding author: luisgard@ucm.es}
        \and M. A. Lara-L\'opez\inst{1,2}
        \and J. Zaragoza-Cardiel\inst{3}
        \and I. A. Zinchenko\inst{4, 5}
        \and L. S. Pilyugin\inst{5, 6}
        \and Vidhi Tailor\inst{7, 8}
        \and Viviana Casasola\inst{7}
        \and Jesús Gallego\inst{1,2}
        \and M. M. Puebla\inst{1,2}
        \and Santiago \'Alvarez-C\'ordoba\inst{9}
        \and G. Val\'e\inst{1, 2}
        \and M. E. De Rossi\inst{10, 11}
        \and Sami Dib\inst{12}
        \and Jacopo Fritz\inst{7,9,13}
        \and Omar L\'opez-Cruz\inst{14}
        \and M. Valerdi\inst{15}
        %\and \inst{1}
        %\and \inst{1}
        %\and \inst{2}
        }

   \institute{Departamento de F\'isica de la Tierra y Astrof\'isica, Fac. de C.C. F\'isicas, Universidad Complutense de Madrid (UCM), E-28040 Madrid, Spain
   \and Instituto de F\'isica de Part\'iculas y del Cosmos, IPARCOS, Fac. C.C. F\'isicas, Universidad Complutense de Madrid (UCM), E-28040 Madrid, Spain
   \and Centro de Estudios de F\'isica del Cosmos de Arag\'on (CEFCA), Plaza San Juan 1, 44001 Teruel, Spain
   \and Astronomisches Rechen-Institut, Zentrum f\"ur Astronomie der Universit\"at Heidelberg, M\"onchhofstra\ss e 12-14, D-69120 Heidelberg, Germany
   \and Main Astronomical Observatory, National Academy of Sciences of Ukraine, 27 Akademika Zabolotnoho St., 03143, Kyiv, Ukraine
   \and Institute of Theoretical Physics and Astronomy, Vilnius University, Saul\.etekio av. 3, 10257, Vilnius, Lithuania
   \and INAF - Istituto di Radioastronomia, Via Gobetti 101, 40129 Bologna, Italy
   \and Dipartimento di Fisica e Astronomia, Alma Mater Studiorum Universit\`a di Bologna, Via Piero Gobetti 93/2, I-40129, Bologna, Italy
   \and Instituto de Radioastronom\'ia y Astrof\'isica, UNAM, Campus Morelia, A.P. 3-72, Morelia 58089, Mexico
   \and Universidad de Buenos Aires, Facultad de Ciencias Exactas y Naturales y Ciclo B\'asico Com\'un, Buenos Aires, Argentina
   \and CONICET-Universidad de Buenos Aires, Instituto de Astronom\'ia y F\'isica del Espacio (IAFE), Buenos Aires, Argentina
   \and Max Planck Institute for Astronomy, K\"onigstuhl 17, 69117, Heidelberg, Germany
   \and INAF–Osservatorio Astronomico di Brera, Via Brera 28, 20121 Milano, Italy
   \and Instituto Nacional de Astrof\'isica, \'Optica y Electr\'onica (INAOE), Luis E. Erro No. 1, Sta. Ma. Tonantzintla, Puebla, C.P. 72840, Mexico
   \and Departamento de Astronomía, Universidad de Chile, Camino del Observatorio 1515, Las Condes, Santiago, Chile
   }
   
   \date{Received 09/2026; Accepted 11/2026}

% \abstract{}{}{}{}{}
% 5 {} token are mandatory

%

  \abstract
   {{Although the global relations between the dust-to-gas ratio (DGR), the CO-to-H$_2$ conversion factor (\alphaCO), and the gas-phase oxygen metallicity are well established, their spatially resolved counterparts on scales of $\sim$100--1000 pc remain poorly constrained, particularly in low-mass, low-metallicity galaxies.}}
   %{The established relation on global galactic scales between the dust-to-gas ratio (DGR), the CO-to-H$_2$ conversion factor (\alphaCO) and the gas-phase oxygen metallicity remains poorly constrained on spatially resolved scales ($\sim$ 100 - 1000 pc) within individual galaxies, particularly in the low-mass and low-metallicity regime.}}
   %Previous studies have established the global relation between the dust-to-gas ratio (DGR), the CO-to-H$_2$ conversion factor (\alphaCO) and the gas-phase oxygen metallicity, but the spatially resolved behavior across individual galaxies remains poorly constrained, particularly in the low-mass and low-metallicity regime.}
   {We aim to characterize the local correlations between these three interstellar medium (ISM) diagnostics on the scales $\sim$250-550\,pc for a sample of five nearby low-mass star-forming galaxies (NGC\,4214, NGC\,2976, NGC\,3077, NGC\,2403, and NGC\,925). Our sample spans distances $\sim 2.9\text{-}6.4\,\mathrm{Mpc}$, $\mathrm{SFR} \sim 0.05\text{-}1.09\,\mathrm{M}_{\odot}\,\mathrm{yr}^{-1}$, and $12+\log(\mathrm{O/H}) \sim 8.1\text{-}8.45\,$.}
   {Oxygen metallicities were derived from optical Integral Field Unit spectroscopy using the S-calibration. Spatially resolved dust masses were estimated by fitting the spectral energy distribution on a pixel-by-pixel basis with CIGALE code, using multiwavelength photometry from DustPedia. Atomic gas masses were derived using HI 21-cm maps from the THINGS survey, and the CO(2-1) luminosities from the HERACLES survey. The spatially resolved DGR and $\alpha_{\rm CO}$ were solved simultaneously by relating the dust mass to the total gas mass, combining the atomic and molecular gas components.}
   {We find a positive correlation between the DGR and metallicity in all five galaxies. Radial negative gradients of DGR are detected in most galaxies except for NGC\,4214 (flat) and NGC\,3077 (positive). We recover an anti-correlation between \alphaCO\ and metallicity for NGC 2403 and NGC 925, while lower-mass galaxies show a flat relation. 
   Positive radial gradients in \alphaCO\ are detected in NGC\,2403 and NGC\,925, while the remaining galaxies exhibit flat profiles.}
  % conclusions heading (optional), leave it empty if necessary as
   {Our spatially resolved analysis supports the existence of the DGR-metallicity and \alphaCO -metallicity correlations on sub-kiloparsec scales, while revealing significant galaxy-to-galaxy scatter and local deviations that cannot be explained by metallicity alone. The results highlight the critical importance of spatially resolved, multi-tracer observations to accurately analyzing the ISM conditions and for deriving reliable galaxy properties in low-mass galaxies across a range of environments.}

   \keywords{galaxies: abundances -- galaxies: dwarf -- galaxies: irregular -- galaxies: starburst -- galaxies: star formation -- galaxies: statistics
               }

   \titlerunning{spatially resolved analysis of the DGR, DMR, $\alpha_{\rm{CO}}$ and Z.}
   \authorrunning{The Metal-THINGS team}
   \maketitle

%%%%%%%%%%%%%%%%%%%%%%%%%%%%%%%%%%%%%%%%%%%%%%%%%%%%%%%%%%%%%%
\section{Introduction}\nolinenumbers

The interstellar medium (ISM) of galaxies is a complex, multiphase environment where stars form from molecular gas, dust grains regulate the thermal balance, and metals produced by stellar nucleosynthesis shape the physical conditions for subsequent generations of star formation \citep{McKee2007,Bolatto2013,Tumlinson2017}. Among the key observables that characterize the ISM are the dust-to-gas ratio (DGR), the dust-to-metal mass ratio (DMR), and the CO-to-H$2$ conversion factor (\alphaCO, defined as $\alpha_{\rm CO} \equiv \Sigma_{\rm mol}/W_{\rm CO}$, where $\Sigma_{\rm mol}$ is the total molecular gas surface density, including helium, and $W_{\rm CO}$ is the integrated CO line intensity). These quantities are closely linked to gas-phase metallicity and provide valuable constraints on dust evolution, molecular cloud shielding, and star formation. {While the DGR traces the overall dust content of the ISM, a more complete picture of dust evolution is provided by the dust-to-metal mass ratio (DMR), which quantifies the fraction of heavy elements locked into dust grains \citep[e.g.,][]{Dwek1998,Jenkins2009,Galliano2021}.}

%The interstellar medium (ISM) of galaxies is a complex, multiphase environment where stars form from molecular gas, dust grains regulate the thermal balance, and metals produced by stellar nucleosynthesis shape the physical conditions for subsequent generations of star formation \citep{McKee2007,Bolatto2013,Tumlinson2017}. Among the key observables that characterize the ISM are the dust-to-gas ratio (DGR), the dust-to-metal mass ratio (DMR), and the CO-to-H$2$ conversion factor (\alphaCO\ , also denoted $X_{\rm CO}$), all of which are closely linked to gas-phase metallicity and provide valuable constraints on dust evolution, molecular cloud shielding, and star formation.

%{Galaxies' interstellar medium (ISM) shows considerable complexity. It comprises a multiphase environment suitable for star formation from molecular gas; dust grains regulate thermal balance, and metals produced by stellar nucleosynthesis (e.g. via supernovae) influence the physical conditions of subsequent star formation \citep{McKee2007,Bolatto2013,Tumlinson2017}. This study aims to investigate how the dust-to-gas ratio (DGR), the dust-to-metal mass ratio (DMR), and the CO-to-H$_2$ conversion factor (\alphaCO) vary with gas-phase metallicity in star-forming galaxies. The objectives are to quantify the interconnections between these properties and to assess how they determine the efficiency of star formation and the lifecycle of dust in the ISM.}

Characterizing these relations is essential for interpreting observations of star-forming galaxies across cosmic time and for constraining models of galaxy evolution \citep{Kennicutt2012,Madau2014,Peroux2020,Tacconi2020}. 

%{Understanding the relationships between gas-phase metallicity, dust abundance, and molecular gas tracers is necessary for interpreting observations of galaxies with ongoing star formation across cosmic time and for constraining scenarios for galaxy evolution \citep{Kennicutt2012,Madau2014,Peroux2020,Tacconi2020}. Two key observational quantities, the DGR and the CO-to-H$_2$ conversion factor (\alphaCO, also denoted X$_{\text{CO}}$), serve as essential diagnostics of ISM conditions, as both exhibit strong metallicity dependencies  that reflect the basic physics of dust creation, grain growth, and molecular cloud shielding \citep{Bolatto2013}. In this study, we report the analysis of multiwavelength observations of star-forming (SF) galaxies from current surveys, combining archived data on atomic and molecular gas with far-infrared and submillimeter dust measurements. Additionally, we incorporate gas-phase metallicities derived from optical emission line spectroscopy. The analysis is grounded in a comparison of these observational datasets, with supplemental insights drawn from relevant literature and recent numerical simulations where applicable.}

%{Beyond the standard DGR framework, an in-depth understanding of the interstellar dust lifecycle requires investigating the DMR, which quantifies the fraction of interstellar heavy elements that have successfully condensed into solid grains \citep[e.g.,][]{Dwek1998,Jenkins2009,Galliano2021}}

Observations from past two decades have established that the DGR increases with gas-phase oxygen abundance \citep{Lisenfeld1998, Galliano2005,Draine2007,Engelbracht2008}, following an approximately linear or super linear trend at intermediate to high metallicities (12 + log(O/H) $\geq$ 8.0) but exhibits a steeper rise and larger scatter at lower metallicities \citep{Remy2014,DeVis2019}. This behavior is consistent with theoretical models in which dust grains are produced in stellar ejecta and subsequently grow through accretion of gas-phase metals in the dense ISM \citep{Asano2013,Zhukovsk2016}. Spatially resolved studies using \textit{Herschel}'s far infrared imaging combined with CO and HI mapping have demonstrated that the DGR-metallicity correlation holds on kiloparsec scales within individual galaxies, with \cite{Sandstrom2013} presenting resolved DGR maps across 26 nearby galaxies from the KINGFISH \citep{Kennicutt2011} and HERACLES \citep{Leroy2009} surveys and finding a tight correlation between the local DGR and metallicity \citep[see also][]{Casasola2022}. Recently, the TYPHOON survey \citep{Park2024} provided sub-kiloparsec resolution measurements of the DGR versus gas-phase metallicity across 11 galaxies, revealing substantial scatter at intermediate metallicities {(8.0 $\leq$ 12 + log(O/H) $\leq$ 8.3)} that may reflect variations in local gas density, star formation history, and the starting of efficient grain growth. At the lowest metallicities (12 + log(O/H) $\leq$ 8.0), the DGR drops steeply, and the ISM becomes increasingly porous and transparent to ultraviolet (UV) radiation, with important consequences for molecular gas chemistry and observability \citep{Madden2016,Madden2020}. 

Complementary to the DGR–metallicity relation (hereafter the DGR-Z relation), \alphaCO\ exhibits a well established anti-correlation with gas-phase metallicity (hereafter the \alphaCO-Z relation),  increasing by factors of 10 or more as metallicity drops below approximately half-solar values \citep{Leroy2011,Schruba2012,Bolatto2013,Accurso2017}. This trend is understood physically as a consequence of reduced dust shielding and enhanced CO photodissociation in low-metallicity environments, where lower dust column densities allow UV photons to penetrate deeper into molecular clouds and dissociate CO molecules, leaving behind substantial reservoirs of CO-dark H$_2$ gas that is molecular but not traced by CO emission \citep{Wolfire2010,Glover2011}. The canonical Milky Way disk value of \alphaCO\ $\sim$ 4.4 \Msun\ pc$^{-2}$ (K km s$^{-1}$)$^{-1}$ \citep{Bolatto2013} represents only one point along a continuum of ISM conditions, and accurate molecular gas mass estimates in metal-poor galaxies require either direct measurements of \alphaCO\ via dust-based methods or the application of metallicity-dependent prescriptions. Observational constraints on \alphaCO\ at low metallicity have been provided by dust-based estimates in Local Group dwarf galaxies \citep{Bolatto2011,Leroy2011}, and, more recently, by high-resolution ALMA \citep{Hodge2013} observations of low metallicity starbursts, which reveal that \alphaCO\ can vary by an order of magnitude depending on spatial scale sampled and the fraction of CO-dark gas included within the beam \citep{Hunt2023}.

Historically, pioneering observational studies of nearby galaxies established a roughly linear correlation between DGR and gas-phase metallicity, interpreted as being consistent with a nearly constant DMR across a wide range of galaxy morphologies \citep[e.g.,][]{Issa1990,Lisenfeld1998}. This assumption of a fixed DMR was reinforced by early analytical chemical evolution frameworks, which argued that the global dust-metallicity scaling could be interpreted as a steady evolutionary sequence where galaxies preserve a nearly constant fraction of condensed heavy elements throughout their lifecycles \citep[e.g.,][]{Dwek1998,Hirashita1999,Edmunds2001,James2002,Inoue2003}. Consequently, a universal DMR became a standard simplifying benchmark implemented across many large galaxy-scale chemical evolution models and cosmological hydrodynamical simulations \citep[e.g.,][]{Silva1998}. However, this paradigm of universality was challenged as observations pushed into more extreme, low-metallicity environments, such as local dwarf irregulars and high-redshift damped Lyman-$\alpha$ (DLA) absorbers, revealing severely depressed dust fractions that deviated by orders of magnitude from standard Local Group values \citep[e.g.,][]{DeCia2013, Zafar2013, Remy2014}. These discoveries shifted the consensus toward a variable, non-linear DMR-Z relation, highlighting a critical metallicity threshold below which interstellar grain growth is severely inefficient.

Despite the well established correlations between metallicity and both DGR and \alphaCO,\ the physical connection between them and their covariance on spatially resolved scales within galaxies remain poorly explored. Theoretical models and simulations predict that dust abundance directly influences CO shielding: a higher DGR at a given metallicity should lead to more effective shielding of CO from UV photodissociation and thus lower \alphaCO,\ while lower DGR should result in enhanced CO photodissociation and higher \alphaCO\ \citep{Wolfire2010,Glover2010}. Recent high-resolution galaxy simulations incorporating time-dependent chemistry and dust evolution have demonstrated that dust growth in dense molecular gas is essential to reproduce observed CO luminosities at low metallicity, providing a direct link between local dust abundance, CO survival, and \alphaCO\ \citep{Hu2023}. Observationally, \cite{Chiang2025} have explored the relationship between \alphaCO\ and grain size distributions (traced by polycyclic aromatic hydrocarbon emission) on resolved scales, suggesting that grain processing and evolution affect both dust abundance and CO shielding.

%However, simultaneous spatially resolved measurements of metallicity, DGR, and \alphaCO\ within the same galaxies are necessary to test these predictions and to disentangle the effects of metallicity, dust abundance, and local ISM conditions are still limited in the literature. {By scaling out the absolute mass of gas-phase metals, the DMR helps disentangle the inherent degeneracy between absolute chemical enrichment and localized interstellar processing. It serves as a direct diagnostic of the competing lifecycles of dust, tracking the efficiency of secondary interstellar grain growth via gas-phase accretion against grain destruction driven by supernova shocks \citep{Asano2013,Zhukovska2014}. Characterizing the behavior of the DMR on spatially resolved scales within low-mass and low-metallicity galaxies is particularly critical \citep[e.g.,][]{Popping2017,DeVis2019,DeLooze2020,Casasola2022}, as these regimes are important not only for understanding high-z environments, but also capture the exact physical transition where dust evolution shifts from stellar-ejecta-dominated production to rapid interstellar growth.}

Nevertheless, simultaneous spatially resolved measurements of metallicity, DGR, and $\alpha_{\rm CO}$ within the same galaxies remain scarce, limiting our ability to disentangle the relative roles of metallicity, dust abundance, and local ISM conditions. By scaling out the absolute mass of gas-phase metals, the DMR helps break the degeneracy between global chemical enrichment and local interstellar processing. It provides a direct probe of the competing dust lifecycle processes, tracing the balance between grain growth through gas-phase accretion and grain destruction by supernova shocks \citep{Asano2013,Zhukovska2014}. Characterizing the DMR on spatially resolved scales is particularly important in low-mass, low-metallicity galaxies \citep[e.g.,][]{Popping2017,DeVis2019,DeLooze2020,Casasola2022}, as these systems not only provide nearby analogs of high-redshift galaxies but also probe the transition from dust production dominated by stellar ejecta to rapid grain growth in the ISM.

Spatially resolved observations are particularly well suited to address these questions because they probe variations in ISM properties across radial gradients and local environments within individual galaxies, reducing the degeneracies inherent to galaxy-integrated measurements. Integral field unit (IFU) spectroscopy enables gas-phase metallicity maps at sub-kiloparsec resolution from strong optical emission lines, while complementary multiwavelength observations (e.g., infrared dust continuum, CO, and HI 21-cm emission) provide spatially resolved measurements of the DGR and \alphaCO\ on matched physical scales \citep{Bundy2015}.

Pioneering spatially resolved studies by \citet{Sandstrom2013} established that both the DGR and \alphaCO\ exhibit systematic radial trends and correlate with local metallicity. However, their metallicity estimates relied primarily on smooth radial abundance gradients derived from H II regions \citep{Moustakas2010}, preventing them from capturing local or azimuthal variations across galaxy disks. More recent IFU surveys, such as TYPHOON \citep{Park2024} and Metal-THINGS, overcome this limitation by directly mapping gas-phase metallicity at sub-kiloparsec scales. In parallel, the COMING survey \citep{Yajima2021,Yasuda2023} has shown that the DGR--Z and \alphaCO\--Z relations are further modulated by local gas density and star formation rate surface density, while \citet{Chiang2024} demonstrated additional correlations with galaxy properties such as stellar mass, star formation rate, and radiation field strength.

Despite these advances, several key gaps remain. Most spatially resolved studies have focused on either the DGR–Z relation or the \alphaCO–Z relation in isolation, with relatively few works examining all three quantities simultaneously on matched spatial scales within the same galaxies. Also, the interplay between radial gradients in metallicity, DGR, and \alphaCO, and the extent to which these gradients are driven by stellar mass, star formation history, or dynamical processes, has not been systematically characterized across a range of galaxy types and masses. Finally, the role of local variations in UV radiation fields, gas density, and dust grain properties in producing scatter around the mean correlations remains poorly constrained observationally, although simulations suggest these factors are critical \citep{Hu2023,Thompson2024}. The implications of these spatially resolved relationships for global galaxy properties, such as total molecular gas mass estimates and star formation efficiency, have not been fully explored.

{In this study, we investigate how the DGR, DMR, and $\alpha_{\rm CO}$ vary with gas-phase metallicity by combining multiwavelength observations of nearby star-forming galaxies, including archival data of atomic and molecular gas, far-infrared and submillimeter dust measurements, and gas-phase metallicities derived from optical emission-line spectroscopy. The aim is to quantify the interplay between these ISM properties and assess their role in regulating star formation efficiency and the dust lifecycle.} We present a spatially resolved analysis of the relationships between gas-phase metallicity, DGR, DMR, and \alphaCO\ in five nearby SF galaxies: NGC 4214, NGC 2976, NGC 3077, NGC 2403, and NGC 925. Our analysis focuses on how the metallicity, DGR, DMR, and \alphaCO\ vary with galactocentric radius, and whether these radial gradients exhibit systematic trends with galaxy properties. We explore the strengths and scatter of the DGR-Z, DMR-Z and \alphaCO-Z correlations on resolved scales, and we compare these results with those reported in the literature. Finally, we highlight cases in which galaxies in our sample deviate from the established relations and explore how these deviations relate to their physical conditions.

%Finally, we highlight the cases in which our sample do not follow the known relations and how their shape could be associated to their physical condition.

This paper is structured as follows. In Sect. 2 we describe the sample selection, the GCMS IFU observations, and the multiwavelength data compiled from the literature. In Sect. 3, we present the derivation of the galaxy properties used in this work. Sect. 4 describes the main observational results, including radial profiles and the correlations between metallicity and $\alpha_{\rm CO}$, and DGR for each galaxy. In Sect. 5 we compare our results with previous studies and discuss the observed trends. Finally, Sect. 6 summarizes our main conclusions and outlines possible future directions.

\section{Data}\label{SampleSelection}

\subsection{Metal-THINGS}

The IFU data presented in this work are taken from the Metal-THINGS survey \citep{LaraLopez2021}, an ongoing IFU spectroscopic survey whose primary goal is to observe the 34 nearby galaxies of the Very Large Array (VLA) mapped in HI by the THINGS survey \citep{Walter2008}. Our sample comprises five nearby star-forming galaxies selected from Metal-THINGS: NGC~4214, NGC~2976, NGC~3077, NGC~2403, and NGC~925. These galaxies represent the low-mass end of the parent sample and have homogeneous optical IFU, HI, CO, and infrared observations required to derive spatially resolved maps of gas-phase metallicity, DGR, and $\alpha_{\rm CO}$.

\begin{figure}[htbp]
    \centering
    \includegraphics[width=\columnwidth]{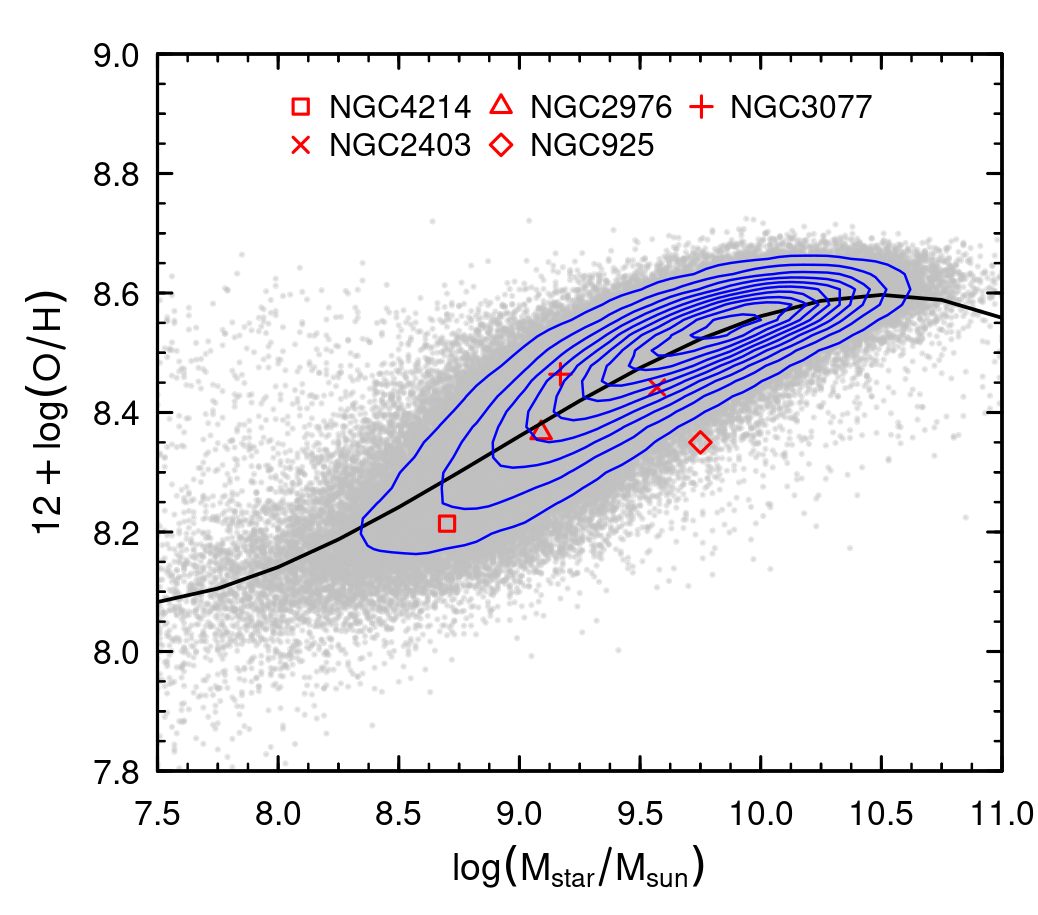}
    \caption{Location of the five Metal-THINGS galaxies on the local stellar mass--metallicity relation. The grey points and blue density contours correspond to the SDSS reference sample, while the solid black curve shows the polynomial fit. The red symbols represent the galaxies analyzed in this work, which populate the low-mass, low-metallicity regime explored in this study.}
    \label{fig:mz_relation}
\end{figure}

Table~\ref{table:sample_properties_part1} summarizes the main properties of the sample. The galaxies span nearly one order of magnitude in stellar mass ($\log M_{\rm star}\sim8.7$--9.8), cover morphologies ranging from dwarf irregulars to late-type spirals, and exhibit different global star formation rates and metallicity gradients. Figure~\ref{fig:mz_relation} shows that all five galaxies lie on the low-mass end of the mass--metallicity relation \citep{Lara-Lopez2013b}, providing an ideal laboratory to investigate the dependence of DGR and $\alpha_{\rm CO}$ on metallicity in the low-metallicity regime. At the common angular resolution adopted in this work, the physical resolution ranges from $\sim$250 to 800 pc, allowing us to probe the ISM on sub-kiloparsec scales.

%The IFU data for this galaxies are taken from the Metal-THINGS survey \citep{LaraLopez2021}, an ongoing IFU spectroscopic survey whose main objective is to observe the 34 nearby galaxies of the Very Large Array (VLA) mapped in HI \citep[The THINGS survey,][]{Walter2008}. Our galaxy sample includes NGC 4214, NGC 2976, NGC 3077, NGC 2403, and NGC 925, which exhibit some of the lowest global stellar masses in the Metal-THINGS sample and possess available data across optical, HI, and CO wavelengths (see Table~\ref{table:sample_properties_part1}).

\begin{table*}
\caption{Parameters of the sample galaxies.}
\label{table:sample_properties_part1}
\small
\setlength{\tabcolsep}{3.5pt}
\centering
\begin{tabular}{lccccccccc}
\hline\hline
\noalign{\smallskip}
\multicolumn{1}{c}{1} & 2 & 3 & 4 & 5 & 6 & 7 & 8 & 9 \\
Galaxy & $D$ & $\log M_{\mathrm{star}}$ & $\mathrm{SFR}_{\mathrm{global}}$ & Morph. & $\nabla(\mathrm{O/H})$ & $\mathrm{O/H}_0$ & R$_{\mathrm{eff, r}}$ & $\text{Res.}$ \\
& [$\mathrm{Mpc}$] & [$\log M_{\odot}$] & [$\mathrm{M}_{\odot}\,\mathrm{yr}^{-1}$] & & [$\mathrm{dex}/R_{\mathrm{eff}}$] & & [$\mathrm{arcsec}$] & [$\mathrm{pc}$] \\
\noalign{\smallskip}
\hline
\noalign{\smallskip}
NGC 4214 & 2.7 & 8.70 & 0.05 & IAB(s)m  & $-0.024$ & 8.214 & 55.78 & 235.6 \\
NGC 2976 & 3.6 & 9.09 & 0.10 & SAc pec  & $-0.041$ & 8.366 & 68.47 & 262.7 \\
NGC 3077 & 3.66 & 9.17 & 0.09 & I0 pec   & $-0.055$ & 8.464 & 38.26 & 314.2 \\
NGC 2403 & 3.01 & 9.57 & 0.85 & SAB(s)cd & $-0.058$ & 8.442 & 91.89 & 319.4 \\
NGC 925  & 6.40 & 9.75 & 1.09 & SAB(s)d  & $-0.043$ & 8.350 & 67.48 & 558.5 \\
\noalign{\smallskip}
\hline
\end{tabular}
\tablefoot{
The columns list: 
(1) Galaxy name; 
(2) Distance in Mpc; 
(3) Logarithm of the stellar mass in solar masses; 
(4) Global star formation rate in solar masses per year; 
(5) Morphological type; 
(6) Metallicity gradient; 
(7) Central metallicity (zero point); 
(8) Effective radius in $r$-band; 
(9) Spatial resolution in parsecs corresponding to an angular resolution of $18''$. These parameters are extracted from the work of \cite{Vale2025}.
}
\end{table*}

The Metal-THINGS survey is being carried out at McDonald Observatory with the 2.7m Harlan Schmidt telescope using the George and Cynthia Mitchell Spectrograph IFU (GCMS), formerly known as VIRUS-P \citep{Hill2008}. The instrument characteristics, observational setup, observing runs, data reduction, and astrometry  are described in detail in \cite{LaraLopez2021, LaraLopez2023, Garduno2023, Pilyugin2025, Vale2025}, and the same procedures are applied to the galaxy sample analyzed here.

The GCMS row-stacked spectra (RSS) are transformed into a three-dimensional data cube using the Python routine 3DCubeGen \citep{IbarraMedel2024_3DCubeGen}. Following previous papers, we use STARLIGHT \citep{Cid2005, Mateus2006, Asari2007} to model the stellar continuum in each spaxel of the flux-calibrated data cube. The best-fitting continuum is then subtracted to isolate the emission-line spectrum, which is analyzed using Gaussian profile fitting following the methodology of \citet{Zinchenko2016, Zinchenko2019a}. Similar to \cite{Garduno2023}, the emission-line flux maps are convolved and reprojected to the resolution of the \textit{Herschel}-SPIRE 250\ $\mu$m band with a {FWHM of 18 arcsec} (see section \S \ref{dustsec} below), using the kernels of \cite{Aniano2011}. SF spaxels are identified using standard BPT diagnostics \citep{BPT1981}, adopting the classification  of \citet{Kauffmann2003} and \citet{Kewley2001}. The resulting BPT diagrams and classification maps are consistent with those presented in Appendices C and D of \citet{Vale2025}.

%We transform the VIRUS-P Row Stacked Spectra (RSS) into a tridimensional data cube using the Python routine called 3DCubeGen (cite). 
%This tool is developed by Ibarra-Medel, H. and is described in his GitHub profile. 
%Similarly to the other papers of the series, we use STARLIGHT \citep{Cid2005, Mateus2006, Asari2007} to fit the stellar continuum to the flux calibrated spectrum in each spaxel of the data cube. After, such stellar continuum is subtracted to the spectrum to measure the emission lines, using Gaussian line profile fittings with the methodology of \citet{Zinchenko2016,Zinchenko2019a}. 

\subsection{THINGS, HERACLES and DustPedia}\label{Dustpedia_section}

We use data at radio wavelengths provided by the THINGS Survey \citep{Walter2008} (FWHM = 7") and The HERA CO Line Extragalactic Survey \citep[HERACLES,][]{Leroy2009} for the $ \prescript{12}{}{{\rm CO}} $ J = (2–1) molecule (FWHM = 13"). 

We take images from DustPedia \citep{Davies2017}, which provides access to multiwavelength imagery and photometry for nearby galaxies. We select the images from the UV band (152.8, 227.1 nm, GALEX) to the mid and far-infrared (3.6, 4.5, 5.8, 8, 24 $\mu$m Spitzer; 70, 100, 160 $\mu$m \textit{Herschel}-PACS; 250 $\mu$m \textit{Herschel}-SPIRE). 

\section{Methodology}\label{physicalproperties}

\subsection{Metallicity estimation}

The gas metallicities are estimated using the S-calibration described in \citet{Pilyugin2016}. We use the term metallicity specifically to refer to the gas-phase oxygen abundances. {The final metallicity maps are shown in Fig. \ref{Z_plots} and preserve the pixel size of 6 arcsec, which is the minimum valid pixel size derived for our dust mass maps.}

\subsection{Dust mass estimation}\label{dustsec}

Dust masses are estimated following the procedure described in \citet{Garduno2023}, which is based on the methodology of \citet{Calzetti2018}. This approach derives spatially resolved dust masses by fitting the spectral energy distribution (SED), constrained by the global dust properties of each galaxy. The method is presented in detail in \citet{Garduno2023}. In brief, we use the Code Investigating GALaxy Emission \citep[CIGALE,][]{Burgarella2005,Nersesian2019} to fit the complete SED from UV to the far infrared using observations of the bands described in section \ref{Dustpedia_section}. 

%For the dust mass estimates, we replicate the same procedure made in \citet{Garduno2023}. This process relies in the methodology of \citet{Calzetti2018} who take the global dust properties of a galaxy to compute the resolved dust masses, fitting the Spectral Energy Distribution (SED).

Prior to the SED fitting, we apply color corrections to the Spitzer, \textit{Herschel}-PACS, and \textit{Herschel}-SPIRE bands. The Spitzer data are additionally corrected for aperture effects. Galactic extinction corrections are applied to the \textit{GALEX} bands using the extinction curve of \citet{Cardelli1989}, and to the Spitzer bands using the extinction law of \citet{Indebetouw2005}.

%Previously, we apply color corrections to the Spitzer, \textit{Herschel}-PACS and \textit{Herschel}-SPIRE bands. The Spitzer bands are also corrected by aperture effects. We also made corrections for galactic extinction to the GALEX bands using the extinction curve of \citet{Cardelli1989}, and for the Spitzer bands we use the curve extinction of \citet{Indebetouw2005}.

To ensure a common spatial resolution, all images are reprojected and convolved to match the resolution of the \textit{Herschel}-SPIRE 250 $\mu$m band. {While our maps are gridded onto a pixel scale of $\sim$6\arcsec (corresponding to the SPIRE $250\,\mu\text{m}$ pixel sampling), the effective angular resolution of our analysis is constrained by the SPIRE $250\,\mu\text{m}$ FWHM of $\sim$18\arcsec. This translates to physical resolution scales ranging from $\sim$ 250 pc to $\sim$ 800 pc across our galaxy sample.} The convolutions are performed using the kernels of \citet{Aniano2011}.

%To have a common spatial resolution, we made a reprojection and convolution to the resolution band of \textit{Herschel}-SPIRE (250 $\mu$m) to all images, with a final pixel size of $\sim$6". The convolutions are performed using the kernels from \citet{Aniano2011}.

CIGALE provides the best-fitting to the SED from which it computes the dust mass (pixel by pixel), adopting the IMF of \cite{Chabrier2003}, the stellar population synthesis models of \cite{BC2003}, the dust attenuation law of \cite{Calzetti2000} and \cite{Leitherer2002}, and the dust emission models of \cite{Draine2014}. The specific parameters adopted are listed in Table \ref{CIGALE_parameters}  {in the Appendix}. The same configuration is used for all galaxies, except for the dust emission parameters, which are adjusted individually {(see  Table \ref{table:dust_grids} in the Appendix.)} {We also adopt this parameter setup following \cite{Garduno2023}, where it was shown to yield high-quality SED fits for the low-mass, low-metallicity galaxy NGC 1569. A discussion on how the selection of alternative dust emission models may influence the derived results is presented in Section \ref{alpha_DGR_SR}.}

%The values reported in Table~\ref{CIGALE_parameters} correspond to NGC~4214; those for the remaining galaxies are provided in the Appendix.

%The values in the modules are the same for all the galaxies except for the Dust emission values that are different for each galaxy (the presented values in the table are for NGC 4214, and the rest for each galaxy can be seen in the Appendix).

%\begin{figure*}
%    \centering
%    \includegraphics[trim={0 1.1cm 0 0cm},clip,width=0.30\linewidth]{Figs/Mdust_repro_r250.png}
%    \includegraphics[trim={0 1.1cm 0 0cm},clip,width=0.30\linewidth]{Figs/Mdust_repro_r250.png}
%    \includegraphics[trim={0 1.1cm 0 0cm},clip,width=0.30\linewidth]{Figs/Mdust_repro_r250.png}
%    \includegraphics[trim={0 1.1cm 0 0cm},clip,width=0.30\linewidth]{Figs/Mdust_repro_r250.png}
%    \includegraphics[trim={0 1.1cm 0 0cm},clip,width=0.30\linewidth]{Figs/Mdust_repro_r250.png}
%    \includegraphics[trim={0 1.1cm 0 0cm},clip,width=0.30\linewidth]{Figs/Mdust_repro_r250.png}
%    \caption{The dust mass maps. The black contours correspond to the galaxy structure in the J band from 2MASS. Each spaxel in the three maps corresponds to a scale of $\sim$6" ($\sim$90 pc).}\label{Mdustmaps}
%\end{figure*}

\subsection{HI and \texorpdfstring{$L_{\rm CO}$}{L CO} estimation}

From the THINGS survey \citep{Walter2008}, we take the HI intensity, which is converted to HI gas mass using equation (5) of \cite{Garduno2023} which considers the flux measured from the moment-0 map (F$_{\rm{HI}}$) and the distance of the galaxy.

%\begin{equation}
%\frac{\rm M_{\rm{HI}}}{\rm{M_{\odot}}}=2.36\times10^5 \left(\frac{\rm D}{\rm{Mpc}}\right)^2\times \frac{\rm F_{\rm{HI}}}{\rm{\rm{Jy\thinspace km/s}}}
%\end{equation}

We estimate the CO luminosity using equation (1) of \cite{Solomon1992}.  This equation considers the rest frequency of the line $230.53\thinspace \mathrm{GHz}$ for CO(2-1) ($\nu_{\mathrm{rest}}$), the distance to the galaxy, the redshift of the galaxy (negligible due to the proximity of the galaxies), and F$_{\mathrm{CO}}$, the velocity-integrated flux measured from the moment-0 map.  We use the CO(2–1) moment-0 maps from the HERACLES survey \citep{Leroy2009}, which are publicly available in processed form.

%As mentioned, we take the CO(2-1) moment-0 map from the data of the HERACLES survey \citep{Leroy2009}, which has already been processed.

%Esta parte no es necesaria
%to use them for science (cite webpage).

%\begin{equation}
%\begin{split}
% \frac{\rm L_{\rm{CO}}}{\mathrm{K}\thinspace \mathrm{km}\thinspace \mathrm{s}^{-1}\thinspace \mathrm{pc}^2} & =
% 3.25\thinspace \times 10^7 \left(\frac{\nu_{\rm{rest}}}{\mathrm{GHz}}\right)^{-2}(1+\rm z)^{-1} \\
% &\times \left(\frac{\rm D}{\mathrm{Mpc}}\right)^2 \left(\frac{\rm F_{\rm{CO}}}{\mathrm{Jy}\thinspace \mathrm{km}\thinspace \mathrm{s}^{-1}}\right),
%\end{split}
%\end{equation}

Both the CO and HI flux maps are reprojected and convolved to the worst resolution when deriving the dust mass (\textit{Herschel}-SPIRE, 250 $\mu$m with a {FWHM $\sim$18\arcsec)} adopting a final pixel size of $\sim$6".

\subsection{DGR and \texorpdfstring{$\alpha_{\rm CO}$}{alpha CO} estimation}\label{gasmassestimates}

This approach combines the atomic (M$_{\rm{HI}}$), molecular (M$_{\rm{H_2}}$), and dust (M$_{\rm{dust}}$) masses based on the relation between these components:

%We follow the method presented in \citet{Leroy2011} and \citet{Sandstrom2013} to derive spatially resolved DGR and $\alpha_{\rm CO}$. This approach combines the atomic (M${\rm HI}$), molecular (M${\rm H_2}$), and dust masses, based on the relation between these components:

\begin{equation}
\frac{\text{M}_{\rm{dust}}}{\rm{DGR}}={{\mu}_{\rm{gal}}(\text{M}_{\rm{HI}}+\alpha_{\rm{CO}}\ \text{L}_{\rm{}CO}}), 
\label{eq_masses}
\end{equation}

where $\mu_{\rm{gal}}$ is the mean atomic weight. This term includes the contribution of hydrogen and other elements to the total gas mass, as defined in \citet{Remy2014}, for which we use the Galactic mass fraction of Helium, Y$_{\odot}=0.270$ \citep{Asplund2009}.

We follow the method presented in \citet{Leroy2011} and \citet{Sandstrom2013} to derive spatially resolved DGR and $\alpha_{\rm{CO}}$. This method relies on the fundamental assumption that the DGR is locally uniform over a given spatial scale. Briefly, the technique solves for $\alpha_{\rm CO}$ (Eq. \ref{eq_masses}) by evaluating the scatter ($\nabla~{\log \mathrm{DGR}}$) across a joint set of pixels as a function of trial $\alpha_{\rm CO}$ values. The final $\alpha_{\rm CO}$ measurement is defined as the minimum of the resulting $\nabla~{\log \mathrm{DGR}}$ curve, corresponding to the value that minimizes variations in the DGR ratio within that region.

This methodology has already been applied to compute global values of the DGR and \alphaCO\ in \cite{Garduno2023}, although in this work we extend such methodology to find spatially resolved values. To compute the spatially resolved maps, we adopt a sliding-window approach with a step size of 1 pixel ($6''$). For each individual pixel, the minimization procedure is performed over a local $3 \times 3$ pixel box ($18'' \times 18''$, matching the FWHM resolution of the dust maps), and the resulting log($\alpha_{\rm CO}$) and log(DGR) values are assigned to the central pixel. This spatial scale ($\sim$250 - 550 pc at the distance of our target galaxies) provides an equilibrium between preserving local structural variations and maintaining the minimum number of pixels to achieve a stable minimization of $\nabla~{\log \mathrm{DGR}}$. Sensitivity tests using $4 \times 4$ pixel boxes yield log($\alpha_{\rm CO}$) and log(DGR) maps nearly identical to our baseline results. 
%However, increasing the box size to $5 \times 5$ or $6 \times 6$ pixels starts to introduce differences in the local values, this could be due to the fact that in larger physical scales the assumption of a uniform local DGR begins to break down.

\begin{figure}[ht!]
    \centering
    {\includegraphics[scale=0.27] {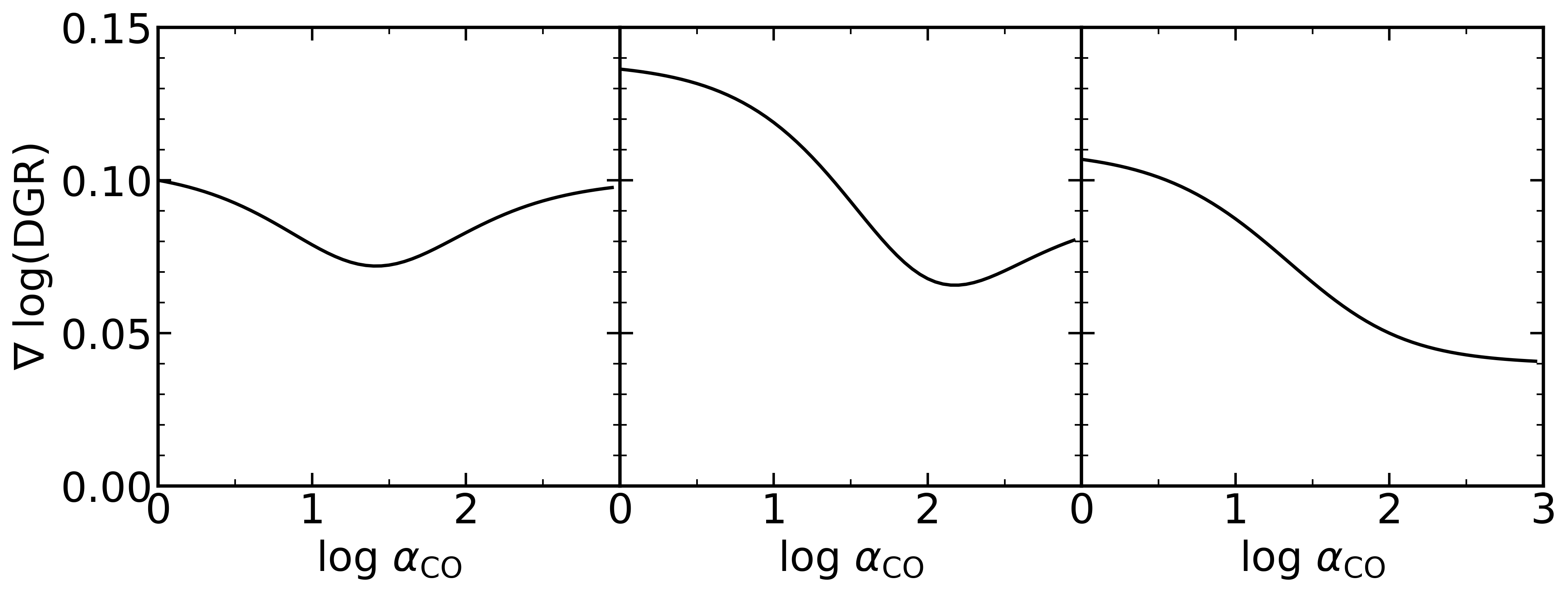}}
     \caption{Pixel solution examples by solving Eq. \ref{eq_masses}. Left and central panels show a proper solution to find the log(\alphaCO) value, while the right panel shows an example in which no solution was found.}
      \label{Pixel_solutions_plot}
\end{figure}

{Fig.~\ref{Pixel_solutions_plot} shows representative minimization curves for selected pixels in NGC 2403 across different galactic regions: a pixel near the galaxy center (left panel), one in the upper-central disk (middle panel), and one in the upper-right outer region (right panel). Pixels for which no valid $\mathrm \alpha_{\text{CO}}$ solution could be obtained are predominantly located in the outer regions of the galaxies, where both CO line emission is faint and dust surface densities ($\mathrm M_{\text{dust}}$) are low. Mathematically and physically, the minimization of $\nabla~{\log \mathrm{DGR}}$ fails to converge to a local minimum in these areas due to two main reasons. First, when CO emission is faint or undetected relative to $\mathrm M_{\text{HI}}$, the $\mathrm \alpha_{\text{CO}} \mathrm L_{\text{CO}}$ term contributes negligibly to the total gas mass. In this regime, increasing $\mathrm \alpha_{\text{CO}}$ merely amplifies observational noise, causing the dispersion to rise monotonically without forming a minimum valley. Second, the technique relies on local spatial variation and contrast between the HI and CO maps to balance against $\mathrm M_{\text{dust}}$. Where dust correlates almost exclusively with HI, or where both gas components present flat, featureless spatial distributions within the fitting box, adjusting $\mathrm \alpha_{\text{CO}}$ cannot reduce the overall scatter, resulting in a flat or non-convergent curve.}

{About 95\% of our spatially resolved solutions in each galaxy have the form of the left and central panel of Fig. \ref{Pixel_solutions_plot}, while for the remaining 5\% no solution was found (right panel of Fig. \ref{Pixel_solutions_plot}).}

\begin{sidewaysfigure*}
    \centering
    \begin{minipage}{0.19\linewidth}\centering \hspace{0.5cm} NGC 4214\end{minipage}
    \begin{minipage}{0.19\linewidth}\centering \hspace{0.5cm} NGC 2976\end{minipage}
    \begin{minipage}{0.19\linewidth}\centering \hspace{0.5cm} NGC 3077\end{minipage}
    \begin{minipage}{0.19\linewidth}\centering \hspace{0.5cm} NGC 2403\end{minipage}
    \begin{minipage}{0.19\linewidth}\centering \hspace{0.5cm} NGC 925\end{minipage}
    \vspace{0.2cm} % Pequeño espacio vertical entre el texto y las imágenes
    
    % Primer bloque: Mapas de Z
    \includegraphics[clip,width=0.19\linewidth]{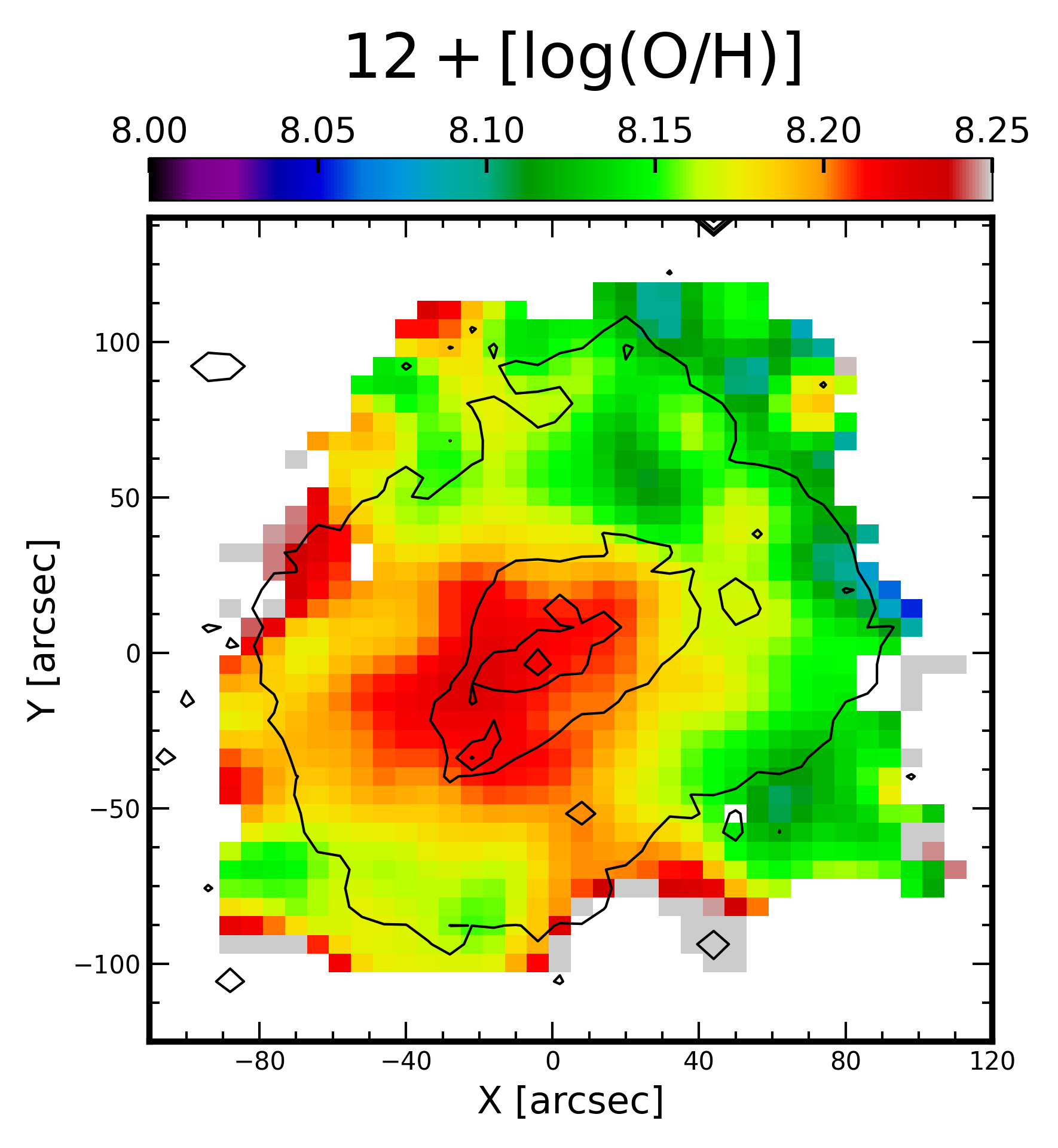}
    \includegraphics[clip,width=0.19\linewidth]{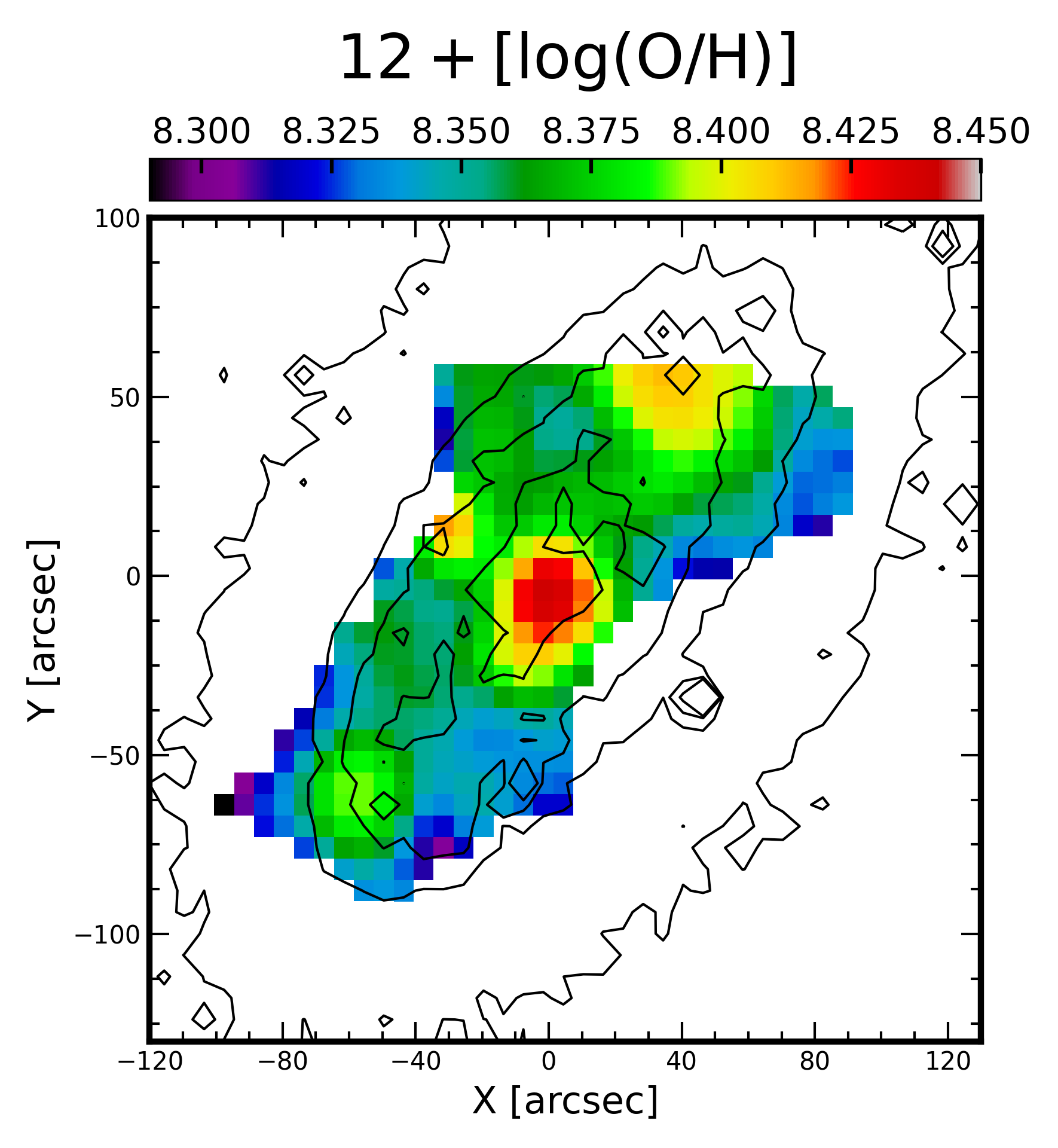}
    \includegraphics[clip,width=0.19\linewidth]{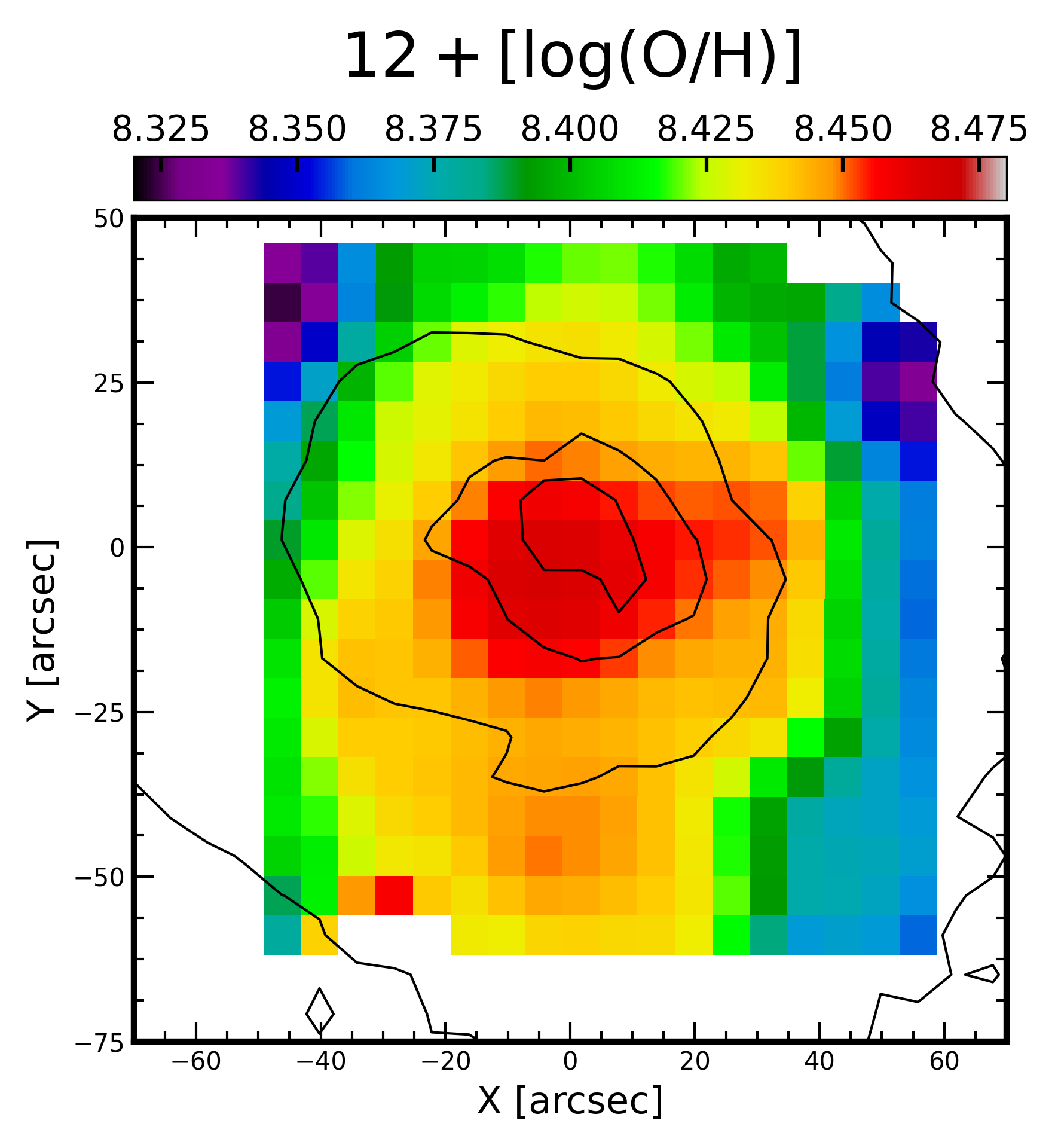}
    \includegraphics[clip,width=0.19\linewidth]{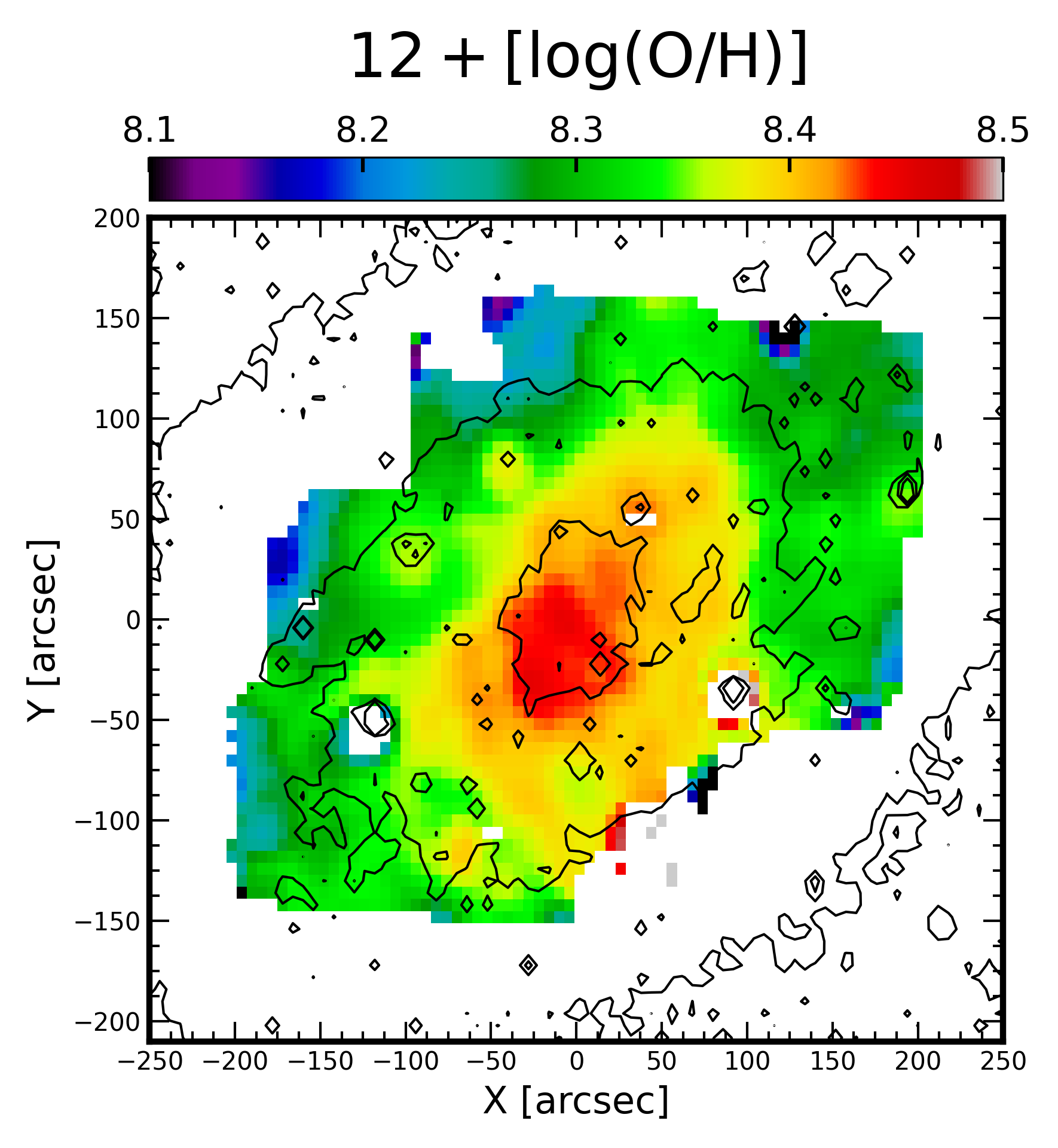}
    \includegraphics[clip,width=0.19\linewidth]{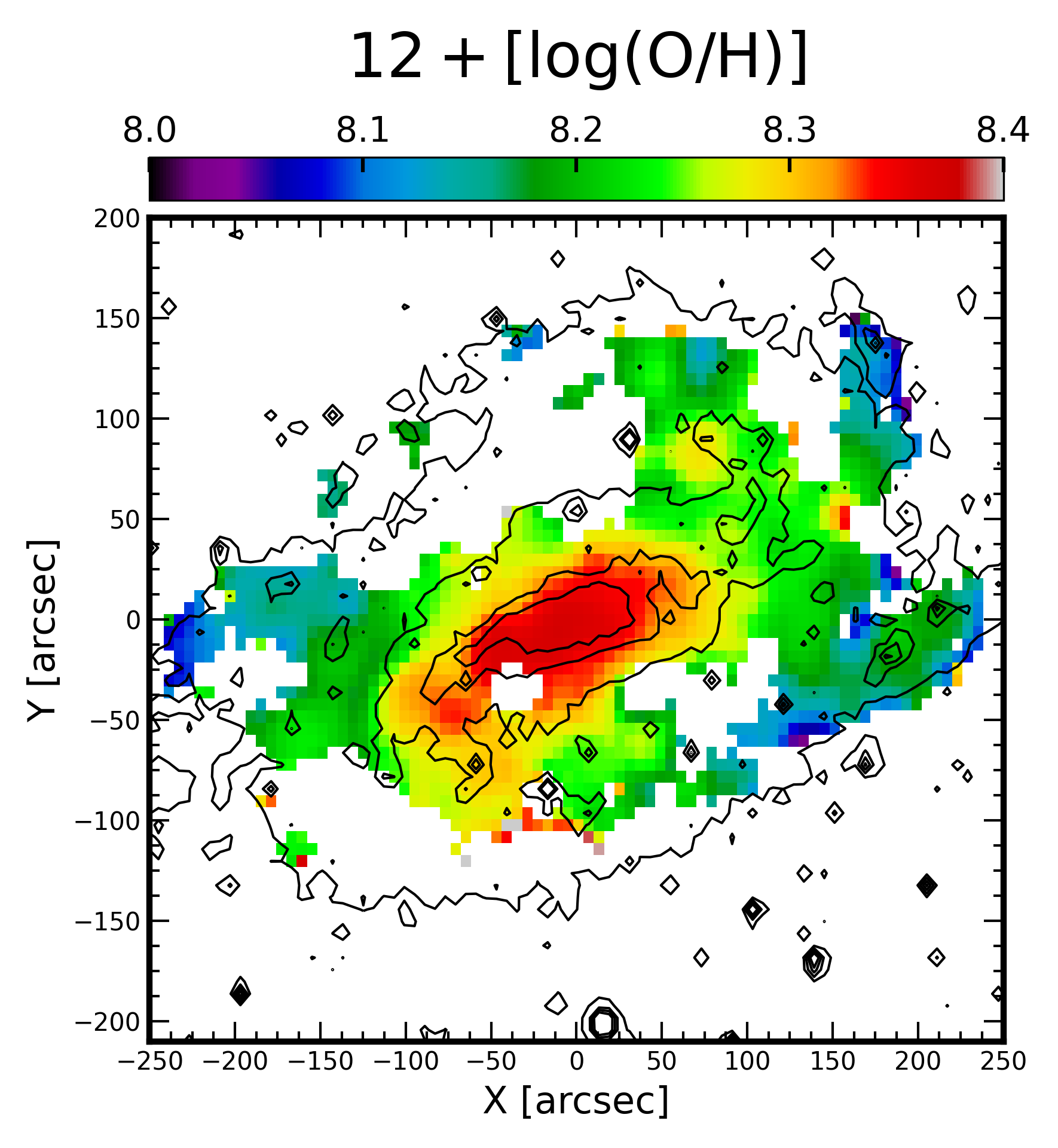}
    \caption{Spatially resolved gas-phase oxygen metallicity maps for the galaxy sample. Each spaxel in the maps has a size of $\sim$6", corresponding to a physical scale of approximately 75-264 pc. The black contours are derived from SDSS g-band image as a reference of the morphology in each galaxy.}
    \label{Z_plots}

    \vspace{1cm} % Espacio entre los dos grupos de mapas
    
    % Primer bloque: Mapas de alpha_CO
    \includegraphics[clip,width=0.19\linewidth]{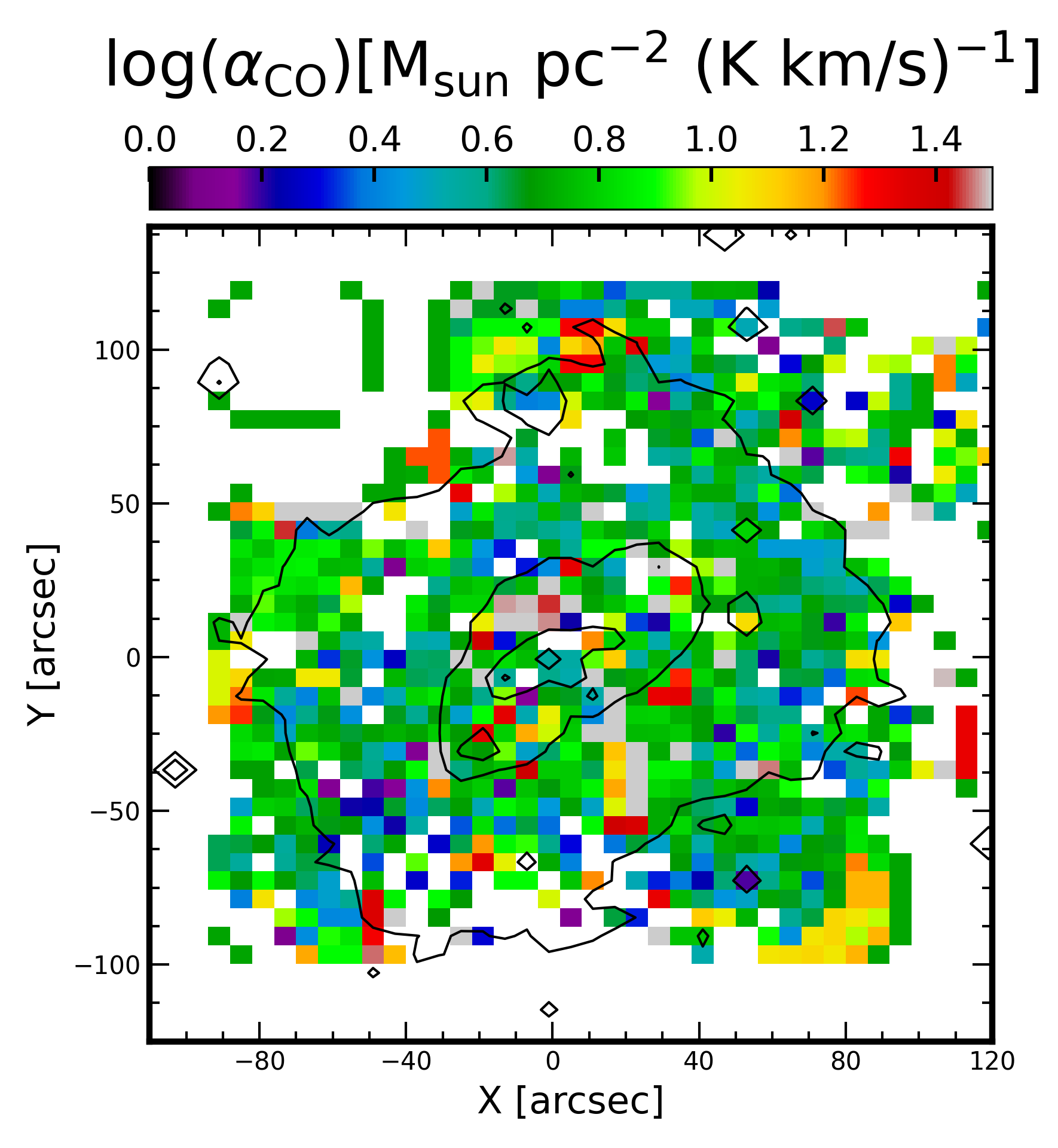}
    \includegraphics[clip,width=0.19\linewidth]{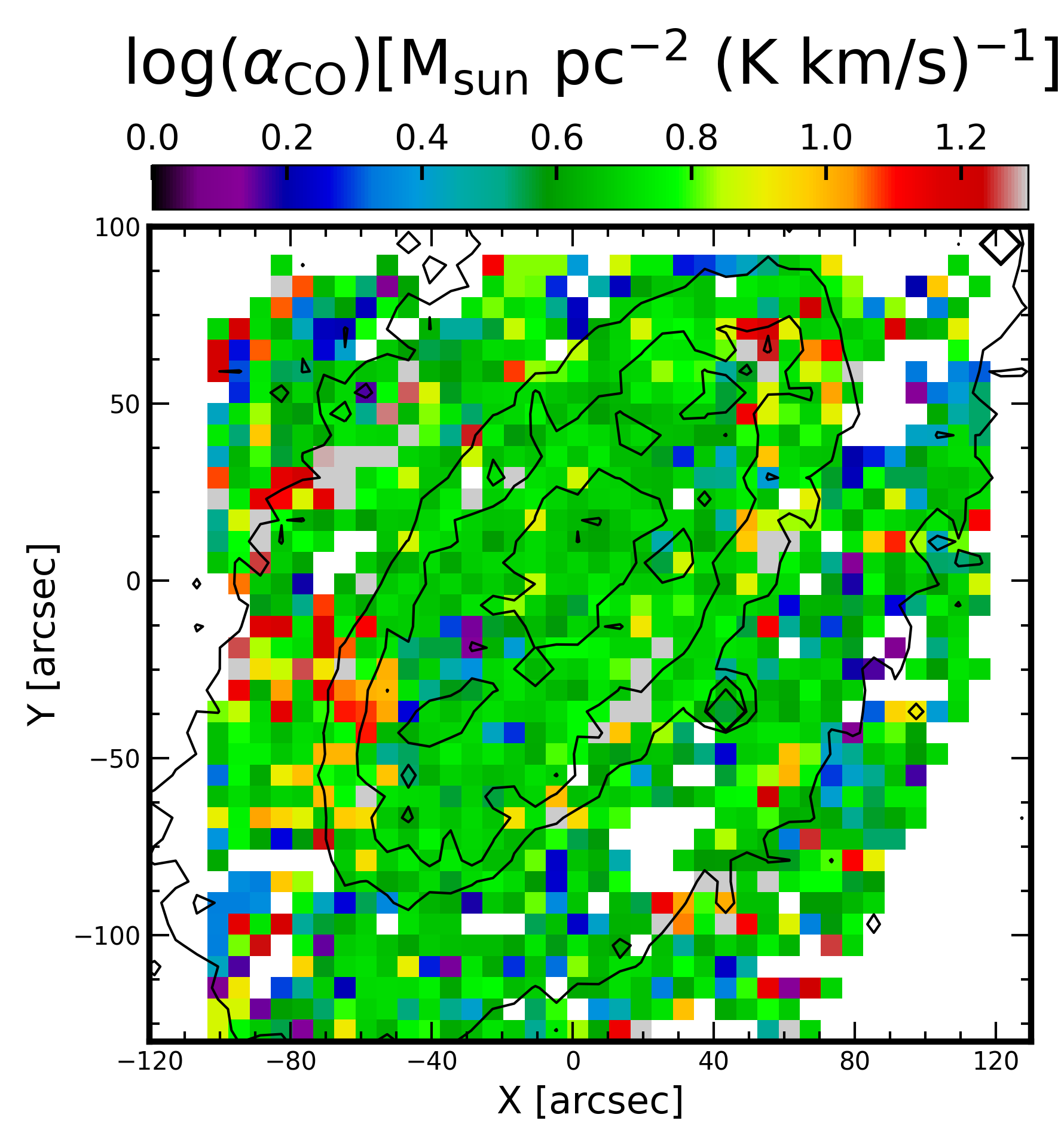}
    \includegraphics[clip,width=0.19\linewidth]{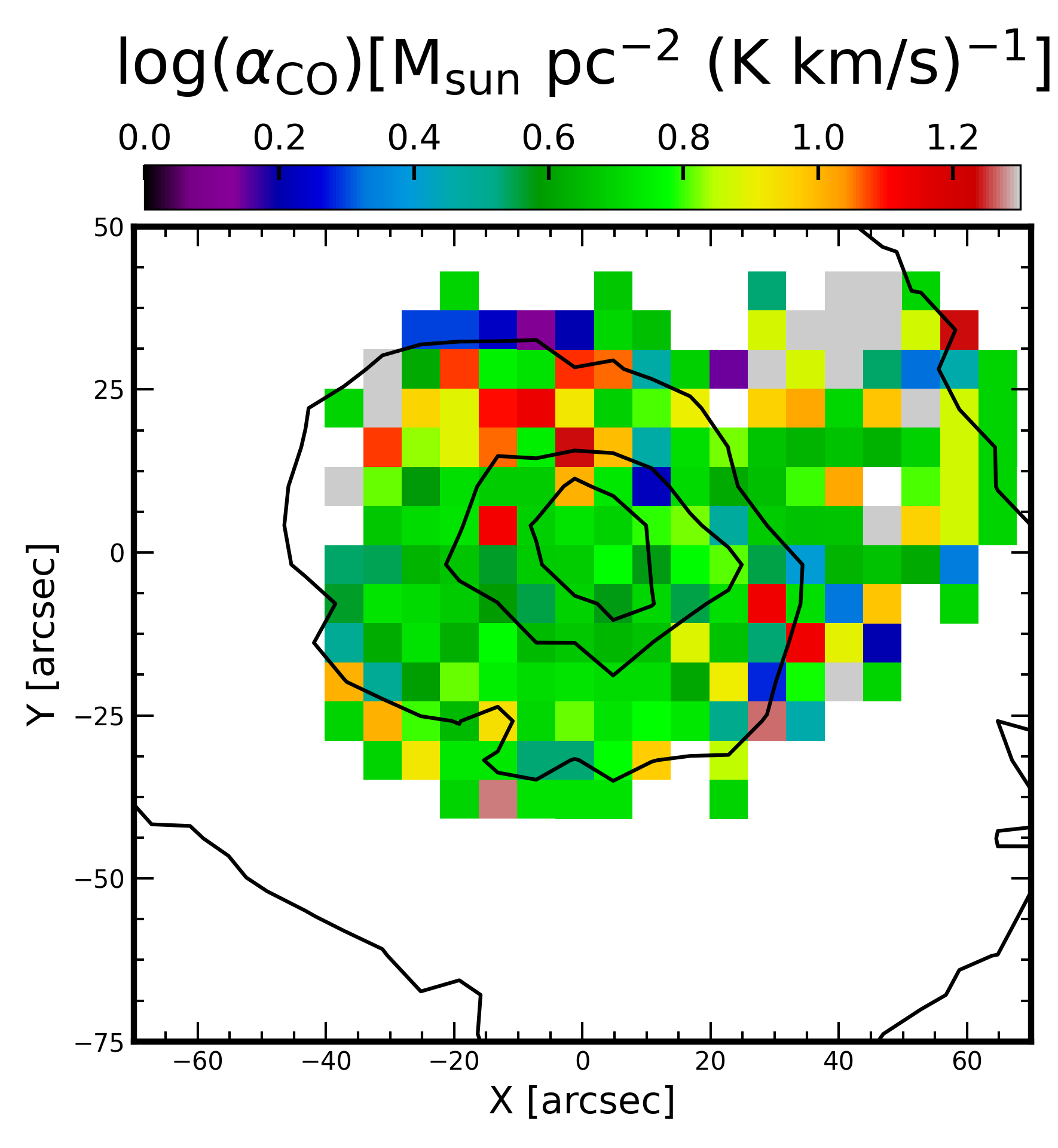}
    \includegraphics[clip,width=0.19\linewidth]{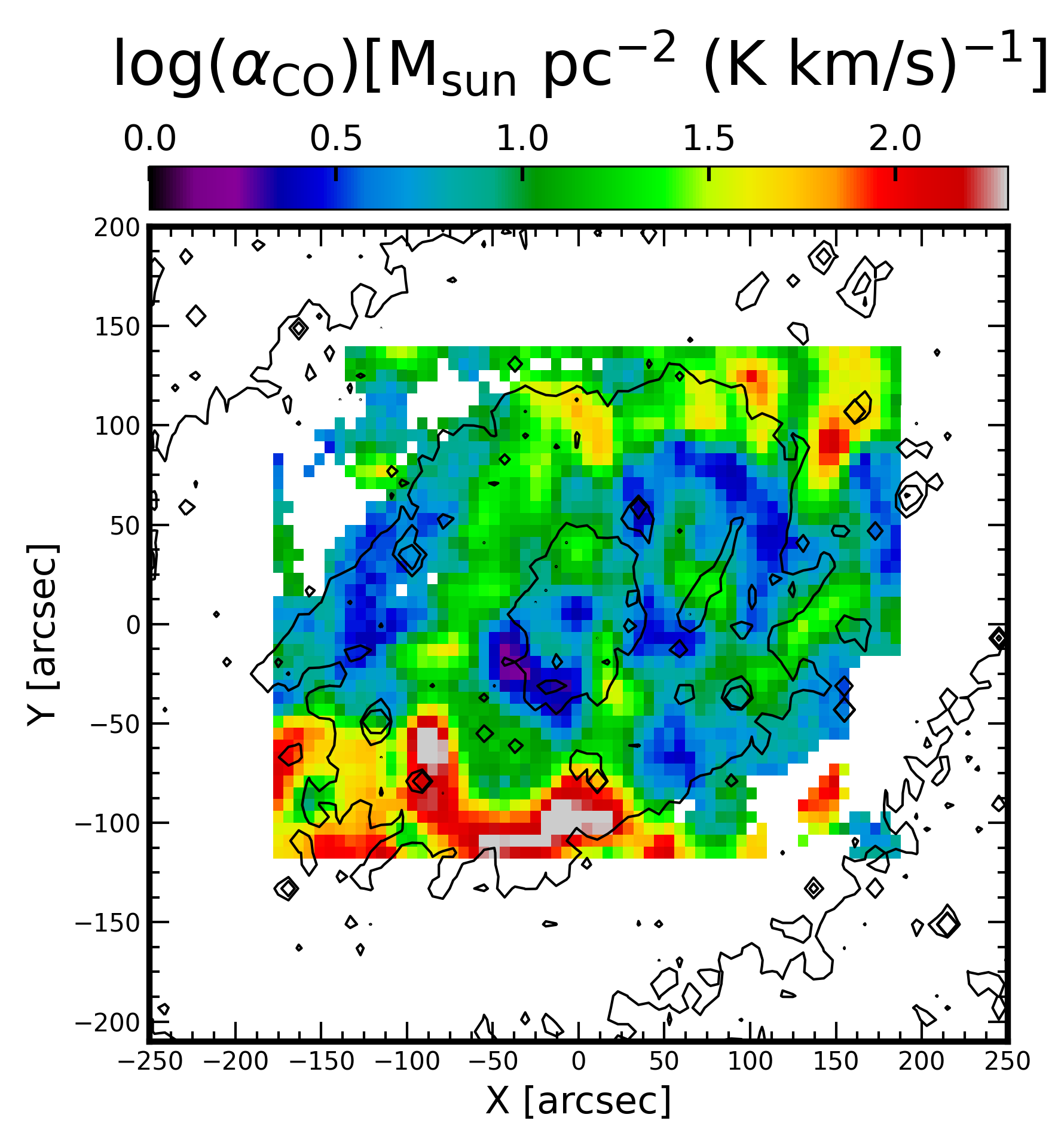}
    \includegraphics[clip,width=0.19\linewidth]{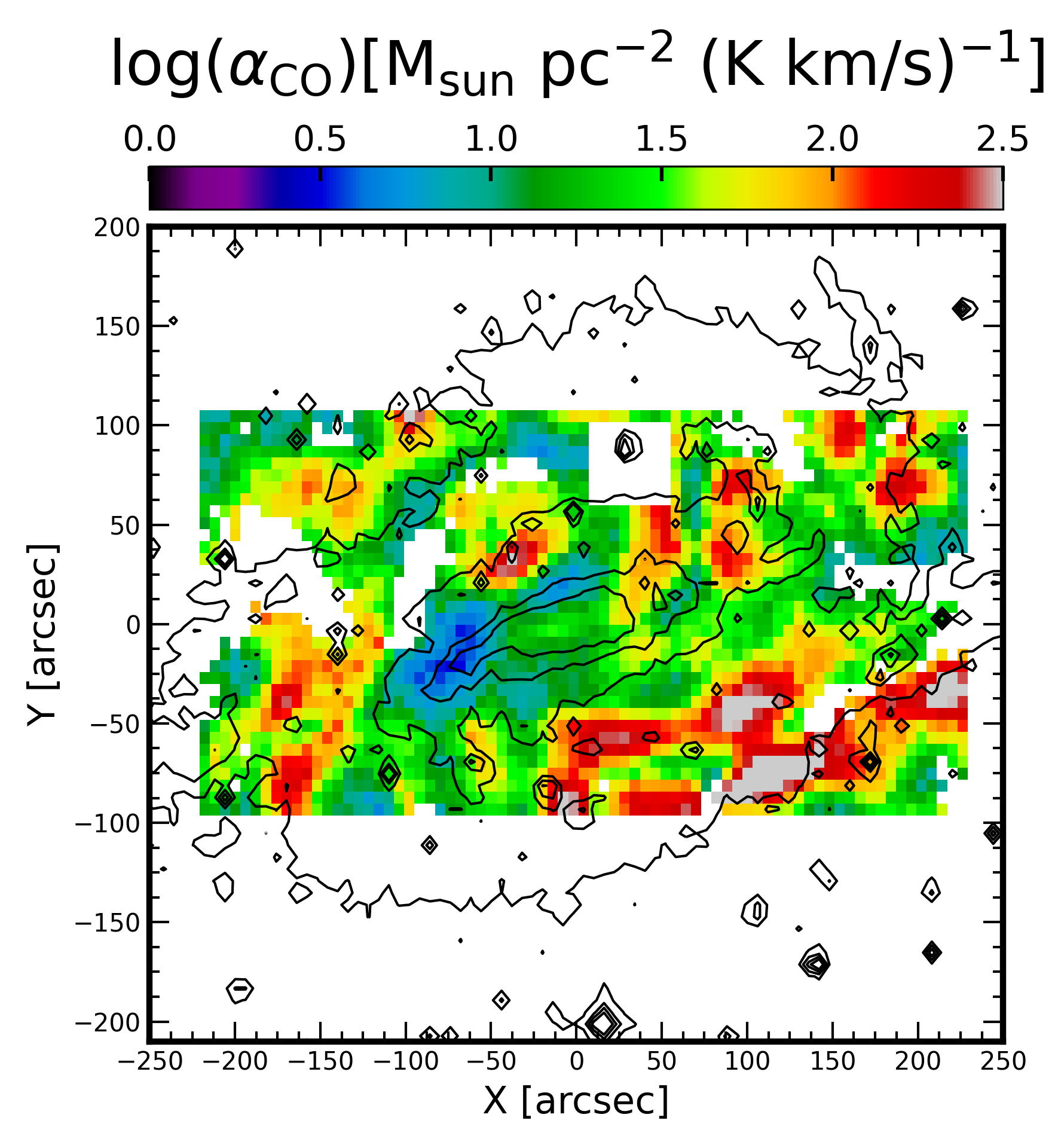}
    \caption{Spatially resolved $\alpha_{\rm{CO}}$ maps. The spatial resolution and reference contours are identical to those described in Fig. 1.}
    \label{alphaCO_plots}

    \vspace{1cm} % Espacio entre los dos grupos de mapas

    % Segundo bloque: Mapas de DGR
    \includegraphics[clip,width=0.19\linewidth]{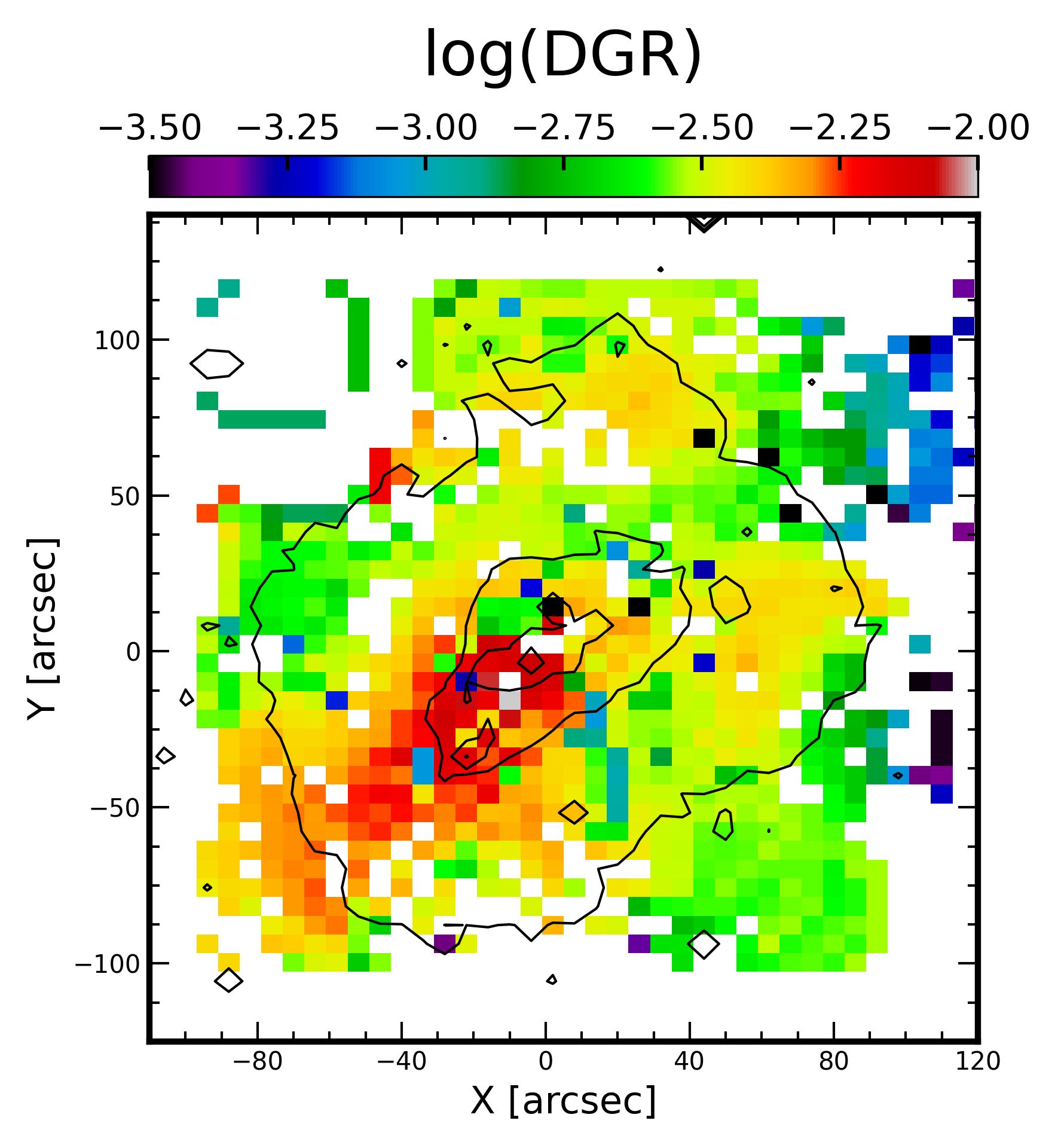}
    \includegraphics[clip,width=0.19\linewidth]{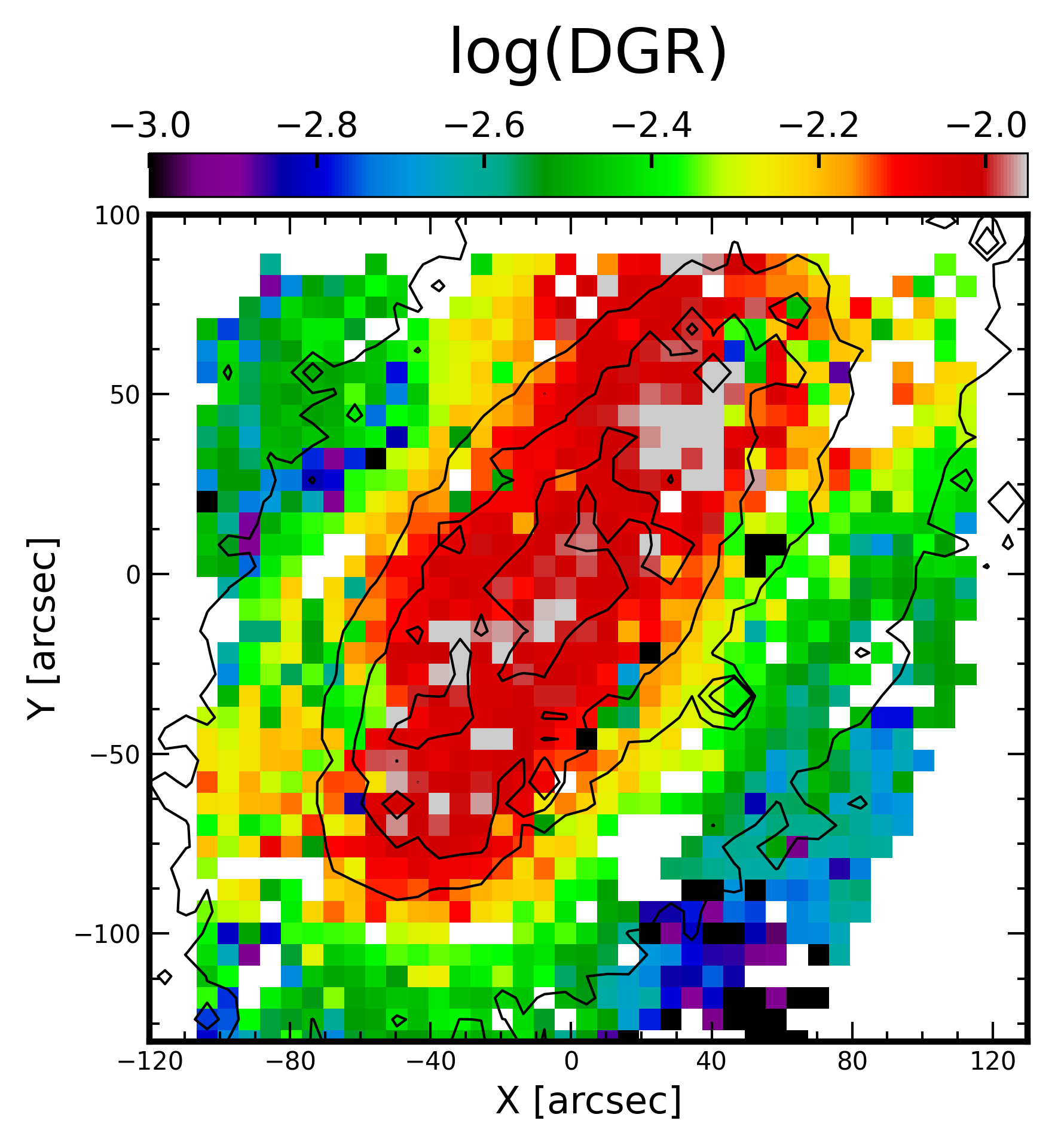}
    \includegraphics[clip,width=0.19\linewidth]{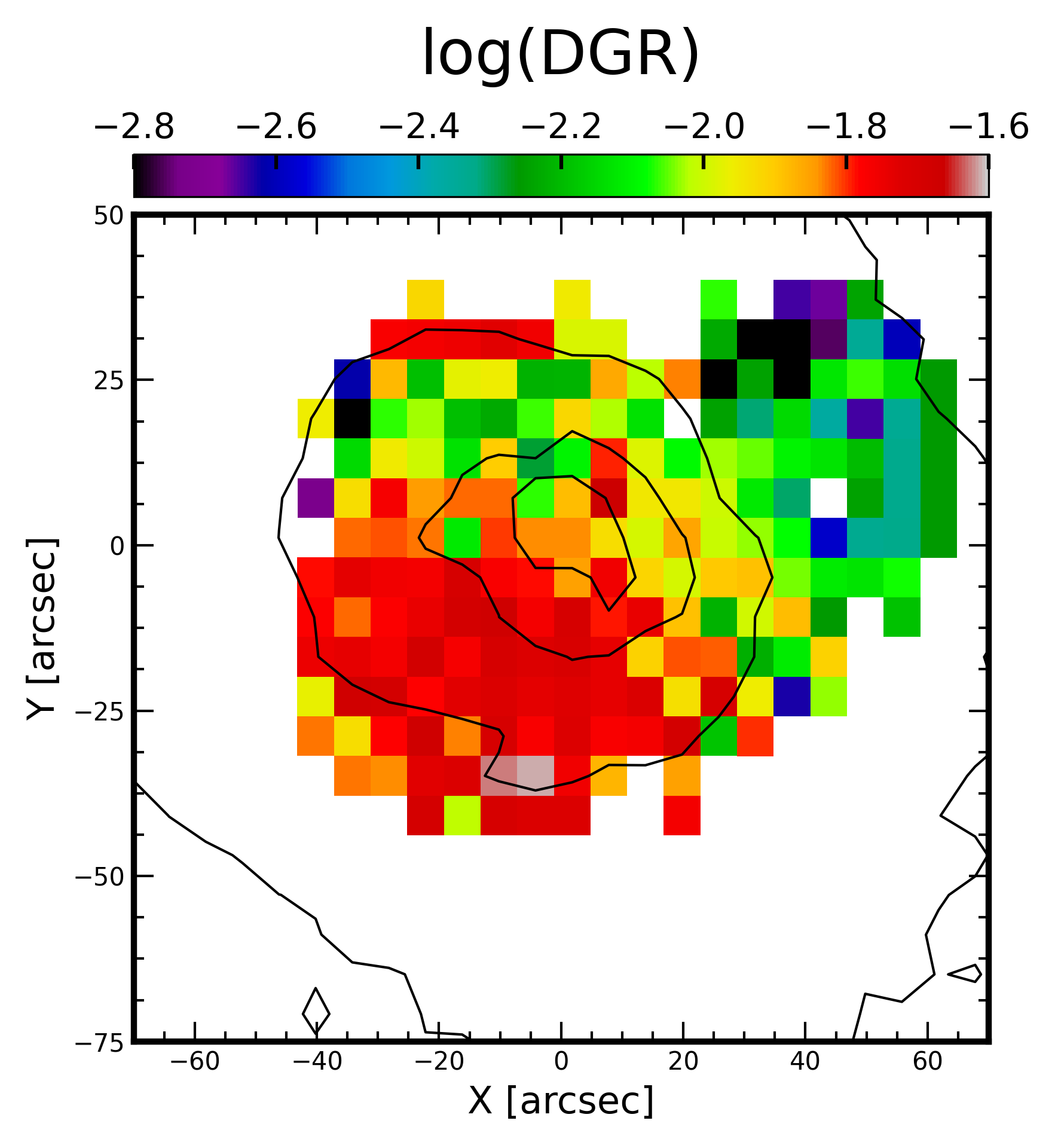}
    \includegraphics[clip,width=0.19\linewidth]{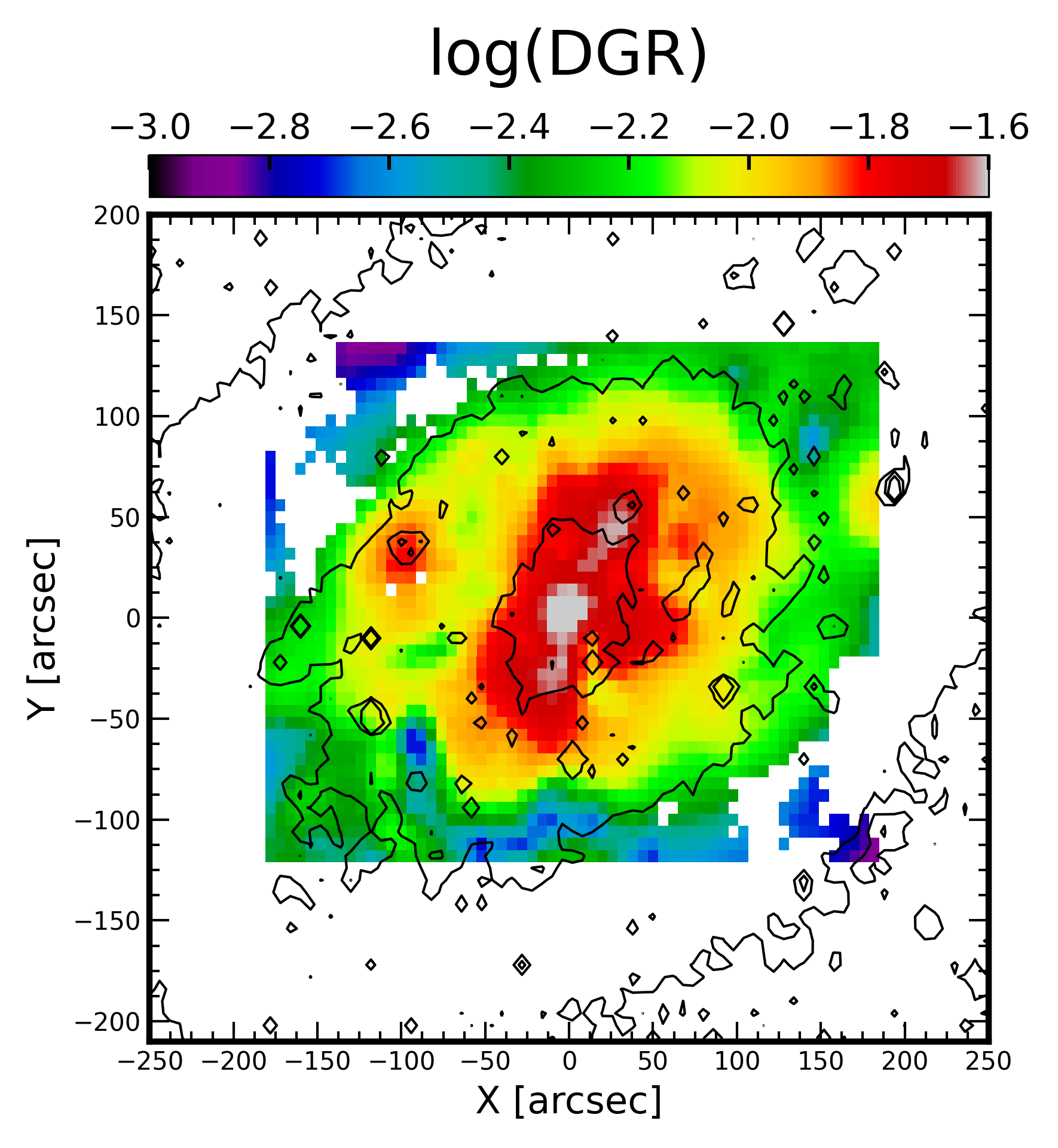}
    \includegraphics[clip,width=0.19\linewidth]{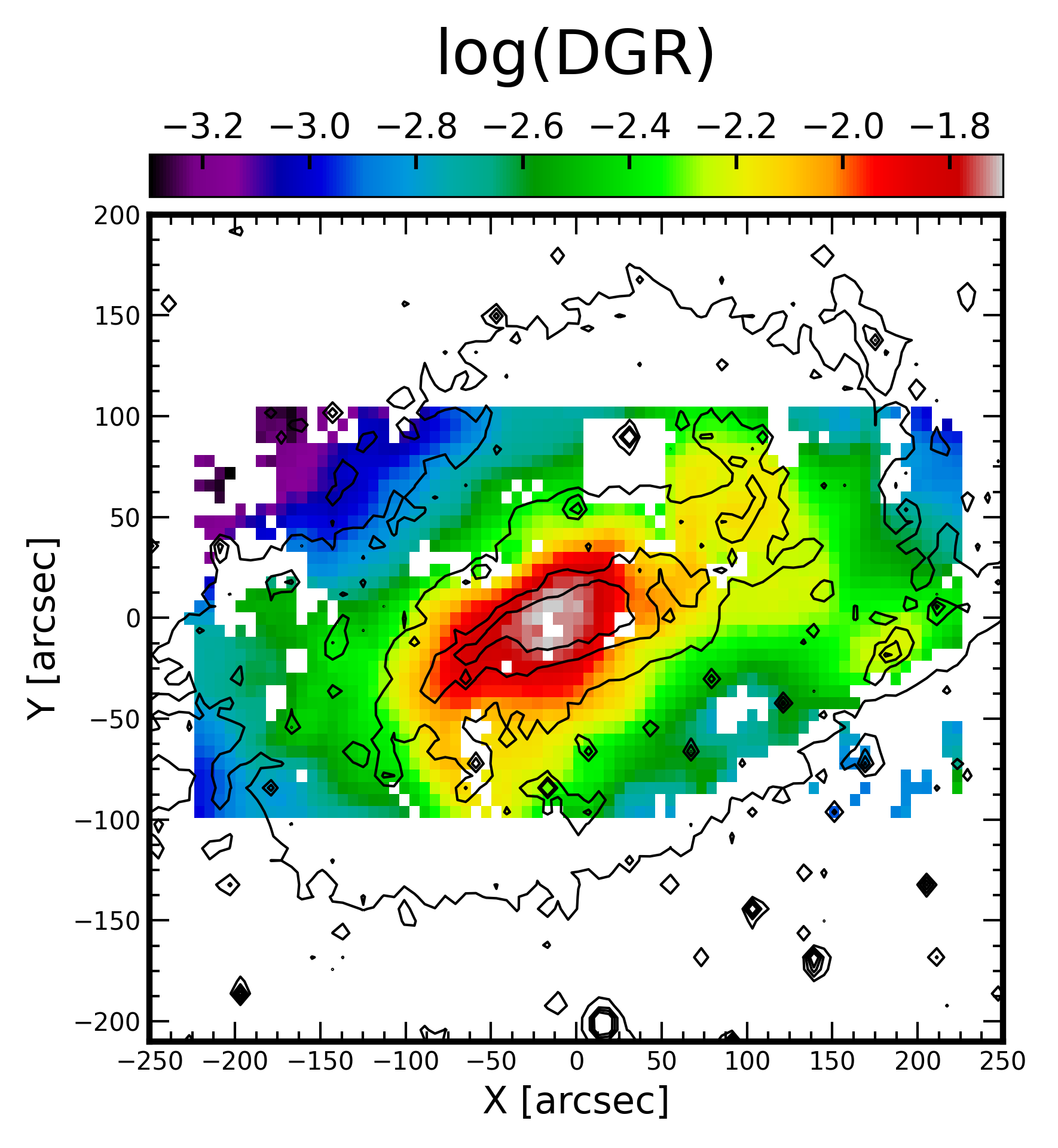}
    \caption{Spatially resolved DGR maps. Spaxel scales and galactic structure contours follow the format established in Fig. 1.}
    \label{DGR_plots}

\end{sidewaysfigure*}

The \alphaCO\ and DGR maps are displayed in Fig. \ref{alphaCO_plots} and \ref{DGR_plots}, respectively. 
For these plots, all DGR and $\alpha_{\rm CO}$ values are shown without applying any cuts based on uncertainties. In contrast, for the derivation of the scaling relations, we retain only measurements that satisfy the following median uncertainty threshold.
{The median uncertainty for the \alphaCO\ values is 0.28~dex (NGC~4214), 0.30~dex (NGC~2976), 0.34~dex (NGC~3077), 0.20~dex (NGC~2403) and 0.23~dex (NGC~925). For the DGR values, the median uncertainty is 0.09~dex (NGC~4214), 0.10~dex (NGC~2976), 0.11~dex (NGC~3077), 0.06~dex (NGC~2403) and 0.06~dex (NGC~925). Additionally, both parameters show a strong anti-correlation, with median Pearson coefficients ($r$) and estimated median covariances of $r = -0.93$, $\text{Cov} = -0.021$ (NGC~4214); $r = -0.97$, $\text{Cov} = -0.022$ (NGC~2976); $r = -0.94$, $\text{Cov} = -0.021$ (NGC~3077); $r = -0.98$, $\text{Cov} = -0.022$ (NGC~2403); and $r = -0.94$, $\text{Cov} = -0.021$ (NGC~925).}

%For these plots, all the values in DGR and \alphaCO\ are shown with any cut in the uncertainties. Later, for the computing of the SR, we select the values with a cut in the uncertainties. 

\subsection{DMR estimation}\label{DMRestimates}

We compute the spatially resolved DMR on a pixel-by-pixel basis, using our previous estimation of spatially resolved dust and gas masses. To ensure a complete census of the chemical enrichment, we determine the DMR by adopting the formulation of { \cite{DeVis2019} (Eq. \ref{eq:dmr}), which accounts for the total metal budget by combining gas-phase metals with those locked inside dust grains ($f_{Z} = 27.36 \times 10^{(12 + \log(\text{O/H}) - 12)}$)}. The underlying gas-phase metal mass fraction is tracked via local oxygen abundances and normalized using the solar mass fraction reference benchmark from \cite{Asplund2009}.

\begin{equation}
\mathrm{DMR} = \frac{\mathrm{\Sigma}_{\mathrm{dust}}}{\mathrm{\Sigma}_{\mathrm{metals}}} = \frac{\mathrm{\Sigma}_{\mathrm{dust}}}{ f_{\mathrm{Z}} \times\mathrm{\Sigma}_{\mathrm{gas}}   + \mathrm{\Sigma}_{\mathrm{dust}}  }
\label{eq:dmr}
\end{equation}

\section{Results}

{Our sample consists of galaxies with diverse morphologies, sizes, galaxy properties, and environmental settings, despite all belonging to the low stellar mass regime (see Table~\ref{table:sample_properties_part1}). Specifically, NGC 4214, NGC 2403, and NGC 925 are isolated galaxies, whereas NGC 2976 and NGC 3077 form part of the M81 group}. However, some patterns appear to be similar, particularly in how the metallicity, $\alpha_{\rm{CO}}$ and DGR are distributed in the galaxies. Figures~\ref{alphaCO_plots} and \ref{DGR_plots} shows the resolved maps of these quantities, providing a first view of the spatial variations that are further quantified through the radial gradients analyzed below.

The metallicity maps show centrally enhanced abundances that decrease toward larger radii, consistent with the negative gradients reported by \citet{Vale2025}. The DGR distributions broadly follow a similar behavior, with decreasing radial trends in NGC~2976, NGC~2403, and NGC~925, while NGC~4214 and NGC~3077 show flatter or more irregular distributions. In contrast, the $\alpha_{\rm CO}$ maps exhibit a stronger galaxy-to-galaxy diversity: NGC~4214, NGC~2976, and NGC~3077 present relatively uniform distributions, whereas NGC~2403 and NGC~925 show more pronounced spatial variations. These trends are quantified through the radial gradients analyzed in the following subsection. 

\subsection{The local relationships of \texorpdfstring{$\alpha_{\rm CO}$}{alpha CO}, DGR and DMR with metallicity} \label{alpha_DGR_SR}

Given the large uncertainties associated with the method used to derive $\alpha_{\rm CO}$ and the DGR (Sect.~\ref{gasmassestimates}), we apply a cut based on measurement uncertainties. For the scaling relations presented below, we retain only values of $\alpha_{\rm CO}$ and DGR with uncertainties below their respective median values (see Sect.~\ref{gasmassestimates}). For the DMR values, we preserve the same spaxels as for the DGR analysis. All the fittings to our data are performed using Bayesian models from the Python packages PyMC v5.16.2 \citep{abril2023pymc} and BAMBI v0.14.0 \citep{JSSv103i15}, with four MCMC chains and 3000 iterations each. {In the Discussion, we also present a Bayesian survival analysis to compare against our baseline methodology and account for non-detections as upper limits. } 

%Because of the method used, to compute the \alphaCO\ and DGR (Sect.\ref{gasmassestimates}), provides very high uncertainties, we applied a cut in the uncertainties. To plot the following scaling relations, we choose the values in \alphaCO\ and DGR with the uncertainty less than their median value (see the values at the end of Sect. \ref{gasmassestimates}).

\begin{figure*}[ht!]
    \centering
    {\includegraphics[scale=0.6] {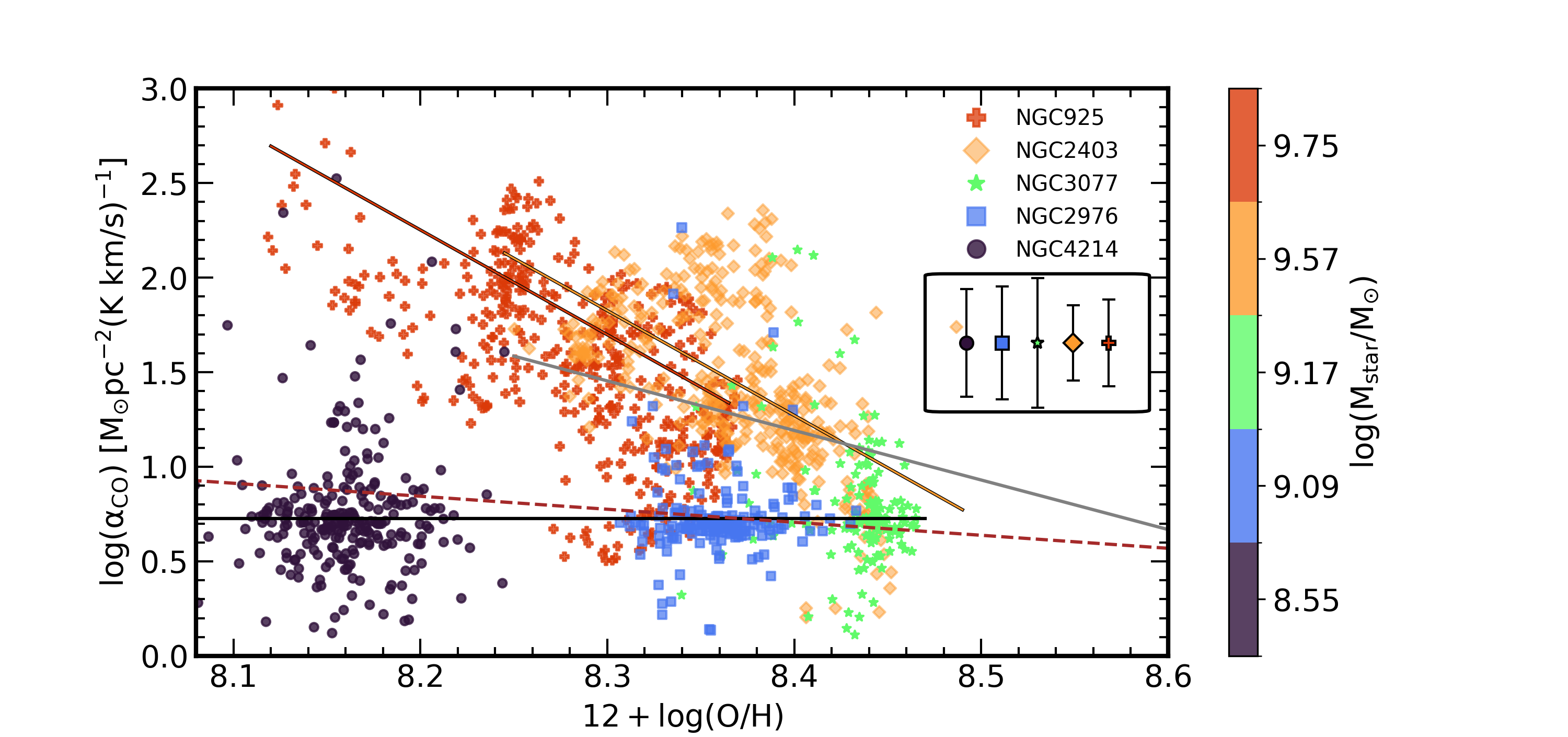}}
     \caption{Spatially resolved \alphaCO\ as a function of metallicity. Data points are color-coded by the global stellar mass of each galaxy, as indicated in the color bar. The solid orange and red lines denote the linear fits for NGC 2403 and NGC 925, respectively. The horizontal black line represents the fitting observed for the lower-mass galaxies NGC 4214, NGC 2976, and NGC 3077. The dashed brown line corresponds to the re-calibrated relation from \cite{Sandstrom2013} and the solid gray line corresponds to the fitting of \cite{Chiang2024}. For clarity, the transparent box shows the median \alphaCO\ error bar for each galaxy.}
      \label{Z_alpha_plot}
\end{figure*}

The relationship between gas metallicity and \alphaCO\  (Fig. \ref{Z_alpha_plot}) reveals clear patterns. For NGC 4214, NGC 2976 and NGC 3077, a flat relation is seen around a median value of log(\alphaCO) $\sim$ 0.73 across all the metallicity range, with an RMSE of 0.32 dex.
In contrast, with NGC 2403 and NGC 925 show a negative correlation is found, with an RMSE of 0.39 and 0.44 dex, respectively. Over the full metallicity range (12 + log(O/H) $\sim$ 8.05 – 8.5), log($\alpha_{\rm CO}$) decreases from $\sim$2.5 at the lowest metallicities to $\sim$0.5 at the highest, corresponding to a variation of $\sim$2 dex. At fixed metallicity, individual galaxies also exhibit a large dispersion.

%In a global overview, across the full metallicity range (12+log(O/H) $\sim$ 7.9-8.5), the log(\alphaCO) decreases from $\sim$ 2.5 at the lowest metallicities to $\sim$ 0.5 at the highest metallicities, representing a factor of variation $\sim$ 5 ($\Delta$ log(\alphaCO) $\sim$ 2). For individual galaxies, there is also a high dispersion which is more evident. 

Each galaxy traces a distinct sequence in the $\alpha_{\rm CO}$-Z relation. For example, NGC~4214 (the lowest-mass galaxy) occupies the lowest metallicity range (12 + log(O/H) $\sim$ 8.05 – 8.33), while NGC~2403 (the second most massive galaxy in the sample) spans the highest metallicity range (12 + log(O/H) $\sim$ 8.25 – 8.5).

%Each galaxy trace a distinct sequence in the metallicity - \alphaCO\ relation. For example, NGC 4214 (with the lowest global stellar mass) occupies the lowest metallicity range (12 + log(O/H) $\sim$ 7.9-8.2) but NGC 2403 (with the highest global stellar mass) spans the highest metallicity range (12 + log(O/H) $\sim$ 8.3-8.5). 

The dashed brown line in Fig.~\ref{Z_alpha_plot} shows the fit from \citet{Sandstrom2013}, recalibrated to the metallicity scale adopted in this work (see {Appendix \ref{recal}}). Overall, this relation is consistent with the trends observed for NGC~4214, NGC~2976, and NGC~3077. The gray line corresponds to the fit from \citet{Chiang2024}, which better matches the slope observed in the higher-mass galaxies of our sample. 
Regarding the level of scatter, our results are consistent with both \citet{Sandstrom2013} and \citet{Chiang2024}, depending on the stellar mass range considered within our sample.

%\textcolor{orange}{Indeed, the log(\alphaCO) values in the sample of \citet{Chiang2024} remain in the range of log(\alphaCO) $>$ 1, while our values populate the range of 0.0 $>$ log(\alphaCO) $>$ 1 in which our low-mass galaxy sample are. Regarding to the level of scattering, we are in agreement both \citet{Sandstrom2013} and \citet{Chiang2024}.}

%The dotted brown line in  Fig. \ref{Z_alpha_plot} corresponds to the fitting of \cite{Sandstrom2013} which was re-calibrated to the metallicity estimations of this work (see Apendix XX). In general, it can be seen an agreement between the fitting of \cite{Sandstrom2013} and the galaxies NGC 4214, NGC 2976 and NGC 3077. Also, the dotted gray line is the fitting of \cite{Chiang2024} which seems to be in agreement with the slope of the high stellar mass galaxies in our sample.

%In Fig. \ref{Z_DGR_plot} can be seen the relation between metallicity and DGR that shows a positive correlation in the galaxy properties.

\begin{figure*}[ht!]
    \centering
    {\includegraphics[scale=0.6]{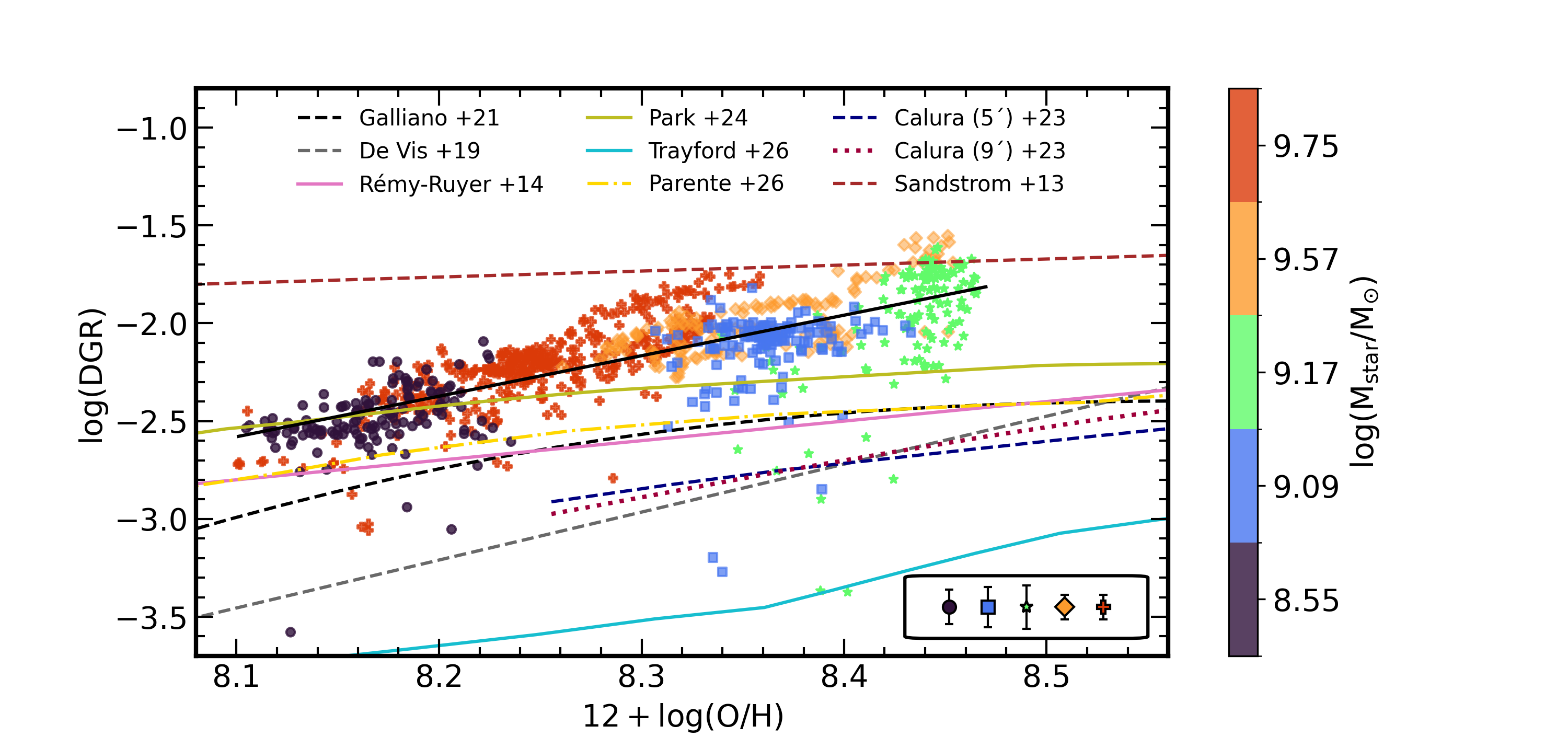}}
     \caption{Spatially resolved DGR versus metallicity. Symbols and color-coding follow the same convention as in Fig. 5. The solid black line indicates our best-fit relation for the full sample. To contextualize these results, several observational, theoretical, and semi-analytical relations from the literature are overplotted (indicated by the various lines; see legend for references). Similar to previous figures, the error bars correspond to the median uncertainty in log(DGR).}
      \label{Z_DGR_plot}
\end{figure*}

Figure~\ref{Z_DGR_plot} shows the relation between the gas metallicity and the DGR, revealing a positive correlation across all galaxies. Over the full metallicity range sampled, log(DGR) increases from the values $\sim$-3 to $\sim$-1.5, corresponding to a variation of $\sim$1.5 dex (a factor of $\sim$30). Overall, the correlation exhibits a moderate dispersion, with an RMSE of 0.18 dex.

%Figure~\ref{Z_DGR_plot} shows the relation between metallicity and DGR, which shows a positive correlation in which all galaxies align. In the full metallicity range, log(DGR) increases globally from $\sim$-3 to -1.5, representing an enhancement factor of $\sim$2. Globally, the strong correlation presents a moderate dispersion, \textcolor{red}{with an RMSE of XX}

%although each galaxy has some values far from the tend (e.g. the values from $\sim$-3.5 to -3.0) 

%Esta te la comento por que no entiendo bien lo que quieres decir. En todo caso, te pongo una frase mas corta
%NGC 4214, NGC 2976 and NGC 3077 keep a position in the relation from the lowest to the highest values in metallicity while their global stellar mass enhance. For the case of NGC 925 and NGC 2403, both also follow the correlation, but their position in the plot does not have a relation with the increment in their global stellar mass.

As before, the dashed brown line represents the \citet{Sandstrom2013} fit, re-calibrated to the metallicity scale adopted in this work. In this case, their relation deviates significantly from our data, with the largest discrepancies occurring at low metallicities (12 + log(O/H) $<$ 8.3).

To provide broader context for this scaling relation, the data displayed in Fig. \ref{Z_DGR_plot} compare our findings against a comprehensive compilation of existing literature. The trends shown in the DGR-Z relation draw upon a wide variety of methodologies. In particular, our data are in agreement with the observational results of \citet{Sandstrom2013} and \citet{Park2024}, which used empirical calibrations of gas tracers, multi-wavelength dust continuum emission modeling and extensive local galaxy surveys. \cite{Sandstrom2013} test most of the THINGS sample, while the galaxies analyzed in \cite{Park2024} are part of the TYPHOON survey with a global stellar mass range between 7.90 $<$ log (\Ms/M$_\odot$) $<$ 10.83. 

%Again, the dashed brown line is the fitting of \cite{Sandstrom2013} (re-calibrated to the metallicity estimations of this work). For this case, such fitting is far from the data, showing the strongest deviation in the low metallicity range of $\sim$ 7.9-8.2)

%The dashed brown line represents the relation from \cite{Sandstrom2013} re-calibrated to our metallicity scale.

\begin{figure*}[ht!]
    \centering
    {\includegraphics[scale=0.6]{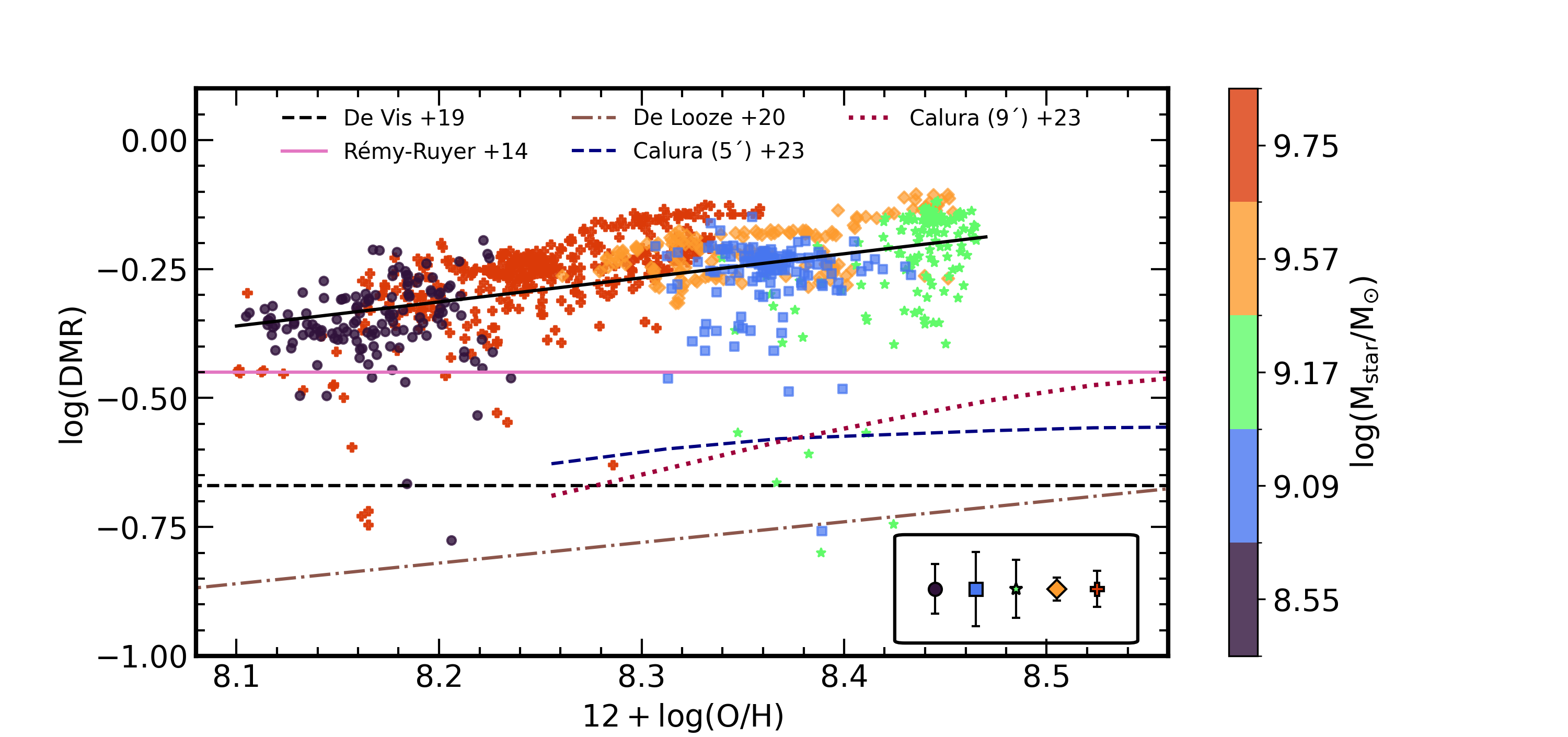}}
     \caption{Spatially resolved DMR versus metallicity. Symbols and color-coding follow the same convention as in Fig. 5. The solid black line indicates the best-fit relation for the full sample. Similar to previous figure several models are plotted in order to compare with the literature. {Similar to previous figures, the error bars correspond to the median uncertainty in  log(DMR).}}
      \label{Z_DMR_plot}
\end{figure*}

Fig.~\ref{Z_DMR_plot} displays the spatially resolved relationship between the log(DMR) and the gas-phase oxygen abundance across our full galaxy sample. We find a clear and statistically significant positive correlation between local metallicity and the fraction of metals locked into dust grains, with a small scatter RMSE = 0.11 dex. In this case, the dashed black line represents the \citet{DeVis2019} fitting, computed for the DustPedia galaxies. As seen in the figure, our DMR values lie above not only of such fitting, but also with the other references in the literature. Later in the Discussion, we comment the origin of this difference in comparison with previous works.

We refer to \cite{Tailor2026}, who present a comprehensive and detailed compilation of the existing literature, which serves as the motivation for our data comparison in Fig. \ref{Z_DGR_plot} and \ref{Z_DMR_plot}. As a short summary, the works in such figures can be broadly divided into empirical scaling relations derived from galaxy surveys and theoretical evolutionary frameworks. As seen along this work, \cite{Sandstrom2013} represents a baseline linear DGR–$Z$ relation computed on kiloparsec scales for SF disks, which assumes a fixed fraction of condensed metals. In contrast, \cite{Remy2014} and \cite{Park2024} implement a broken power-law relation to capture a sharp, non-linear drop in dust abundance at sub-solar metallicities ($12+\log(\text{O/H}) \lesssim 8.0$), highlighting a regime where dust production is dominated by stellar sources due to inefficient interstellar grain growth. With a different approach \cite{DeVis2019} use a large DustPedia sample to model the DGR–$Z$ relation as a smooth single power law, while illustrating that the DMR increases with enrichment before saturating into a plateau for metal-rich environments. This high-metallicity saturation and flat DMR trend is similarly supported by the unresolved global datasets of \cite{DeLooze2020} and the multi-wavelength systematic calibrations of \cite{Galliano2021}, both reflecting a phase where dust destruction and gas-phase accretion reach an interstellar equilibrium.

On the theoretical side, the multi-zone chemical evolution dust models of \cite{Calura2023}, track an equilibrium lifecycle parameterization optimized for spiral galaxies like NGC 628 (Model 5’), while Model 9’ represents a rapid, growth-dominated accretion regime that predicts a steeper scaling slope. The predictions from recent hydrodynamical and cosmological simulations, the COLIBRE simulation block \citep{Trayford2026}, incorporates an explicit element-by-element multi-species grain lifecycle (accounting for seeding, shattering, and sputtering), while the co-evolutionary dust-chemistry models of \cite{Parente2026} track how local gas density and supernova feedback regulate localized DGR ratios across cosmic time. {Predictions from such cosmological simulations (e.g., EAGLE, COLIBRE) carry systematic uncertainties, as nucleosynthetic yields are uncertain by factors of a few and elemental abundance evolution is heavily dependent on the specific choice of yield tables, this may partially account for discrepancies between model predictions and observational data.}

{When comparing our spatially resolved DGR and DMR ratios to global literature trends, we observe a systematic offset, with our values sitting higher than several established relations \citep[e.g.,][]{DeVis2019,Galliano2021}. This variation is driven primarily by differences in the chosen dust mass modeling methodologies and wavelength coverage. Our methodology utilizes CIGALE with a Draine dust prescription and relies on {\it{Herschel}}-SPIRE bands up to $250\,\mu\text{m}$ to maximize spatial resolution, making our measurements highly sensitive to the warmer dust components. In contrast, many literature scaling relations incorporate sub-millimeter data up to $500\,\mu\text{m}$ or $850\,\mu\text{m}$ to trace the colder dust reservoirs. Furthermore, the systematic differences between underlying dust models are significant, our approach utilizes a CIGALE/Draine prescription, whereas many comparative studies employ other models, e.g. the THEMIS model, which is known to yield dust masses approximately a factor of 2 to 3 lower \citep{Nersesian2019}.}

Because our calculated DMR is a monotonically increasing function of the DGR ($\text{DMR} = \text{DGR} / (\text{DGR} + f_Z)$), the higher dust masses inferred from our models naturally propagate into higher localized DMR values. Our resolved maps recover local DMR values reaching up to approximately 60\% in certain regions. This approaches the theoretical maximum limit ($\text{DMR} \approx 0.6$) for the fraction of metals that can be physically locked into dust grains \citep{Konstantopoulou2022,Palla2024}. Reaching this asymptotic limit highlights the efficiency of dust condensation in these specific environments, though it also underscores the sensitivity of these fractions to the adopted dust model. {In particular, low-mass galaxies frequently exhibit a sub-millimeter excess appearing at wavelengths $\ge 500\,\mu\text{m}$, which is indicative of a dust composition diverging from standard Milky Way properties. If models are constrained only by data up to $\sim 250\,\mu\text{m}$, this long-wavelength excess can be missed, carrying significant implications for the derived total dust mass. Within our sample, evidence of a sub-millimeter or millimeter excess has been reported for NGC~4214 \citep{Kiuchi2004} and NGC~2976 \citep{Ejlali2025}. Furthermore, an excess is potentially present in NGC~3077 but appears absent in NGC~0925 \citep{Kirkpatrick2013}, while it has been speculated in NGC~2403 \citep{Bendo2010}. Consequently, the dust masses and resulting DMR in these specific environments could be subject to variations depending on how this long-wavelength behavior is treated. Finally, we caution that direct comparisons with certain theoretical frameworks, such as the models of \cite{Calura2023} calibrated specifically for more massive, metal-rich spiral galaxies like NGC 628, must be treated carefully, as the interstellar medium conditions, feedback mechanisms, and star formation histories of our low-mass sample differ considerably.}

\subsection{Radial gradients of \texorpdfstring{$\alpha_{\rm CO}$}{alpha CO}, DGR and DMR.} \label{Grad_DGR_alphaCO}

The radial gradients use the same $\alpha_{\rm CO}$, DGR and DMR values (and associated uncertainties) as Sect.~\ref{alpha_DGR_SR}. The normalized radii in {the r band}, $R/R_{\rm eff}$ and $R/R_{25}$, are taken from \citet{Vale2025}. 

%The following radial gradients take the same \alphaCO\ and DGR values (with the associated uncertainties) than in Sect.\ref{alpha_DGR_SR}. Also, the values to compute both the R/R{$\rm_{eff}$} and the R/R{$\rm_{25}$} are taken from \cite{Vale2025}.

\begin{figure*}[ht!]
    \centering
    {\includegraphics[scale=0.6]{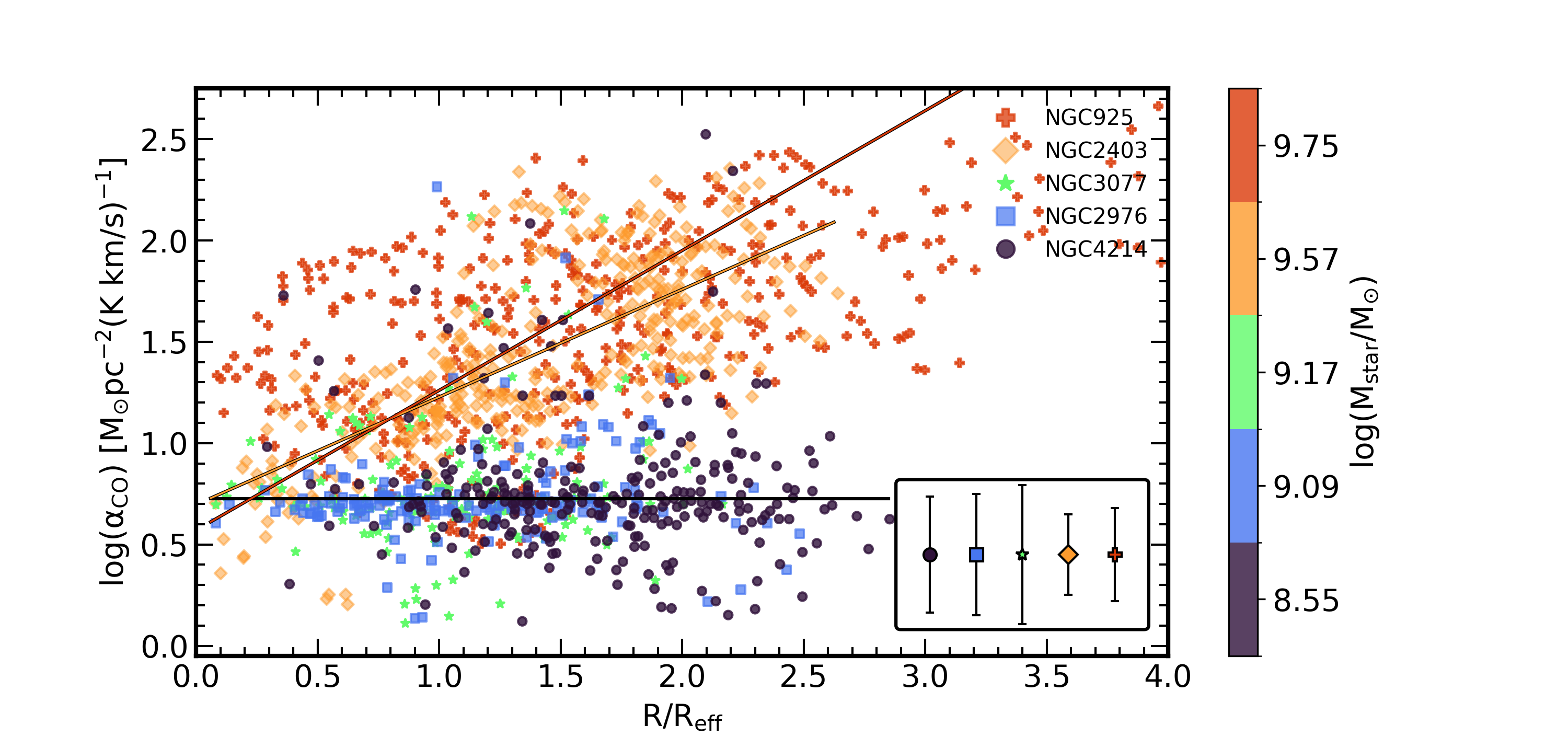}}
     \caption{Radial distribution of the \alphaCO\ as a function of the normalized galactocentric radius R/R{$\rm_{eff}$}. Symbols and color-coding are consistent with previous figures. The solid orange and red lines denote the linear fits for NGC 2403 and NGC 925, respectively. The black line is the fitting for the lower-mass galaxies NGC 4214, NGC 2976 and NGC 3077. Similar to previous figures, the error bars correspond to the median uncertainty in \alphaCO. }
      \label{Reff_alpha_plot}
\end{figure*}

The log($\alpha_{\rm CO}$) gradient (Fig.~\ref{Reff_alpha_plot}) reveals two distinct radial trends that depend on global stellar mass. NGC~4214, NGC~2976, and NGC~3077 exhibit flat radial profiles around the median value log($\alpha_{\rm CO}$) $\sim$ 0.73. In contrast, NGC~2403 and NGC~925 show clear positive radial gradients, with NGC~2403 displaying the steeper slope. NGC~3077 stands out as a peculiar galaxy in this figure, with log($\alpha_{\rm CO}$) spanning $\sim$ 0 – 2.1, and exhibiting a very large dispersion (RMSE $\sim$ 0.33 dex). A similarly high level of scatter is observed in the $\alpha_{\rm CO}$–Z relation. To enable a direct comparison with \citet{Sandstrom2013} and \citet{Chiang2024}, Fig.~\ref{R25_alphaCO_plot} shows the $\alpha_{\rm CO}$ gradient as a function of the normalized radius $R/R_{25}$ (see Appendix \ref{app_BF}). The \citet{Sandstrom2013} relation lies systematically below the values observed for NGC~4214, NGC~2976, and NGC~3077. In contrast, the \citet{Chiang2024} relation matches the highest stellar mass galaxies in our sample, although its slope is broadly consistent with that of NGC~2403.

%The R/R{$\rm_{eff}$} vs log(\alphaCO) plot (Fig. \ref{Reff_alpha_plot}) again reveals peculiar radial trends in the galaxies according to the global stellar mass. For NGC 4212, NGC 2976 and NGC 3077 there is a flat radial profile around log(\alphaCO) $\sim$ 0.70. For NGC 2403 and NGC 925, both show a strong positive radial gradient, NGC 2403 with a steeper slope. 

%Of particular interest are some values in NGC 3077,  with log(\alphaCO) ranging from $\sim$0.8-2.3, but with considerable scatter. 

%Of particular interest are the values in NGC~3077, with log($\alpha_{\rm CO}$) ranging from $\sim$0 to 2.3, and exhibiting a very large dispersion \textcolor{red}{(RMSE $\sim$ XX)}, similar to the relation seen in .
%The high dispersion is similar to the seen in the relation log(\alphaCO) and metallicity.

%In Fig.\ref{R25_alphaCO_plot} is shown the radial \alphaCO\ gradient with the normalized radius using the R{$\rm_{25}$}. This is to compare with the previous work of \cite{Sandstrom2013} and \cite{Chiang2024}. As seen in the plot, the fitting of \cite{Sandstrom2013} is under the values of NGC 4214, NGC 2976 and NGC 3077, while the fitting of \cite{Chiang2024} does not follow particularly some galaxy of our sample, but keeps a similar slope than NGC 925. 

%The relation between R/R{$\rm_{eff}$} and DGR (Fig. \ref{Reff_DGR_plot}) shows negative radial gradients for NGC 2976, NGC 2403 and NGC 925. NGC 4214 reveals a flat radial profile whereas NGC 3077 reveals a positive DGR gradient.

%NGC 2976, NGC 2403 and NGC 925 have less dispersion at short radii but increments towards large radii. For NGC 4214 and NGC 3077 show a constant dispersion at all radii (not taking into account the values that deviate from the fitting).   

\begin{figure*}[ht!]
    \centering
    {\includegraphics[scale=0.6]{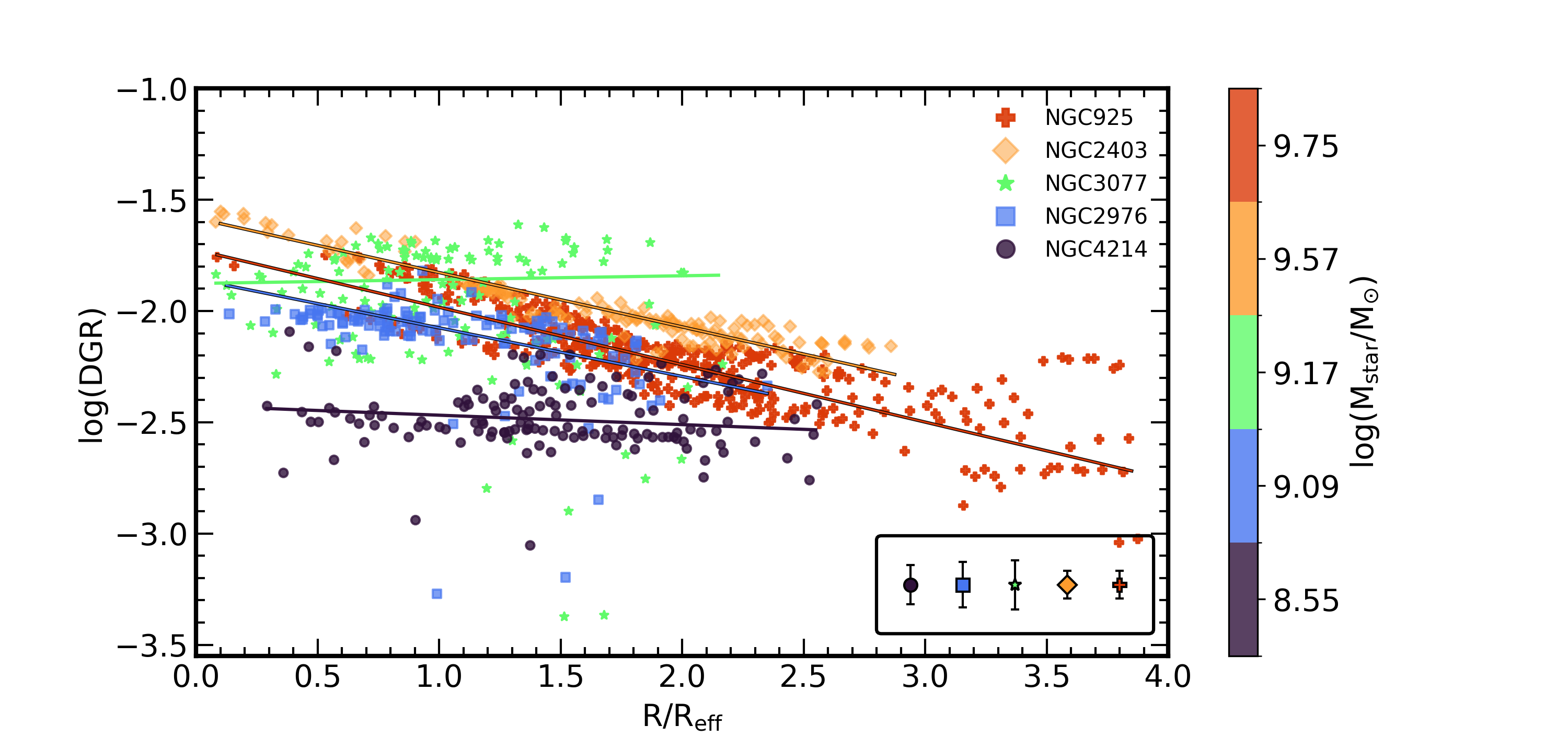}}
     \caption{Radial distribution of the DGR as a function of the normalized galactocentric radius R/R{$\rm_{eff}$}. Symbols and color-coding are consistent with previous figures. The solid lines denote the linear fittings for each galaxy in the sample. Similar to previous figures, the error bars correspond to the median uncertainty in log(DGR).}
      \label{Reff_DGR_plot}
\end{figure*}

The relation between $R/R_{\rm eff}$ and DGR (Fig.~\ref{Reff_DGR_plot}) shows negative radial gradients for NGC~2976, NGC~2403, and NGC~925. NGC~4214 is consistent with a flat profile, whereas NGC~3077 also displays a flat gradient. The RMSE values for each galaxy are reported in Table~\ref{fittings_vals}. Figure~\ref{R25_DGR_plot} shows the radial DGR gradients as a function of the normalized radius $R/R_{25}$, enabling a direct comparison with the \citet{Sandstrom2013} relation (see Appendix \ref{app_BF}). This relation lies systematically above most of the measurements in our galaxies, while broadly preserving the overall shape of the observed trends, with the exception of NGC~3077.

%The Fig.\ref{R25_DGR_plot} shows the radial DGR gradient with the normalized radius using the R{$\rm_{25}$}, again to compare with the fitting of \cite{Sandstrom2013}. Such fitting is above of the most values of our galaxies, but preserves the same shape than the galaxies (except for NGC 3077). 

\begin{figure*}[ht!]
    \centering
    {\includegraphics[scale=0.6]{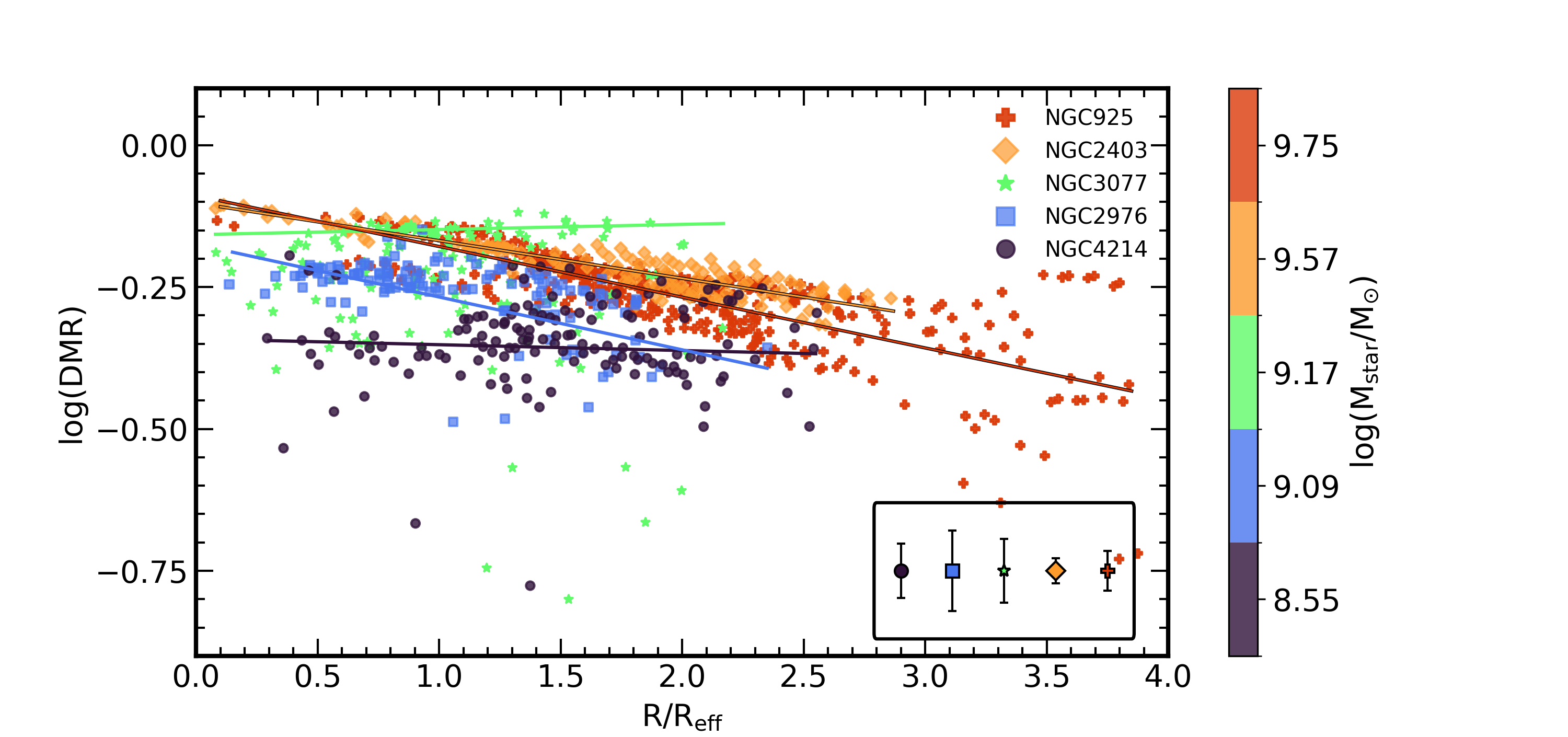}}
     \caption{Radial distribution of the DMR as a function of the normalized galactocentric radius R/R{$\rm_{eff}$}. Symbols and color-coding are consistent with previous figures. The solid lines denote the linear fittings for each galaxy in the sample. Similar to previous figures, the error bars correspond to the median uncertainty in log(DMR).}
      \label{Reff_DMR_plot}
\end{figure*}

Figure \ref{Reff_DMR_plot} illustrates the spatially resolved radial distribution of the DMR as a function of the normalized galactocentric radius ($R/R_{\text{eff}}$) across our galaxy sample. We observe two distinct behaviors within our sample. First, a clear negative radial gradient is present in four of the five galaxies: NGC 2976, NGC 3077, NGC 2403, and NGC 925. The measured slopes for these systems exhibit a highly uniform behavior. In contrast, the dwarf irregular galaxy NGC 4214 displays a notably distinct, flat radial profile across its entire disk. The internal dispersion around these linear fits varies significantly from galaxy to galaxy. NGC 2403 exhibits the tightest and most well-defined radial profile with a RMSE = 0.02, whereas the flat profile of NGC 4214 is accompanied by the largest structural scatter in our sample RMSE = 0.35

In addition to the effective radii, we have also evaluated the DMR utilizing the $R_{25}$ radius. We provide these measurements for completeness, as there is currently no broadly accepted consensus in the literature on the optimal radius to use for this in the other derivations. Presenting both approaches allows for broader comparability. The results derived using $R_{25}$ for the DMR are detailed in Appendix \ref{app_BF}, figure~\ref{R25_DMR_plot}.

\section{Discussion}\label{discuss_section}

The anticorrelation between metallicity and log(\alphaCO) (Fig. \ref{Z_alpha_plot}), particularly for NGC 2403 and NGC 925, is consistent with theoretical predictions and previous observational studies showing that CO becomes a better tracer of H$_2$ at higher metallicities due to enhanced dust shielding \citep{Bolatto2011, Bolatto2013,Kepley2016}. 
The lower log(\alphaCO) values ($\leq$ 0.70), corresponding to  galaxies NGC 4214, NGC 2976 and NGC 3077, show flat trends, that are consistent with the \cite{Sandstrom2013} relation.  In general, the observed anticorrelation and the slopes of NGC 2403 and NGC 925 (Table \ref{fittings_vals}) are qualitatively consistent with a metallicity-dependent \alphaCO\ relations from the literature e.g., \alphaCO\ $\propto$ Z$^{-1}$ to Z$^{-3}$\citep{Bolatto2013}. As mentioned in \cite{Bolatto2011} and \cite{Kepley2016} some specific factors can affect the \alphaCO-Z relation, e.g. the gas temperature, the ISM pressure or the dust content in galaxies, which indicates that metallicity alone does not fully determine the \alphaCO. Such factors can contribute significantly to the interplay of both galaxy properties and thus the precise determination of the relation \citep{Bolatto2013,Thompson2024}

Regarding the scatter, log($\alpha_{\rm CO}$) varies by $\sim$ 2 – 3 dex at fixed metallicity. This scatter exceeds the typical measurement uncertainties and likely reflects intrinsic variations in the ISM conditions such as the UV field strength, gas surface density, cloud structure, and the internal geometric structure of the molecular gas that affect \alphaCO\ independently of metallicity \citep{Bolatto2013,Sandstrom2013,Solis-Castillo2020,Ramambason2024}. The positive correlation between the metallicity and log(DGR) (Fig. \ref{Z_DGR_plot}) confirms the fundamental expectation that dust abundance scales with metal abundance, as dust grains form from heavy elements \citep{Solis-Castillo2020,Madden2016,Park2024}.
Our relation (Table \ref{fittings_vals}) is consistent with a power-law relation DGR $\propto$ Z$^{n}$ with n $\sim$ 1-2. Our data do not populate the regime of low metallicities (12 + log(O/H) $\leq$ 8.0), however a steeper than linear scaling (n $>$ 1) at low metallicities would suggest that dust production efficiency increases super linearly with metallicity, possibly due to enhanced grain growth in molecular clouds at higher metallicity, or reduced destruction efficiency in more metal-rich environments \citep{Remy2014,Madden2016}.

In general, moderate scatter exists around the mean correlation, with a RMSE = 0.11 dex. This scatter may arise from variations in dust temperature (affecting dust mass estimates), differences in DMR due to selective dust destruction or growth, or systematic uncertainties in metallicity calibrations \citep{Sandstrom2013,Remy2014,Solis-Castillo2020,Tailor2025}. These factors could explain the variation between our data and the fitting of \cite{Sandstrom2013}.

%\textcolor{orange}{Aqui no veo ningun aplanamiento en esta relacion}
%There is a suggestion of flattening in the DGR metallicity relation at the highest metallicities (12 + log(O/H) > 8.4), where DGR increases more slowly with metallicity. This behavior could indicate saturation of dust production, where most available metals are already locked in dust grains, or increased dust destruction in high metallicity environments with more vigorous star formation. \citep{Madden2016,Park2024}

The positive slope of $m = 0.46 \pm 0.03$ found in our DMR-Z scaling relation provides critical insights into the localized lifecycle of interstellar grains. Classical galaxy evolution models and simpler numerical recipes often implement a fixed, universal DMR across all cosmic epochs \citep{Feldmann2015}. However, our spatially resolved data clearly show a linear increase of the DMR in this mass regime, showing that more metal-enriched regions possess a higher fraction of their metal budget locked up in the dust phase. This behavior is consistent with theoretical frameworks where dust mass growth at intermediate metallicities ($12+\log(\text{O/H}) \sim 8.1 - 8.5$) is driven predominantly by secondary grain growth via the accretion of gas-phase volatiles within dense cloud cores \citep{Asano2013,Zhukovska2014}. Our empirical findings track the resolved trends observed in wider galaxy surveys, which show that as the local metal supply increases, the interstellar conditions become rapidly more favorable for grain growth, outpacing localized supernova destruction mechanisms \citep[e.g.,][]{Popping2017,DeVis2019,DeLooze2020,Galliano2021}.

Regarding the radial gradients of $\alpha_{\rm CO}$ (Fig.~\ref{Reff_alpha_plot}), a positive gradient indicates that CO becomes an increasingly inefficient tracer of H$2$ toward the outer disk, consistent with the combined effects of decreasing metallicity and reduced dust shielding at larger radii \citep{Sandstrom2013, Lee2024}. An intermediate behavior may point to a transition regime in which metallicity effects on \alphaCO\ become significant, but are still modulated by local ISM conditions \citep{Sandstrom2013}.

%Regarding to the gradients of \alphaCO\ (Fig. \ref{Reff_alpha_plot}), a positive gradient indicates that the CO becomes a progressively poorer tracer of H$_2$ in outer disk regions, consistent with decreasing metallicity and dust shielding at larger radii \citep{Sandstrom2013,Lee2024}. An intermediate behavior could suggest a transition regime where metallicity effects on \alphaCO\ become significant but are modulated by local ISM conditions \citep{Sandstrom2013}.

However, a flat behavior is not necessarily unexpected in low-mass galaxies. Although their lower metallicities would generally imply higher $\alpha_{\rm CO}$ values, these systems may already lie in a regime where $\alpha_{\rm CO}$ is uniformly elevated across the disk, reducing any radial variation. In this context, the flat profiles may reflect a saturation of metallicity effects, or the increasing importance of local ISM conditions such as gas surface density and cloud structure, which can modulate $\alpha_{\rm CO}$ independently of metallicity \citep{Bolatto2011, Kepley2016}.

%Este parrafo te lo cambie por que un comportamiento flat si seria esperado
%However, a flat behavior is unexpected given that low-mass galaxies typically have lower metallicities, which should increase \alphaCO. The flat profiles may indicate that these galaxies have already reached sufficiently low metallicity that \alphaCO\ is uniformly elevated, or that the previous factors already mentioned (that affect the \alphaCO\ e.g., high gas surface densities, different cloud structures) compensate for metallicity effects \citep{Bolatto2011,Kepley2016}.

The negative radial gradients in the DGR (Fig.~\ref{Reff_DGR_plot}) likely reflect the underlying metallicity gradients, as lower metallicities in the outer regions provide fewer metals available for dust formation. Additional contributions from radial variations in dust destruction and growth processes may also play a role \citep{Solis-Castillo2020, Park2024,Tailor2025}. The metallicity gradients reported by \citet{Vale2025} support this interpretation, showing negative slopes for NGC~2976, NGC~2403, and NGC~925.
Since the DGR is linked to star formation activity through the cycling of dust and gas \citep[e.g.][]{Yasuda2023}, a comparison with radial SFR profiles (Puebla et al., in prep.) may help constrain the physical processes driving these trends.

%The negative gradients in the DGR (Fig. \ref{Reff_DGR_plot}) reflect the combined effects of metallicity gradients (that lower metallicity in outer regions provides fewer metals for dust formation) and potentially radial variations in dust destruction efficiency \citep{Solis-Castillo2020,Park2024}. In \cite{Vale2025} can be seen the metallicity gradients of these galaxies which confirms the negative slope for NGC 2976, NGC 2403 and NGC 925. Because the SFR can be related to the DGR \citep[][e.g.]{Yasuda2023}, the comparison of these results with the SFR radial gradients (Puebla, M.M. et al in prep.) could help to constraint the physical processes that are happening in these galaxies. 

While most galaxies in our sample exhibit negative DGR gradients, NGC~4214 shows a flat profile and NGC~3077 displays a positive gradient. These deviations may be linked to their disturbed morphologies and interaction histories, which can redistribute metals and dust in non-standard ways \citep{LaraLopez2022}, or to enhanced dust production in the outer regions driven by recent star formation \citep{Park2024}. NGC~4214 is the most representative dwarf irregular galaxy in our sample, while NGC~3077 is part of the M81 group and shows clear evidence of tidal interaction \citep{Okamoto2023}. In addition, NGC~3077 appears to exhibit a steep negative SFR gradient in the inner regions that flattens at larger radii (Puebla et al., in prep.), which may contribute to its atypical DGR profile. {The contrasting radial profiles observed across our sample can be contextualized by the distinct morphological and environmental histories of these galaxies. Within the M81 group, NGC 3077 and NGC 2976 exhibit remarkably different behaviors driven by their specific locations in the group. NGC 3077 resides at the dense core of the M81 triplet, where strong tidal interactions have actively stripped and remixed the gas disk. This environmental disruption naturally accounts for its flat or slightly positive DGR radial gradient. Conversely, while NGC 2976 is also a known member of the M81 group, it is located on the periphery at a projected distance of ~200 kpc from M81. This outer position effectively shields NGC 2976 from strong tidal interactions, allowing it to maintain a much quieter environmental history that physically explains the preservation of its regular, negative radial gradient.}

{For the remaining galaxies in our sample, external environmental factors play a negligible role. NGC 4214, NGC 2403, and NGC 925 are all classified as isolated galaxies \citep{Bettoni2003}, ruling out environmental tidal stripping as a mechanism for ISM mixing. In the case of the dwarf irregular NGC 4214, its remarkably flat DGR profile could be driven entirely by internal mechanisms. Intense starburst-induced stellar feedback and large-scale galactic outflows could dominate this system, efficiently mixing gas and dust across the disk and flattening the local gradients independently of environmental interactions. The case of NGC 4214 could be potentially similar to NGC 1569 as seen in \cite{Garduno2023}.}

%The unusual behavior of NGC 4214 or NGC 3077 may reflect the galaxy disturbed morphology and interaction history, which could redistribute metals and dust in non-standard ways, or indicate enhanced dust production in outer regions through recent star formation \citep{Park2024}. Indeed, NGC 4214 is the clearest dwarf irregular galaxy in our sample, while NGC 3077 is known to be part of the M81 group with clear evidence of galaxy interaction \citep{Okamoto2023}. NGC 3077 looks to show a strong negative SFR radial gradient from the center up to a certain distance in which such gradient is truncated to become in a flat profile (Puebla, M.M. et al. in prep.).

Galaxies show systematic vertical offsets in the DGR at fixed radius, correlating with stellar mass, for example at R/R{$\rm_{eff}$} $\sim$ 1.5. Higher-mass galaxies (NGC 2403, NGC 925) have higher DGR values at those radii compared to lower-mass ones (NGC 4214, NGC 2976), consistent with the mass-metallicity relation and the expectation that more massive galaxies have produced more metals and dust over their histories \citep{Lara-Lopez2013b,Madden2016,Park2024}.

Again, the scatter may arise from local variations in ISM or conditions in each galaxy, e.g. the dust temperature (affecting infrared-derived dust masses), uncertainties in gas mass estimates, or real physical variations in DMR ratios due to differential dust growth or destruction \citep{Sandstrom2013,Solis-Castillo2020,Tailor2025}. {However, in our case, temperature-induced uncertainties in infrared-derived dust masses are potentially mitigated because warmer dust components contribute minimally to the total mass budget, whereas single-component SED fits are mostly biased by spatial variations in dust temperature profiles. Spatially resolved studies demonstrate that the dust mass is predominantly dominated by cold dust ($T_c \sim 13\text{--}24\text{ K}$), where local conditions, such as the radial temperature gradients found in M 33 \citep{Kramer2010} or the bar-driven central concentrations observed in NGC 6946 \citep{Alvarez2026}, must be accounted for to prevent systematic biases in global SED modeling.}

The localized efficiency of dust production and survival across galactic disks is clearly shown by the radial distribution of the DMR as a function of the normalized galactocentric radius, as presented in Fig. \ref{Reff_DMR_plot}. Our Bayesian linear analysis reveals a striking homogeneity among the late-type spirals and interacting systems (NGC 2976, NGC 3077, NGC 2403, and NGC 925), which all display well-defined negative radial DMR gradients with slopes that are nearly identical. This systematic radial decline demonstrates that the fraction of metals locked up in solid dust grains decreases toward the outer regions of these galaxies, independent of the absolute drop in gas-phase metallicity.  Physically, this inside-out decrease in condensation efficiency is highly consistent with dust evolution models where secondary grain growth via the accretion of gas-phase volatiles requires a dense, cold, and heavily shielded molecular ISM to outpace localized supernova grain destruction \citep[e.g.,][]{Asano2013,Zhukovska2014,Popping2017}. As the galactocentric radius increases, the systematic drop in gas surface densities and molecular gas fractions suppresses these secondary growth pathways, leaving a larger fraction of heavy elements in the gas-phase within the pristine outer disks \citep[e.g.,][]{DeVis2019,DeLooze2020,Casasola2022}. It is particularly noteworthy that while NGC 3077 exhibits an anomalous positive radial gradient in its DGR, its DMR gradient flips to a standard negative slope, suggesting that its apparent outer dust enhancement is entirely a consequence of localized metal redistribution from its ongoing tidal interactions rather than an increased intrinsic dust condensation efficiency. In contrast to the rest of the sample, the dwarf irregular galaxy NGC 4214 represents a clear exception, displaying a completely flat radial DMR profile. This spatial invariance may indicate that the physical timescales governing the balance between stellar dust injection, interstellar grain growth, and supernova destruction \citep[e.g.,][]{Feldmann2015,Galliano2021} remain exceptionally uniform and well-mixed across its entire disrupted disk, highlighting a structurally distinct evolutionary pathway compared to typical disk-dominated systems. {Another important scenario to take into account is that an elevated specific star formation rate ($\text{sSFR}$) and a low stellar mass (e.g. for both NGC 4214 and NGC 3077) could favor the development of highly efficient galactic winds, which can dynamically redistribute the gas phase and shape their local DGR and DMR distributions. For galaxies similar to NGC 4214, particularly violent feedback-driven outflows may account for a more complex and irregular metal spatial distribution. Together with environmental influences, these intense feedback processes help explain why these low-mass systems display distinct trends compared to their more massive counterparts.}

{The empirical trends derived from our sub-kiloparsec analysis form a natural complement to the spatially resolved ISM modeling of \cite{Tailor2026}. In their resolved analysis of DGR ratios and cold gas budgets, they demonstrated that standard Milky Way conversion factors \citep[$\alpha_{\text{CO, MW}}$,][]{Bolatto2013} severely underestimate molecular gas masses in CO-faint regions of low-mass systems such as NGC 2403 and NGC 925. To reconcile dust continuum models with resolved gas surface densities, their framework adopted a steep, non-linear metallicity dependence \citep[$\alpha_{\text{CO}} \propto Z^{-2}$,][]{Schruba2012} across sub-solar regimes. Our sub-kiloparsec spectroscopic observations provide direct empirical grounding for this modeling approach. By independently solving for $\alpha_{\text{CO}}$ and DGR without imposing prior constraints on the functional form of the conversion factor, our pixel-level IFU data demonstrate that the steep scaling required to balance resolved gas and dust budgets is an intrinsic physical property of the local interstellar medium.}

{Together, these complementary methodologies establish a cohesive framework for mapping cold gas and dust across low-metallicity environments. While \cite{Tailor2026} show the necessity of environment-sensitive $\alpha_{\text{CO}}$ prescriptions for maintaining self-consistent DMR ratios across spatially resolved galaxy disks, our maps reveal the underlying local physical driver—namely, enhanced CO photodissociation occurring within low-extinction, dust-poor envelopes across individual star-forming complexes. This convergence between resolved ISM mass-balancing models and direct sub-kiloparsec spectroscopic measurements confirms that universal or weakly scaling conversion factors systematically underestimate molecular gas reserves in sub-solar regimes ($12+\log(\text{O/H}) \lesssim 8.3$). Synthesizing spatially resolved spectroscopy with resolved dust and gas modeling thus provides a robust, mutually reinforcing foundation for calibrating molecular gas budgets and DMR ratios in nearby dwarf and intermediate-mass galaxies. Furthermore, examining how these properties evolve across redshift reiterate how detailed constraints from local systems can provide critical benchmarks for understanding galaxy evolution in the high-redshift Universe \citep{Popping2022}.}

\subsection{Robustness of Scaling Relations via Survival Analysis}

{To account for low signal-to-noise detections in our datasets without introducing potential selection biases, we implemented a Bayesian linear regression incorporating survival analysis. Previously, given the substantial uncertainties associated with the derivation of $\alpha_{\rm CO}$ and DGR (Sect.~\ref{gasmassestimates}), standard cuts based on median uncertainty thresholds were applied. However, to assess the extent to which the adopted uncertainty cuts may affect our results, we treat data points exceeding uncertainty thresholds as right-censored upper limits using PyMC's censoring capabilities (\texttt{pm.Censored}). For the DMR analysis, spaxels are retained consistently with the DGR classification. All scaling relations were fitted within this probabilistic framework using Bayesian models implemented in PyMC v5.16.2 \citep{abril2023pymc} and BAMBI v0.14.0 \citep{JSSv103i15}, adopting four MCMC chains with 3000 iterations each. The full set of scaling relations derived from the survival analysis is presented in Appendix~\ref{survival_analysis}.}

{Overall, applying the Bayesian survival analysis preserves the slope, scatter, and overall structure of the majority of our scaling relations compared to our baseline fitting methodology. Specifically, our DGR-Z and DMR-Z relations, as well as the radial gradients of $\mathrm{DGR}$ and $\mathrm{DMR}$ remain largely unchanged for all galaxies, with the exception of NGC 3077.}

%, alongside the radial $\mathrm{DGR}$ and $\mathrm{DMR}$ gradients for NGC\,3077, undergo substantial alterations under the survival analysis framework.

%In contrast, significant modifications in slope and scatter emerge in the remaining relations when upper limits are properly accounted for. The $\alpha_{\mathrm{CO}}$--$Z$ relation across all galaxies exhibits notable changes, shifting from steep individual trends to a shallower, unified behavior. Furthermore, the radial profiles of $R/R_{\mathrm{eff}}$--$\alpha_{\mathrm{CO}}$ for all galaxies, alongside the radial $\mathrm{DGR}$ and $\mathrm{DMR}$ gradients for NGC\,3077, undergo substantial alterations under the survival analysis framework.}

{In contrast, accounting for upper limits leads to changes in the $\alpha_{\mathrm{CO}}$--$Z$ relations and radial $\alpha_{\mathrm{CO}}$ gradients, as well as in the radial DGR and DMR gradients of NGC 3077. For the $\alpha_{\mathrm{CO}}$--$Z$ relations, the steep negative slopes initially obtained for NGC 2403 and NGC 925 become  shallower, changing from $m=-5.50$ to $m=-1.24$ and from $m=-5.54$ to $m=-2.65$, respectively. Conversely, the combined low-mass subsample changes from an essentially flat relation ($m=0.016$) to a negative slope ($m=-1.46$). These results suggest a more continuous dependence of $\alpha_{\mathrm{CO}}$ on metallicity across the sampled mass range, although the strength of the anticorrelation varies among galaxies.
The radial $\alpha_{\mathrm{CO}}$ gradients also become flatter when upper limits are incorporated. For NGC 2403 and NGC 925, the initially positive slopes decrease from $m=0.53$ to $m=0.06$ and from $m=0.69$ to $m=0.17$, respectively, while the combined low-mass subsample remains nearly flat ($m=0.020$ to $m=-0.019$). 
A notable change is also observed in NGC 3077, whose radial DGR and DMR gradients shift from slightly positive ($m=0.017$ and $m=0.009$, respectively) to negative ($m=-0.19$ and $m=-0.116$). These revised gradients bring NGC 3077 into closer agreement with the general behavior of the other galaxies. Overall, these differences highlight the sensitivity of the inferred scaling relations to the treatment of low-S/N measurements, particularly for $\alpha_{\mathrm{CO}}$, and emphasize the importance of accounting for observational limitations when interpreting the underlying physical trends.
}

\section{Conclusions}

In this work, we present a spatially resolved, panchromatic analysis of the DGR, the \alphaCO\ conversion factor, and the gas-phase metallicity for a sample of five nearby low-mass star-forming galaxies: NGC~4214, NGC~2976, NGC~3077, NGC~2403, and NGC~925. The sample spans a stellar mass range of log($M_\star/M_\odot$) $\sim$ 8.55 – 9.75 and includes morphological types from dwarf irregulars to late-type spirals, providing a diverse testbed for exploring metallicity–ISM relations. Using GCMS IFU spectroscopy to derive metallicity within the Metal-THINGS survey, CIGALE SED fitting for dust masses, and HI and CO data from THINGS and HERACLES, we derive pixel-by-pixel maps of these ISM diagnostics at a common angular resolution of {$\sim$18\arcsec}, corresponding to sub-kiloparsec scales of $\sim$250-550 pc. Our main results are summarized as follows.

%In this work, we  presented a spatially resolved, phanchromatic analysis of the dust-to-gas ratio (DGR), the CO-to-H$_2$ conversion factor (\alphaCO), and the gas-phase oxygen abundance (metallicity) for a sample of five nearby low-mass star-forming galaxies  NGC 4214, NGC 2976, NGC 3077, NGC 2403, and NGC 925, spanning a stellar mass range of log (\Ms/M$_\odot$) $\sim$ 8.55–9.75, and morphological types, from dwarf irregulars to late-type spirals, providing a diverse testbed for the metallicity–ISM property relations. Using GCMS IFU spectroscopy for metallicity (in the framework of the Metal-THINGS Survey), CIGALE SED fitting for dust masses, and THINGS/HERACLES HI and CO data, we have derived pixel-by-pixel maps of all three ISM diagnostics at a common angular resolution of $\sim$6 arcseconds, corresponding to sub-kiloparsec physical scales. Our main findings are summarized as follows.

\begin{itemize}

\item The DGR-Z relation holds on spatially resolved scales. We confirm a significant positive correlation between the DGR and gas-phase metallicity across all five galaxies. Over the metallicity range $12 + \log(\mathrm{O/H}) \sim 8.1-8.45$, log(DGR) increases from $\sim -3$ to $\sim -1.5$, with an RMSE of  0.18 dex. This behavior is consistent with dust abundance scaling with metal content, as dust grains originate from heavy elements produced by stellar nucleosynthesis (e.g. SN explosions) and subsequently grow through accretion in the dense ISM. Our results are in good agreement with the kiloparsec-scale relations of \citet{Sandstrom2013} and the sub-kiloparsec measurements of \citet{Park2024}, extending this correlation into the low-mass, low-metallicity regime.

\item We confirm a  positive correlation between the DMR and Z across our full sub-kiloparsec dataset, yielding a global linear slope of m = 0.46 and a tight internal dispersion RMSE = 0.11. This non-zero slope directly rules out the traditional paradigm of a constant, universal DMR across all evolutionary stages in low-mass galaxies. Physically, this trend implies that as the local metal supply increases, the capacity of the ISM to lock heavy elements into solid grains becomes systematically more efficient, providing robust empirical evidence for accelerated secondary grain growth via gas-phase accretion in more enriched environments.

\item The \alphaCO–Z anti-correlation is confirmed in the two higher-mass galaxies of our sample. We detect a clear anti-correlation between log(\alphaCO) and metallicity in NGC 2403 and NGC 925. log(\alphaCO) decreases by $\sim$2 dex over the sampled metallicity range, from $\sim$ 2.5 at 12 + log(O/H) $\sim$ 8.05 to $\sim$ 0.5 at 12 + log(O/H) $\sim$ 8.5. This behavior is consistent with  theoretical expectations that CO becomes an increasingly reliable tracer of H$_2$ at higher metallicities, due to enhanced dust shielding that protects CO molecules from photodissociation in the outer envelopes of molecular clouds. The observed slopes are steeper than those reported by \cite{Sandstrom2013} for more massive galaxies, suggesting that the metallicity sensitivity of \alphaCO\ may be enhanced in the low-mass regime, possibly due to the particular ISM structure and stronger radiation fields characteristic of these systems.

\item Radial gradients of DGR, DMR and \alphaCO\ trace the underlying metallicity gradients. We detect significant negative radial gradients in log(DGR) for NGC 2403 and NGC 925, reflecting the inside-out metallicity gradients characteristic of late-type spiral galaxies. We detect persistent, negative radial log(DMR) gradients across the galaxies NGC 2976, NGC 3077, NGC 2403 and NGC 925. Correspondingly, positive radial gradients in log(\alphaCO) are found for NGC 2403 and NGC 925, consistent with the expectation that CO becomes a less efficient molecular gas tracer at larger galactocentric radii where metallicity is lower. NGC 4214 shows a flat DGR and DMR radial profile, consistent with its shallow or flat metallicity gradient, as expected for a dwarf irregular system.

\item Two of the lowest-mass galaxies exhibit distinct behavior compared to the rest in our sample. NGC~4214 and NGC~3077 show flat $\alpha_{\rm CO}$ and DGR trends. In particular, NGC~4214, which occupies the lowest metallicity range in our sample ($12 + \log(\mathrm{O/H}) \sim 8.1$ - 8.3), displays low log($\alpha_{\rm CO}$) values ($\sim$ 0.1 - 1.0), suggesting that local ISM conditions, such as an enhanced interstellar radiation field, elevated gas temperatures, increased turbulence, or high star formation rate surface densities, can suppress $\alpha_{\rm CO}$ independently of metallicity. NGC~3077 also stands out by exhibiting a flat DGR radial gradient in contrast to the negative in the rest of the sample. This behavior is likely linked to its disturbed morphology and tidal interaction history within the M81 group, which may redistribute dust and metals and drive enhanced star formation in the outer regions.

%\item Galaxy-specific factors could produce significant deviations from the global trends. The most significant fact from the metallicity–\alphaCO\ trend is exhibited by NGC 4214, which occupies the lowest metallicity range in our sample (12 + log(O/H) $\sim$ 8.1–8.2) and shows  low log(\alphaCO) values ($\sim$0.6–0.8). This behavior could suggests that galaxy-specific factors, for example an enhanced interstellar radiation field driven by its starburst activity, elevated gas temperatures, increased turbulence, or high star formation rate surface densities, can significantly suppress \alphaCO\ independently of metallicity. NGC 3077 displays an anomalous positive DGR radial gradient, in contrast to the negative gradients seen in the other galaxies, likely reflecting its disturbed morphology, tidal interaction history within the M 81 group, and possible dust redistribution driven by interaction-induced star formation in its outer regions.

\item Significant scatter highlights the multi-parameter nature of ISM diagnostics. At fixed metallicity, log(\alphaCO) spans 2–3 dex across our sample, substantially exceeding typical measurement uncertainties. This large dispersion indicates that metallicity alone does not fully determine \alphaCO, and that local ISM conditions, including UV field strength, gas surface density, molecular cloud structure, and star formation activity, play an important and independent role. Similarly, the DGR-Z relation also exhibits moderate scatter, likely reflecting variations in dust temperature, DMR ratios, and systematic uncertainties in the metallicity calibration. These results reinforce the conclusion that spatially resolved, multi-tracer observations are essential to accurately characterize the ISM conditions in low-mass galaxies.

\item { Bayesian survival analysis was performed to assess the sensitivity of the results to the adopted uncertainty cuts, treating the low-S/N measurements excluded from the original analysis as upper limits. This approach yields a negative $\alpha_{\mathrm{CO}}$--$Z$ relation across the sample, including the low-mass galaxies, and generally shallower radial $\alpha_{\mathrm{CO}}$ gradients. The analysis also produces negative radial DGR and DMR gradients for NGC 3077, in contrast to the slightly positive gradients obtained in the original analysis. These differences highlight the sensitivity of the derived relations to the treatment of low-S/N measurements and suggest that the physical trends should be interpreted with caution.}

%Esta parte la podemos hablar, segun entiendo no se trato non-detections as upper limits, sino mediciones por debajo de tu cut de S/N como upper limits?
%\item {Incorporating a Bayesian survival analysis to properly treat non-detections as upper limits resolves another perspective of analysis. This approach recovers a unified, sample-wide negative correlation for the $\alpha_{\mathrm{CO}}$--$Z$ relation, indicating that the transition to a CO-dark ISM is more gradual than previously thought. Furthermore, it significantly flattens the radial $\alpha_{\mathrm{CO}}$ gradients across all galaxies, suggesting highly homogeneous CO shielding conditions throughout the disks, and corrects the anomalous positive radial dust gradients of NGC 3077 to the expected negative trends.}

\end{itemize}

These results highlight the importance of accounting for local ISM conditions when deriving molecular gas masses from CO observations in low-mass and low-metallicity galaxies, with direct implications for studies of star formation efficiency and galaxy evolution at low metallicity and across cosmic time.

\begin{acknowledgements}
%      Part of this work was supported by \emph{ESO}, project
%      number Ts$\sim$17/2--1.
We thank to the referee for their constructive report and very useful suggestions which have improved the robustness and clarity of this manuscript. LEG acknowledges the support of the Spanish grant PID2024-157374OB-I00 funded by MCIN/AEI/10.13039/501100011033/FEDER, EU. MALL, LEG, MMP, and GV acknowledge grants  PID2024-155875OB-I00 and PID2021-123417OB-I00 funded by MICIU/AEI/10.13039/501100011033/FEDER, EU. JG, MALL and LEG acknowledge the support of the project PCI2022-135023-2 funded by MCIN/AEI/10.13039/501100011033 and Unión Europea “NextGenerationEU” / PRTR.
MALL  acknowledges grant RYC2020-029354-I funded by MICIU/AEI/10.13039/501100011033 by “ESF Investing in your future”, by “ESF+”.
GV acknowledges support from the Comunidad de Madrid through grant PIPF-2024/TEC-34555, and the Atracción de Talento program 2022-T1/TIC-23797, as well as grant PID2023-146372OB-I00 funded by MICIU/AEI/10.13039/501100011033 and by ERDF, EU. IAZ acknowledges funding from the Deutsche Forschungsgemeinschaft (DFG; German Research Foundation) project-ID 550945879. VT and VC acknowledge funding from the INAF Mini Grant 2024 program ‘DustPedia meets Metal-THINGS: Dust-METAL’. L.S.P acknowledges support from the Research Council of Lithuania (LMTLT) (grant no. P-LU-PAR-26-15). SAC and JF acknowledge the support by PAPIIT-UNAM under grant IN102226.

\end{acknowledgements}

% Please note that we have included the references below in
% order to compile the document, but we ask you to:

%%%%%%%%%%%%%%%%%%%%%%%%%%%%%%%%%%%%%%%%%%%%%%%%%%
%\section*{Data Availability}

%The data underlying this article will be shared on reasonable request with the corresponding author. All data cubes, images, and spectra will be made available on the Metal-THINGS website (in preparation).



\bibliographystyle{bibtex/aa}
\bibliography{bibtex/Biblio}

%%%%%%%%%%%%%%%%%%%%%%%%%%%%%%%%%%%%%%%%%%%%%%%%%%

%%%%%%%%%%%%%%%%% APPENDICES %%%%%%%%%%%%%%%%%%%%%

\begin{appendix}

\section{Survival Analysis}\label{survival_analysis}

To properly account for upper limits in non-detections across our observational datasets, we implemented a Bayesian linear regression incorporating survival analysis. This methodology prevents potential truncation biases in the derived slopes ($m$) and intercepts ($b$) of the galaxy scaling relations, providing robust posterior distributions for both individual galaxies and the combined sample. The coefficients of such fitting can be seen in Table \ref{fittings_vals_survival}.

\begin{figure*}[ht!]
    \centering
    {\includegraphics[scale=0.5]{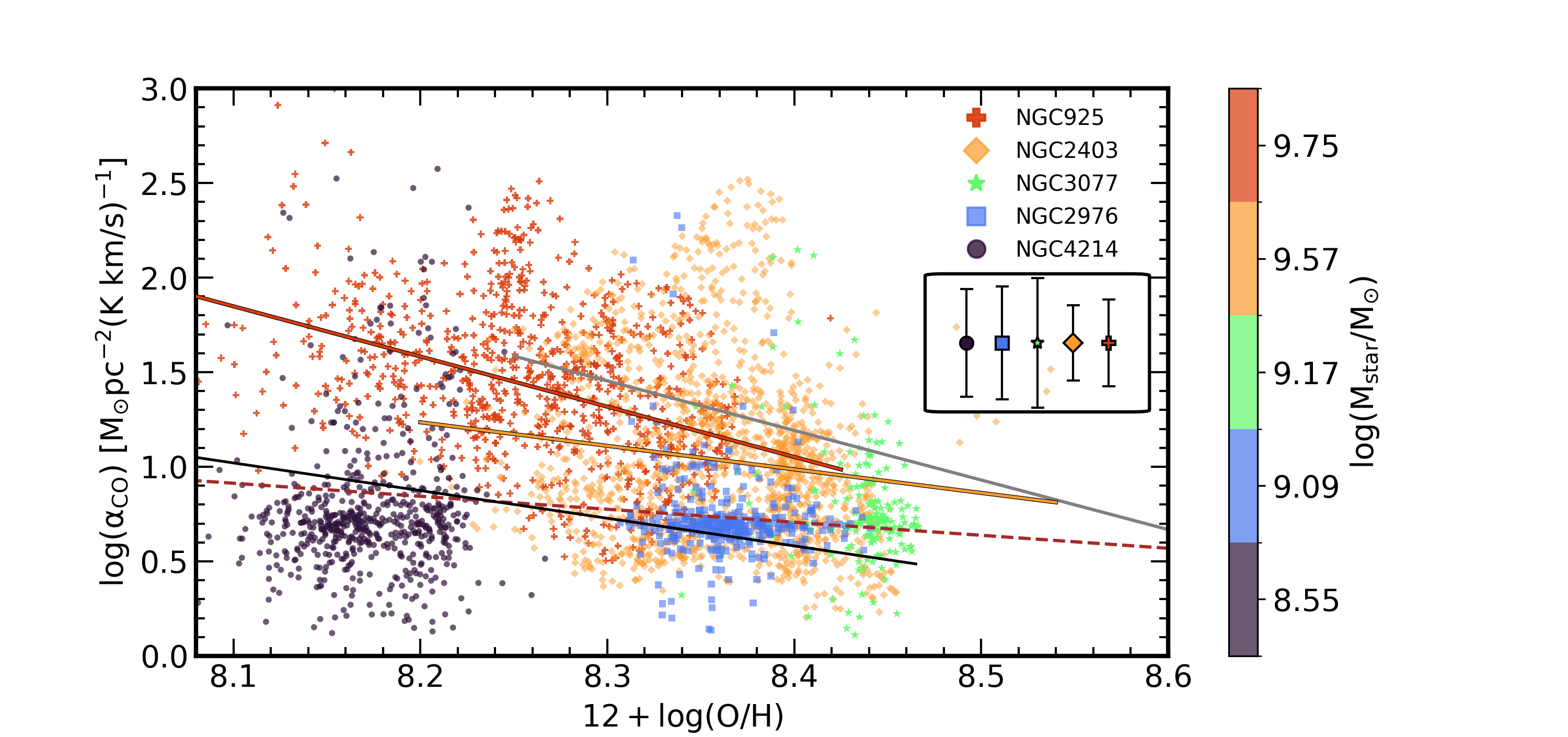}}
     \caption{Spatially resolved \alphaCO\ as a function of metallicity for the Survival Analysis. Data points are color-coded by the global stellar mass of each galaxy, as indicated in the color bar. The solid orange and red lines denote the linear fits for NGC 2403 and NGC 925, respectively. The black line represents the fitting observed for the lower-mass galaxies NGC 4214, NGC 2976, and NGC 3077. The dashed brown line corresponds to the re-calibrated relation from \cite{Sandstrom2013} and the solid gray line corresponds to the fitting of \cite{Chiang2024}. For clarity, the transparent box shows the median \alphaCO\ error bar for each galaxy.}
      \label{Z_alpha_plot_survival}
\end{figure*}

\begin{figure*}[ht!]
    \centering
    {\includegraphics[scale=0.5]{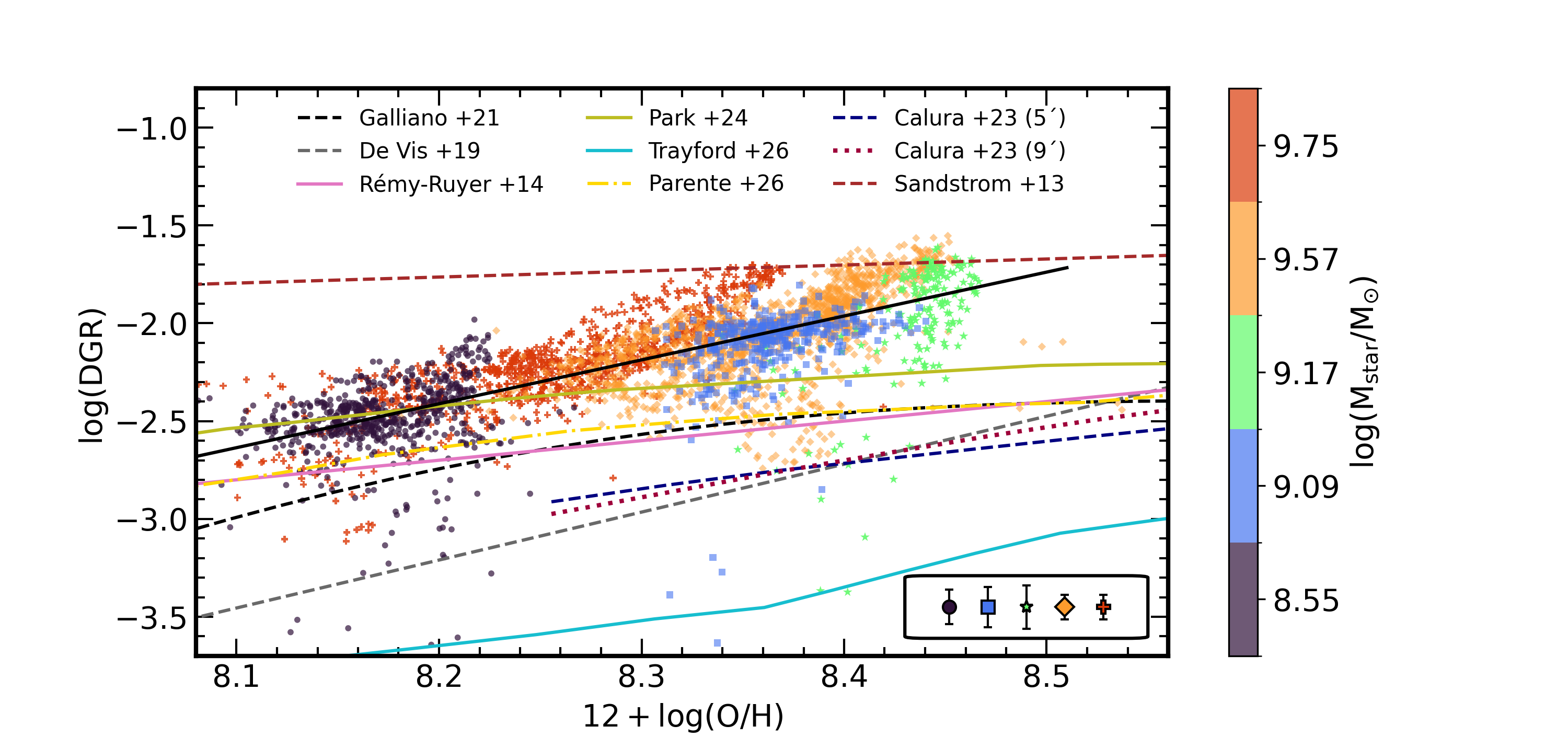}}
     \caption{Spatially resolved DGR versus metallicity for the Survival Analysis. Symbols and color-coding follow the same convention as in Fig. 5. The solid black line indicates our best-fit relation for the full sample. To contextualize these results, several observational, theoretical, and semi-analytical relations from the literature are overplotted (indicated by the various lines; see legend for references). Similar to previous figures, the error bars correspond to the median uncertainty in log(DGR).}
      \label{Z_DGR_plot_survival}
\end{figure*}

\begin{figure*}[ht!]
    \centering
    {\includegraphics[scale=0.5]{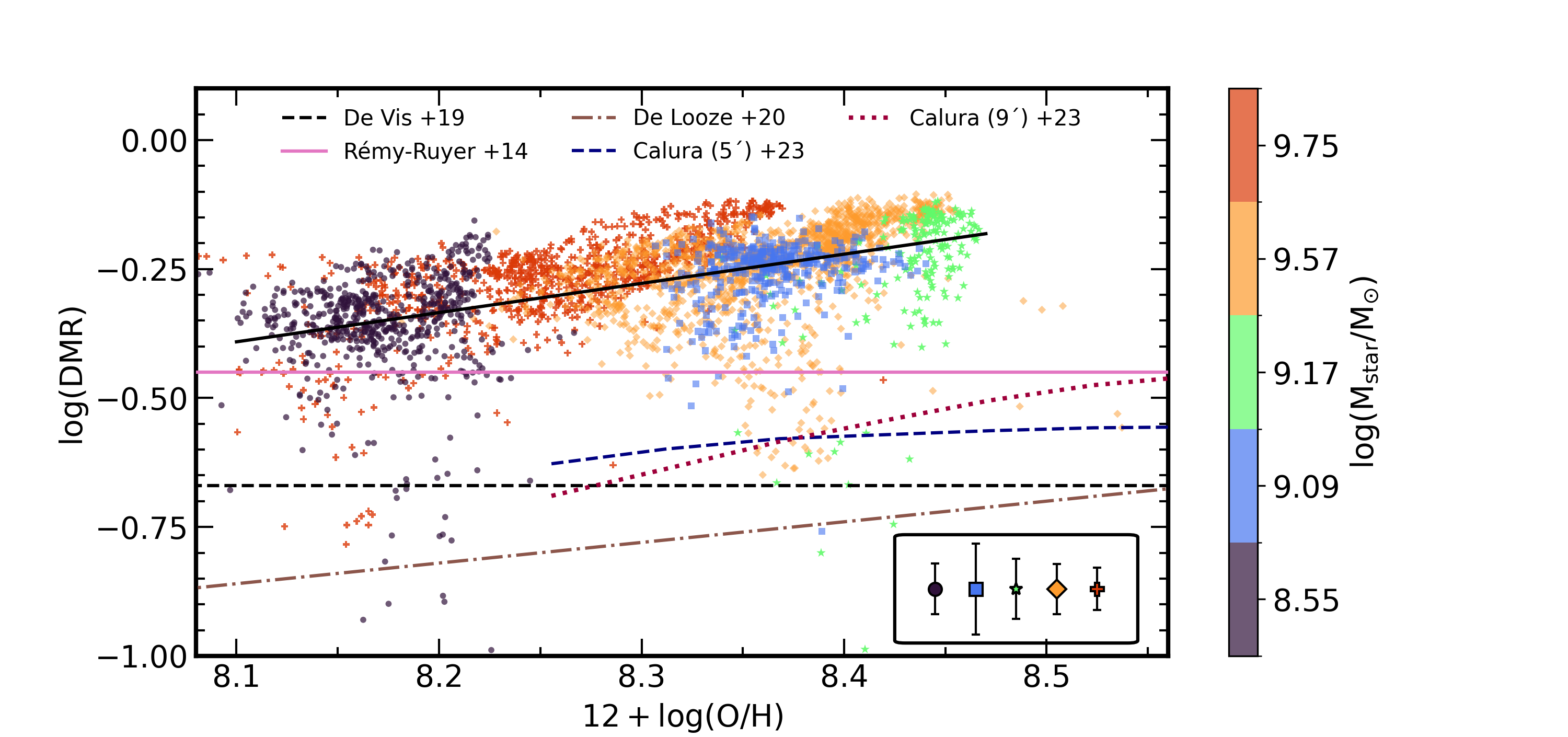}}
     \caption{Spatially resolved DMR versus metallicity for the Survival Analysis. Symbols and color-coding follow the same convention as in Fig. 5. The solid black line indicates the best-fit relation for the full sample. Similar to previous figure several models are plotted in order to compare with the literature. {Similar to previous figures, the error bars correspond to the median uncertainty in  log(DMR).}}
      \label{Z_DMR_plot_survival}
\end{figure*}

\begin{figure*}[ht!]
    \centering
    {\includegraphics[scale=0.5]{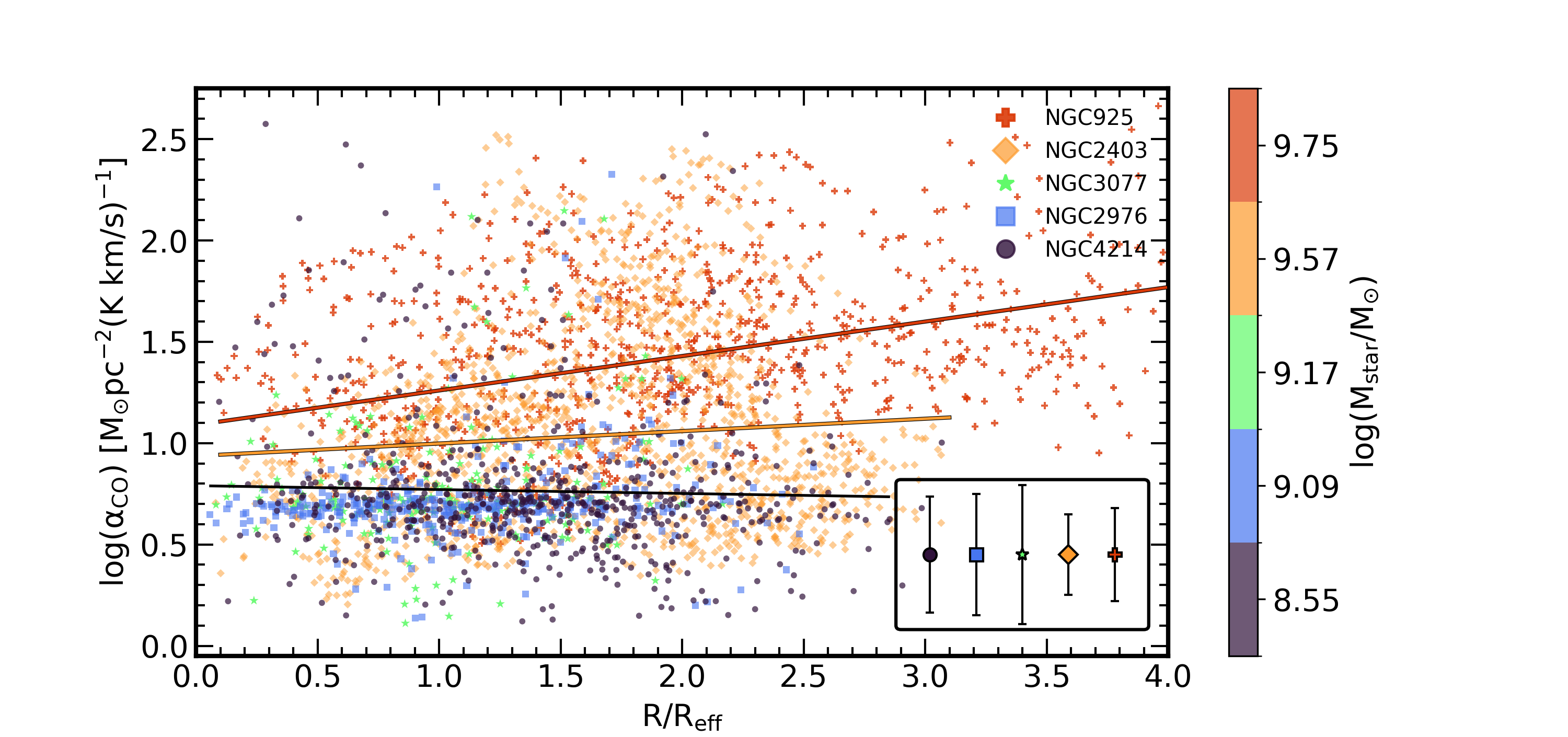}}
     \caption{Radial distribution of the \alphaCO\ as a function of the normalized galactocentric radius R/R{$\rm_{eff}$} for the Survival Analysis. Symbols and color-coding are consistent with previous figures. The solid orange and red lines denote the linear fits for NGC 2403 and NGC 925, respectively. The black line is the fitting for the lower-mass galaxies NGC 4214, NGC 2976 and NGC 3077. Similar to previous figures, the error bars correspond to the median uncertainty in \alphaCO. }
      \label{Reff_alpha_plot_survival}
\end{figure*}

\begin{figure*}[ht!]
    \centering
    {\includegraphics[scale=0.5]{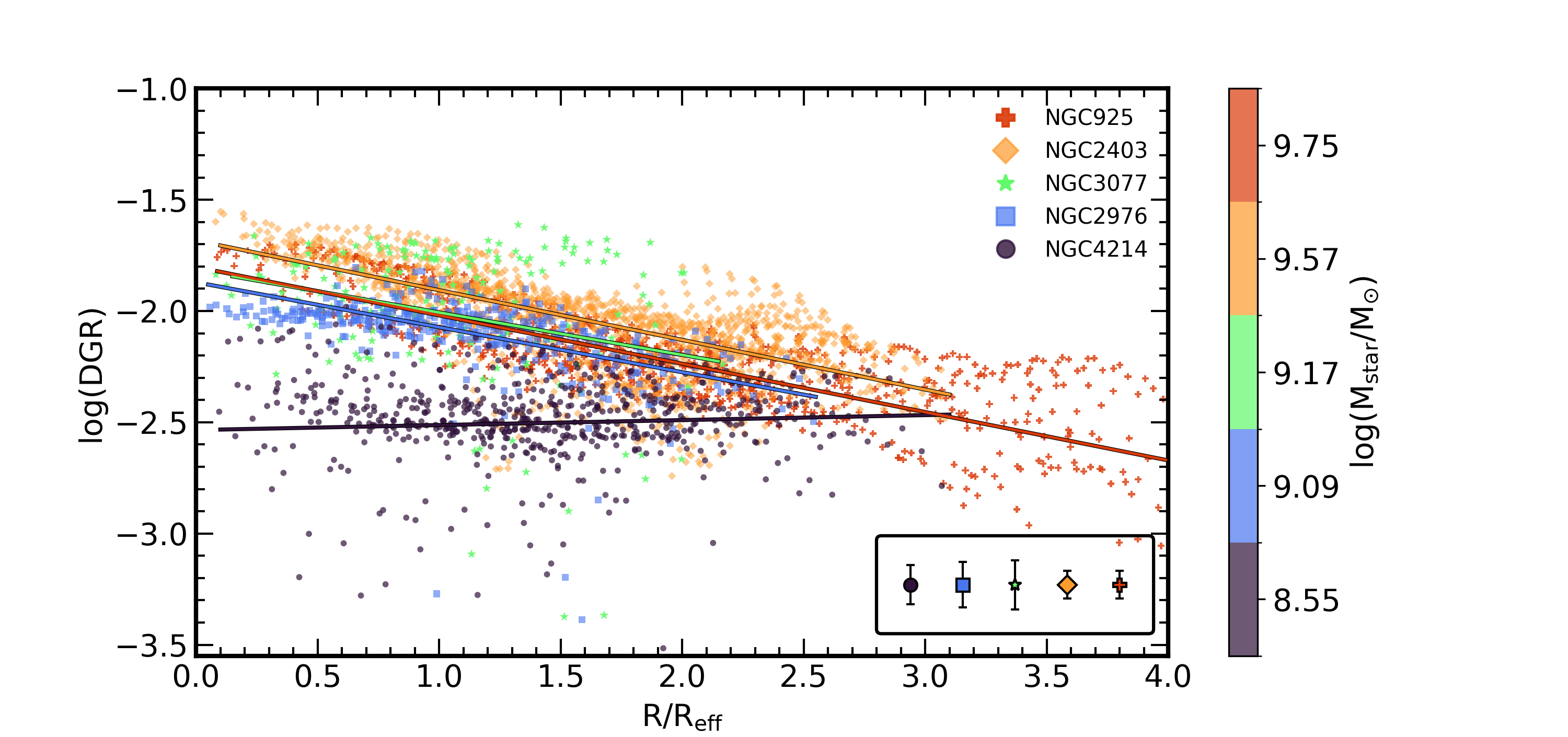}}
     \caption{Radial distribution of the DGR as a function of the normalized galactocentric radius R/R{$\rm_{eff}$} for the Survival Analysis. Symbols and color-coding are consistent with previous figures. The solid lines denote the linear fittings for each galaxy in the sample. Similar to previous figures, the error bars correspond to the median uncertainty in log(DGR).}
      \label{Reff_DGR_plot_survival}
\end{figure*}

\begin{figure*}[ht!]
    \centering
    {\includegraphics[scale=0.5]{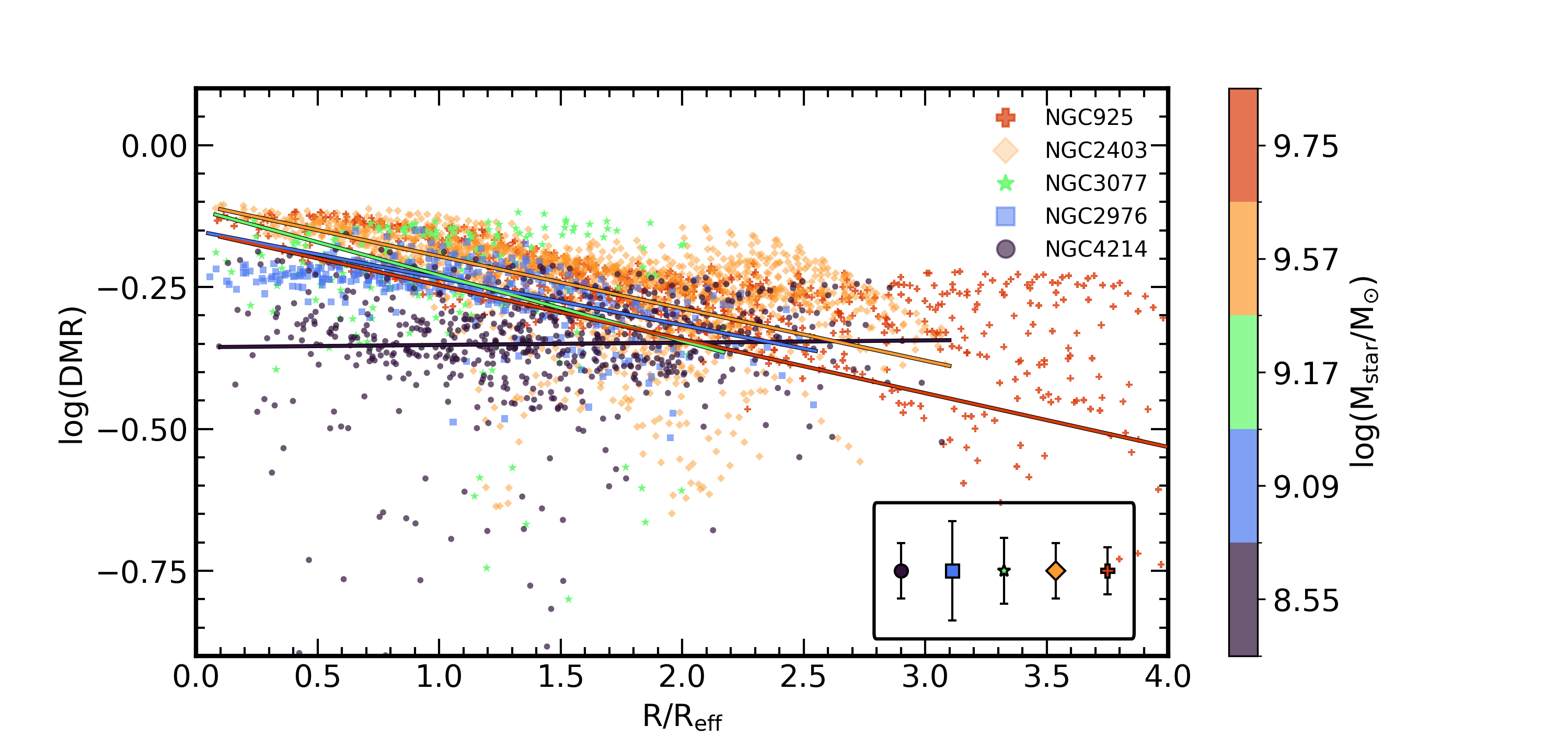}}
     \caption{Radial distribution of the DMR as a function of the normalized galactocentric radius R/R{$\rm_{eff}$} for the Survival Analysis. Symbols and color-coding are consistent with previous figures. The solid lines denote the linear fittings for each galaxy in the sample. Similar to previous figures, the error bars correspond to the median uncertainty in log(DMR).}
      \label{Reff_DMR_plot_survival}
\end{figure*}

\section{Bayesian Fittings}

In this appendix we show all the Bayesian fitting values that we compute for all our scaling relationships. 

\begin{table*}
\caption{Best-fit linear regression parameters for the studied relations.}
\label{fittings_vals}
\centering
\begin{tabular}{llcccc}
\hline\hline
\noalign{\smallskip}
Relation & Sample / Galaxy & Slope ($m$) & Intercept ($b$) & RMSE & $ \text{r}$ \\
\noalign{\smallskip}
\hline
\noalign{\smallskip}
Z-DGR & All & $2.07 \pm 0.04$ & $-19.35 \pm 0.53$ & 0.18 & 0.73 \\
\hline
Z-DMR & All & $0.47 \pm 0.04$ & $-4.14 \pm 0.30$ & 0.11 & 0.39 \\
\hline
$Z$ - $\alpha_{\rm CO}$ & Low mass & $0.016 \pm 0.001$ & $0.73 \pm 0.09$ & 0.32 & 0.02 \\
 & NGC\,2403 & $-5.50 \pm 1.50$ & $48.61 \pm 12.27$ & 0.39 & -0.57 \\
 & NGC\,925 & $-5.54 \pm 0.10$ & $47.68 \pm 0.85$ & 0.44 & -0.60 \\
\noalign{\smallskip}
\hline
R/R$_{\mathrm{eff}}$ - DGR & NGC\,4214 & $0.00 \pm 0.01$ & $-2.50 \pm 0.02$ & 0.17 & -0.13 \\
 & NGC\,2976 & $-0.22 \pm 0.01$ & $-1.86 \pm 0.01$ & 0.18 & -0.45 \\
 & NGC\,3077 & $0.017 \pm 0.005$ & $-1.88 \pm 0.02$ & 0.33 & 0.22 \\
 & NGC\,2403 & $-0.24 \pm 0.01$ & $-1.58 \pm 0.01$ & 0.06 & -0.94 \\
 & NGC\,925 & $-0.26 \pm 0.01$ & $-1.73 \pm 0.01$ & 0.12 & -0.84 \\
\noalign{\smallskip}
\hline
R/R{$\rm_{eff}$} - DMR & NGC 4214 & $-0.003 \pm 0.001$ & $-0.35 \pm 0.02$ & 0.10 & -0.05 \\
 & NGC 2976 & $-0.09 \pm 0.02$ & $-0.17 \pm 0.03$ & 0.14 & -0.34 \\
 & NGC 3077 & $0.009 \pm 0.001$ & $-0.16 \pm 0.04$ & 0.18 & 0.25 \\
 & NGC 2403 & $-0.07 \pm 0.01$ & $-0.10 \pm 0.02$ & 0.10 & -0.92 \\
 & NGC 925 & $-0.09 \pm 0.02$ & $-0.09 \pm 0.01$ & 0.16 & -0.77 \\
\noalign{\smallskip}
\hline
R/R$_{\rm{eff}}$ - $\alpha_{\rm CO}$ & Low mass & $0.020 \pm 0.002$ & $0.73 \pm 0.09$ & 0.34 & 0.01 \\
 & NGC\,2403 & $0.53 \pm 0.03$ & $0.70 \pm 0.04$ & 0.30 & 0.72 \\
 & NGC\,925 & $0.69 \pm 0.07$ & $0.57 \pm 0.01$ & 0.54 & 0.54 \\
\noalign{\smallskip}
\hline
\end{tabular}
\tablefoot{The table lists the coefficients obtained from the linear fits to the relations given in the first column, adopting the functional form $y = mx + b$. RMSE denotes the root mean square error and r the Pearson correlation test. NGC~4214, NGC~2976, and NGC~3077 were combined into a single low-mass subsample to derive a common fit.}

\end{table*}

\begin{table*}
\caption{Best-fit linear regression parameters derived from Bayesian Survival Analysis for the studied relations.}
\label{fittings_vals_survival}
\centering
\begin{tabular}{llcccc}
\hline\hline
\noalign{\smallskip}
Relation & Sample / Galaxy & Slope ($m$) & Intercept ($b$) & RMSE & $ \text{r}$ \\
\noalign{\smallskip}
\hline
\noalign{\smallskip}
$Z$ - DGR & All & $2.24 \pm 0.03$ & $-20.80 \pm 0.29$ & 0.21 & 0.74 \\
\hline
$Z$ - DMR & All & $0.57 \pm 0.02$ & $-4.98 \pm 0.16$ & 0.18 & 0.45 \\
\hline
$Z$ - $\alpha_{\rm CO}$ & Low mass & $-1.46 \pm 0.03$ & $12.86 \pm 0.26$ & 0.75 & -0.13 \\
 & NGC\,2403 & $-1.24 \pm 0.27$ & $11.44 \pm 2.23$ & 0.75 & -0.30 \\
 & NGC\,0925 & $-2.65 \pm 0.24$ & $23.35 \pm 1.97$ & 0.78 & -0.35 \\
\noalign{\smallskip}
\hline
R/R$_{\mathrm{eff}}$ - DGR & NGC\,4214 & $0.02 \pm 0.02$ & $-2.53 \pm 0.03$ & 0.20 & -0.05 \\
 & NGC\,2976 & $-0.20 \pm 0.02$ & $-1.87 \pm 0.03$ & 0.23 & -0.51 \\
 & NGC\,3077 & $-0.19 \pm 0.07$ & $-1.82 \pm 0.08$ & 0.61 & -0.24 \\
 & NGC\,2403 & $-0.22 \pm 0.01$ & $-1.68 \pm 0.01$ & 0.11 & -0.68 \\
 & NGC\,0925 & $-0.22 \pm 0.01$ & $-1.80 \pm 0.01$ & 0.16 & -0.82 \\
\noalign{\smallskip}
\hline
R/R$_{\mathrm{eff}}$ - DMR & NGC\,4214 & $0.004 \pm 0.009$ & $-0.356 \pm 0.015$ & 0.13 & 0.05 \\
 & NGC\,2976 & $-0.083 \pm 0.011$ & $-0.151 \pm 0.013$ & 0.17 & -0.36 \\
 & NGC\,3077 & $-0.116 \pm 0.030$ & $-0.113 \pm 0.035$ & 0.34 & -0.23 \\
 & NGC\,2403 & $-0.092 \pm 0.006$ & $0.037 \pm 0.013$ & 0.12 & -0.51 \\
 & NGC\,0925 & $-0.095 \pm 0.006$ & $0.030 \pm 0.015$ & 0.19 & -0.73 \\
\noalign{\smallskip}
\hline
R/R$_{\mathrm{eff}}$ - $\alpha_{\rm CO}$ & Low mass & $-0.019 \pm 0.006$ & $0.79 \pm 0.07$ & 0.75 & -0.03 \\
 & NGC\,2403 & $0.06 \pm 0.02$ & $0.94 \pm 0.04$ & 0.73 & 0.16 \\
 & NGC\,0925 & $0.17 \pm 0.02$ & $1.09 \pm 0.03$ & 0.68 & 0.35 \\
\noalign{\smallskip}
\hline
\end{tabular}
\tablefoot{The table lists the coefficients obtained from Bayesian linear regression with survival analysis (\texttt{pm.Censored}) to properly account for upper limits, adopting the functional form $y = mx + b$. RMSE denotes the root mean square error and $r$ the Pearson correlation coefficient. All indicates the combined fit of the sample.}
\end{table*}

%\section{The DGR-Z and DMR-Z in comparison with other works}

%\begin{figure*}[ht!]
%    \centering
%    {\includegraphics[scale=0.7]{Figs/A_Z_DGR_all_error_1x1.png}}
%     \caption{Scaling relations of the DGR-Z. Our data points are color-coded  similar with previous figures. To contextualize these results, several observational, theoretical, and semi-analytical relations from the literature are overplotted (indicated by the various lines; see legend for references). This highlights the overall trend of increasing DGR with advancing chemical evolution.}
%      \label{DGR_Z_plot}
%\end{figure*}

%\begin{figure*}[ht!]
%    \centering
%    {\includegraphics[scale=0.7]{Figs/A_Z_DMR_all_error_1x1.png}}
%     \caption{Scaling relation of the DMR as a function of metallicity. The data points from our sample are color-coded similar with previous figures. The lines represent a compilation of empirical calibrations and theoretical models from the literature, illustrating the variations and scatter in the fraction of metals locked into dust grains across different metallicity regimes.}
%\label{DMR_Z_plot}
%\end{figure*}

\section{Bayesian Fittings for gradients in \texorpdfstring{R$_{25}$}{R25}}\label{app_BF}

This appendix presents the Bayesian best-fitting parameters for the radial gradients expressed as a function of $R_{25}$.

\begin{table*}
\caption{Best-fit parameters for the $\alpha_{\rm CO}$, DGR and DMR relations for the R$_{25}$ gradients.}
\label{table:fits_updated}
\centering
\begin{tabular}{llcccc}
\hline\hline
\noalign{\smallskip}
Relation & Sample / Galaxy & Slope ($m$) & Intercept ($b$) & RMSE & $\text{r}$ \\
\noalign{\smallskip}
\hline
\noalign{\smallskip}
R/R{$\rm_{25}$} - $\alpha_{\rm CO}$ & Low mass & $0.017 \pm 0.001$ & $0.73 \pm 0.09$ & 0.32 & 0.02 \\
 & NGC 2403 & $2.72 \pm 0.13$ & $0.70 \pm 0.04$ & 0.30 & 0.72 \\
 & NGC 925 & $3.24 \pm 0.30$ & $0.57 \pm 0.01$ & 0.53 & 0.54 \\
\noalign{\smallskip}
\hline
\noalign{\smallskip}
R/R{$\rm_{25}$} - DGR & NGC 4214 & $-0.00 \pm 0.01$ & $-2.50 \pm 0.02$ & 0.17 & -0.01 \\
 & NGC 2976 & $-0.69 \pm 0.03$ & $-1.86 \pm 0.01$ & 0.18 & -0.45 \\
 & NGC 3077 & $0.015 \pm 0.004$ & $-1.88 \pm 0.02$ & 0.33 & 0.26 \\
 & NGC 2403 & $-1.25 \pm 0.18$ & $-1.58 \pm 0.63$ & 0.06 & -0.94 \\
 & NGC 925 & $-1.22 \pm 0.20$ & $-1.73 \pm 0.90$ & 0.12 & -0.84 \\
\noalign{\smallskip}
\hline

\noalign{\smallskip}
R/R$_{\mathrm{25}}$ - DMR & NGC\,4214 & $-0.028 \pm 0.001$ & $-0.34 \pm 0.03$ & 0.10 & -0.05 \\
 & NGC\,2976 & $-0.30 \pm 0.07$ & $-0.17 \pm 0.03$ & 0.12 & -0.34 \\
 & NGC\,3077 & $0.008 \pm 0.001$ & $-0.16 \pm 0.04$ & 0.17 & 0.24 \\
 & NGC\,2403 & $-0.34 \pm 0.01$ & $-0.10 \pm 0.01$ & 0.02 & -0.92 \\
 & NGC\,925 & $-0.42 \pm 0.02$ & $-0.09 \pm 0.01$ & 0.05 & -0.77 \\
\noalign{\smallskip}
\hline
\end{tabular}
\tablefoot{The table lists the coefficients obtained from the linear fits to the relations given in the first column, adopting the functional form $y = mx + b$.}
\end{table*}

\begin{figure*}[ht!]
    \centering
    {\includegraphics[scale=0.5]{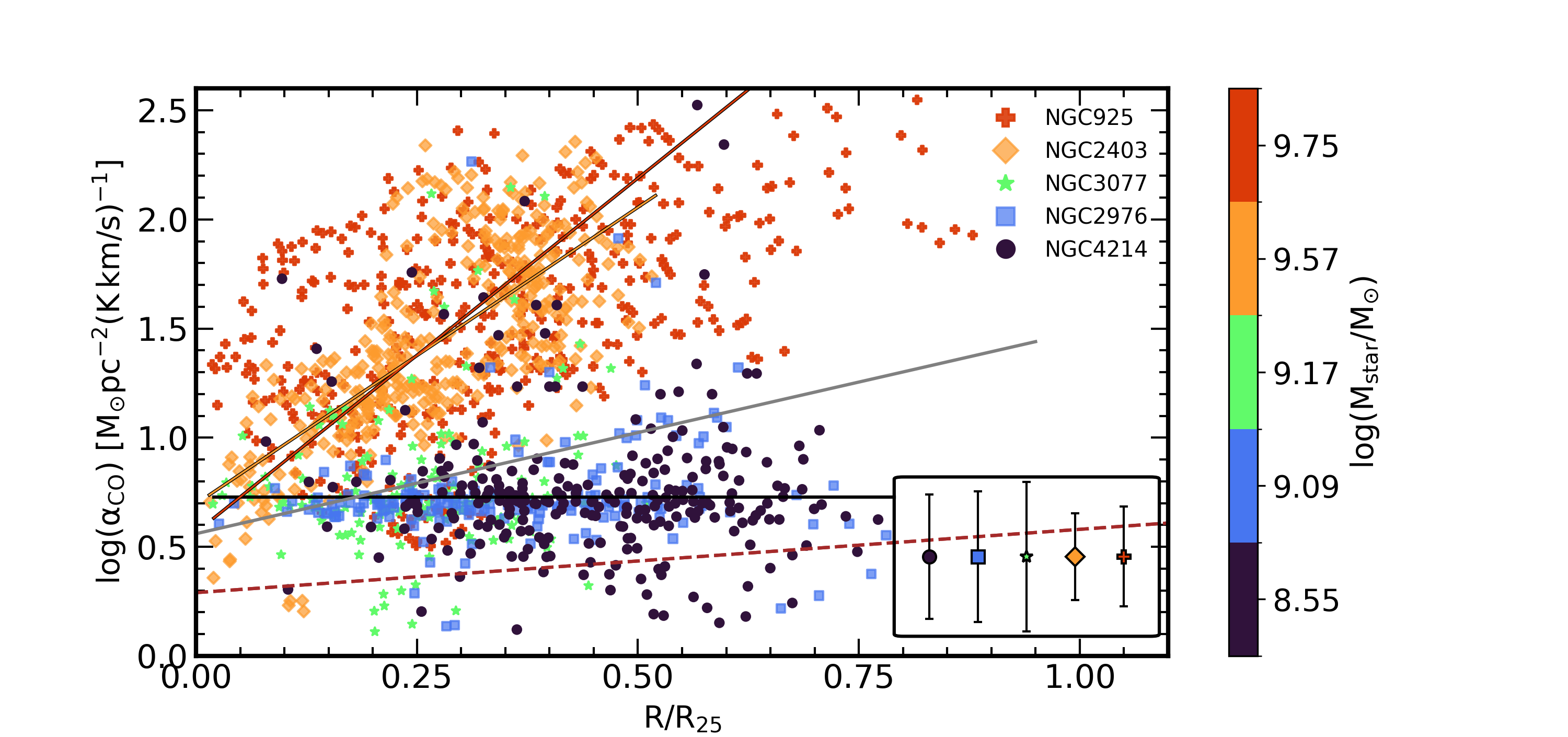}}
     \caption{Radial distribution of the \alphaCO\ as a function of the radius R/R{$\rm_{25}$}. Symbols and color-coding are consistent with previous figures. The solid orange and red lines denote the linear fits for NGC 2403 and NGC 925, respectively. The horizontal black line represents the fitting for the lower-mass galaxies NGC 4214, NGC 2976 and NGC 3077. The brown dashed line is the fitting of \cite{Sandstrom2013} while the gray line is the fitting of \cite{Chiang2024}. Similar to previous figures, the error bars correspond to the median uncertainty in log(\alphaCO ). }
      \label{R25_alphaCO_plot}
\end{figure*}

\begin{figure*}[ht!]
    \centering
    {\includegraphics[scale=0.5]{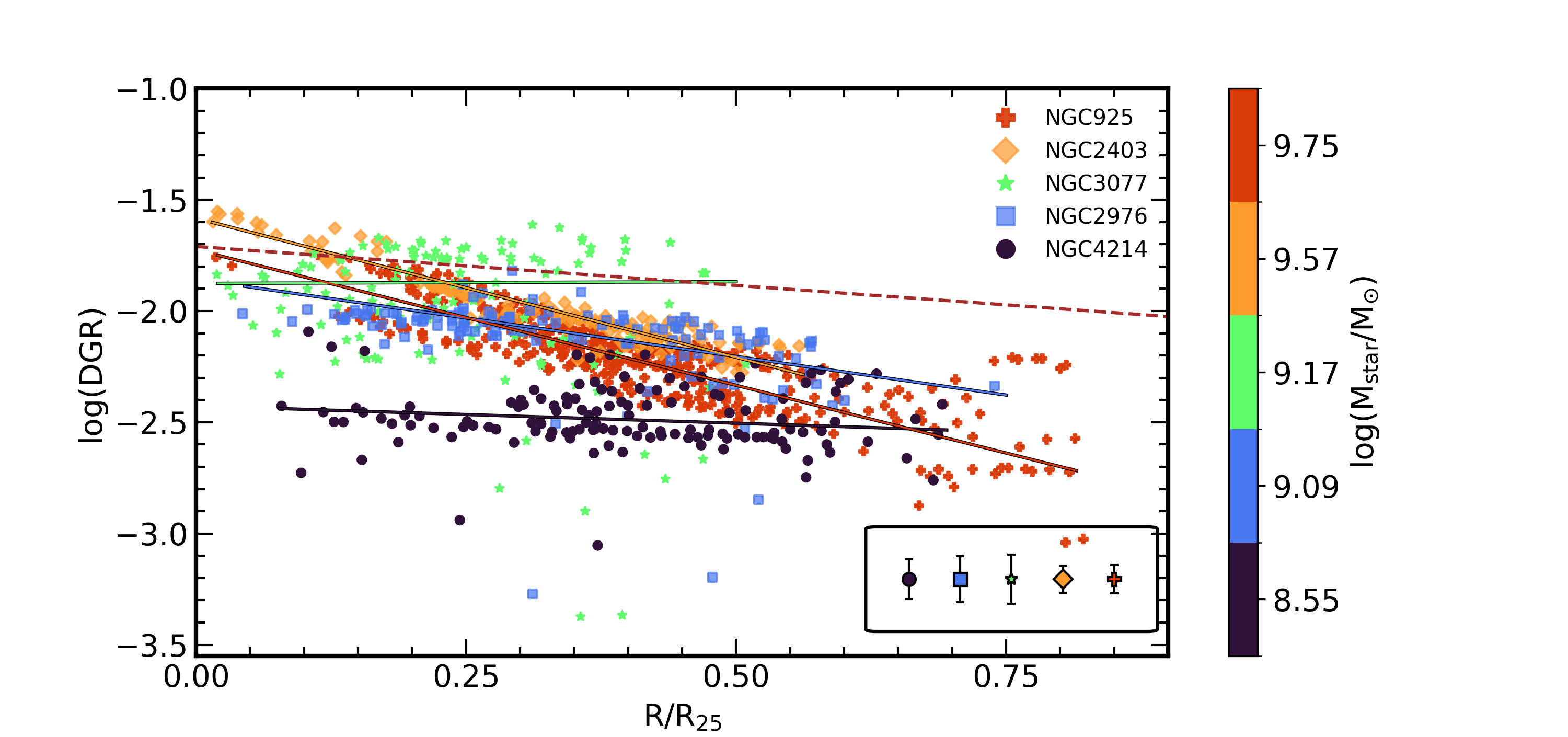}}
     \caption{Radial distribution of the DGR as a function of the radius R/R{$\rm_{25}$}. Symbols and color-coding are consistent with previous figures. The solid lines denote the linear fitting for each galaxy in the sample. The brown dashed line is the fitting of \cite{Sandstrom2013}. Similar to previous figures, the error bars correspond to the median uncertainty in log(DGR).}
      \label{R25_DGR_plot}
\end{figure*}

\begin{figure*}[ht!]
    \centering
    {\includegraphics[scale=0.5]{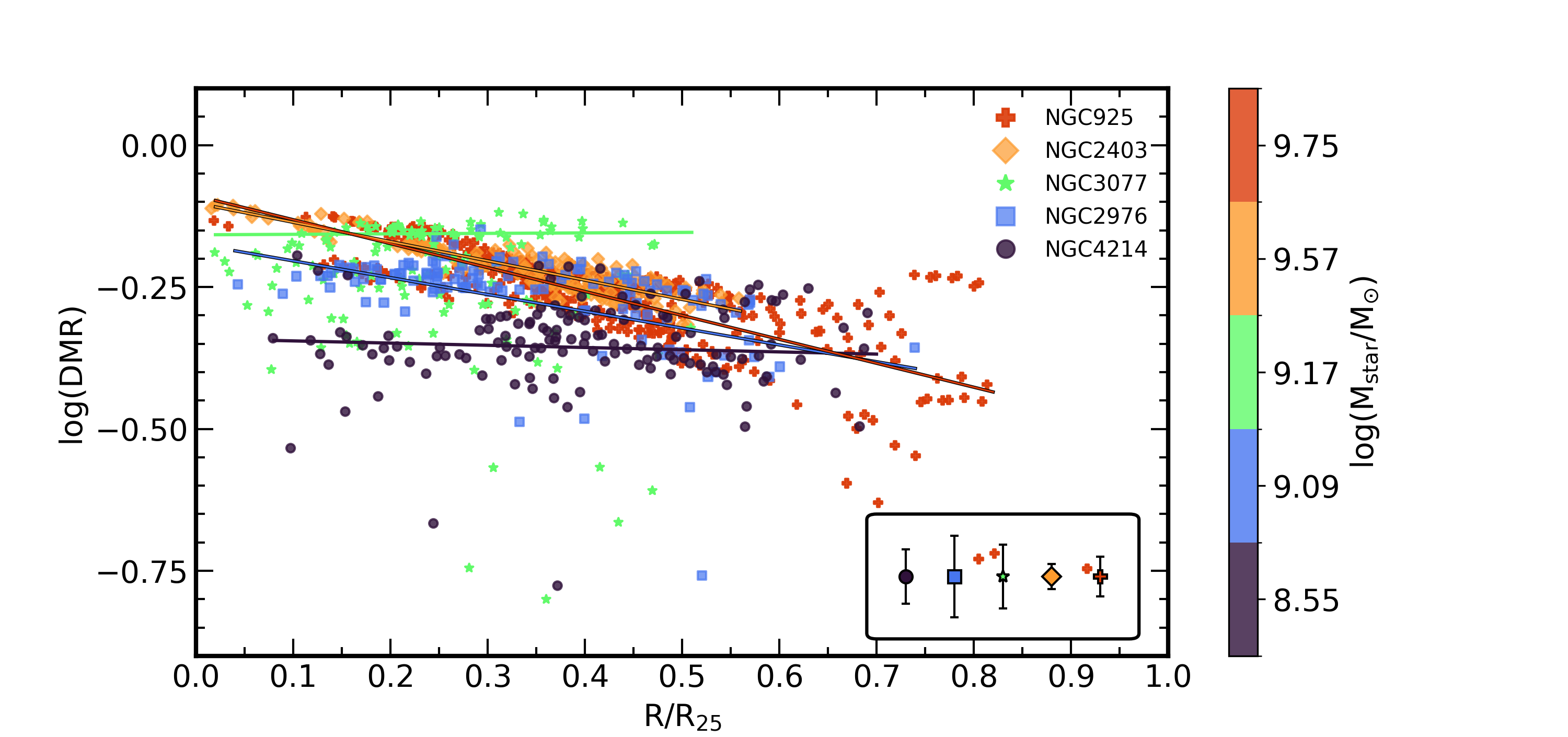}}
     \caption{Radial distribution of the DMR as a function of the radius R/R{$\rm_{25}$}. Symbols and color-coding are consistent with previous figures. The solid lines denote the linear fitting for each galaxy in the sample. Similar to previous figures, the error bars correspond to the median uncertainty in log(DMR).}
      \label{R25_DMR_plot}
\end{figure*}

\section{Metallicity recalibration.}\label{recal}

This appendix presents the calibration used to convert the oxygen abundances adopted by \citet{Sandstrom2013}, based on the \citet{Pilyugin2005} calibration, to the \citet{Pilyugin2016} scale used throughout this work. The resulting best-fitting relation is $Z_{\text{PT05}} = 0.36 Z_{\text{PG16}}$.

   \begin{figure}
   \includegraphics[clip,width=1\linewidth]{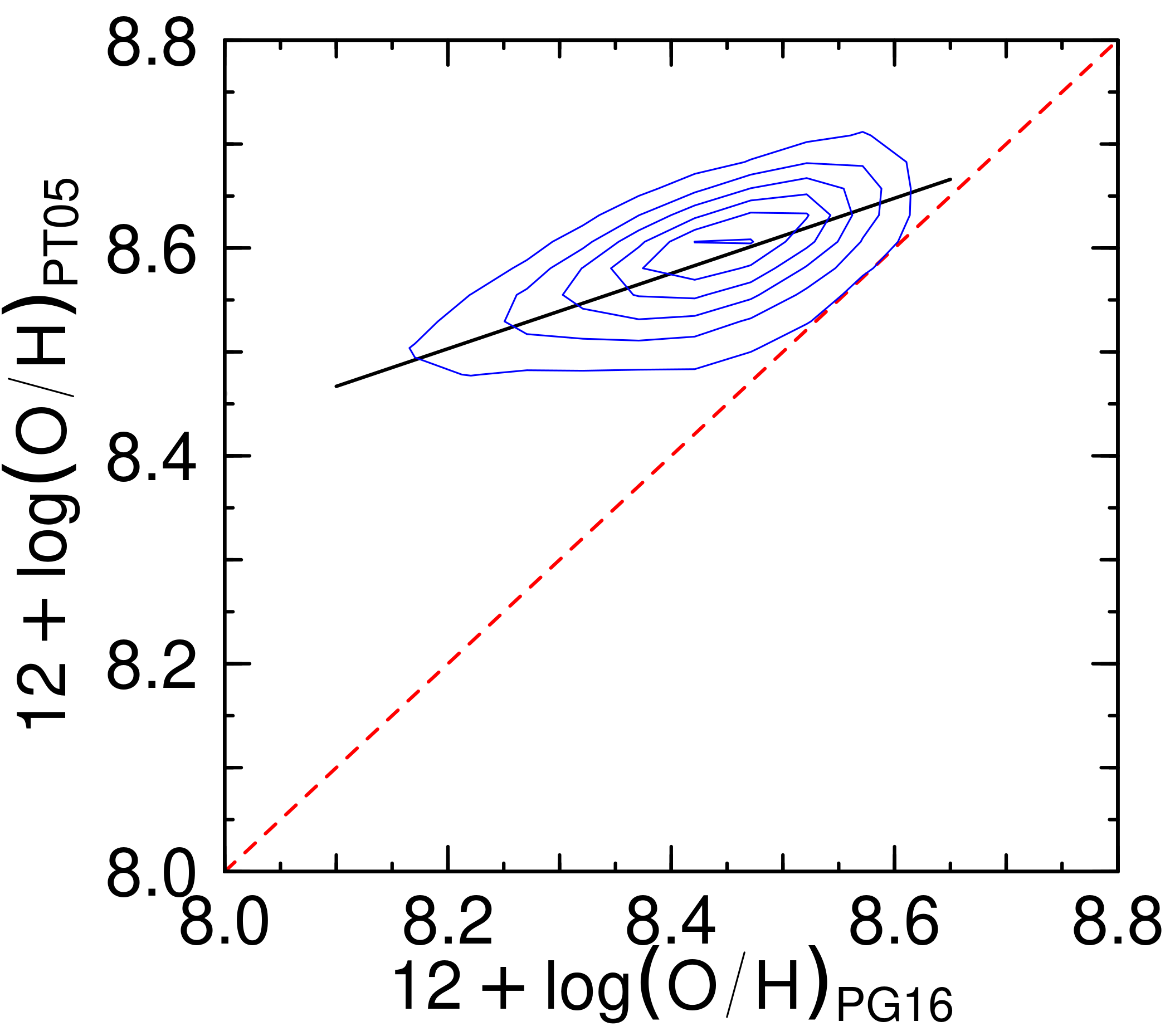}
      \caption{Comparison between the metallicity calibrations of \citet{Pilyugin2005} and \citet{Pilyugin2016}. The blue contours show the data density distribution, and the black line represents the best-fitting linear relation used to convert metallicities from the \citet{Pilyugin2005} calibration to the \citet{Pilyugin2016} scale. The red dashed line denotes the one-to-one relation.}
         \label{fig1}
   \end{figure}

\section{Parameters CIGALE code}

In this appendix we show the input modules that we establish in the CIGALE code (Table \ref{CIGALE_parameters}) and particularly for the dust emission models for each galaxy (Table \ref{table:dust_grids}).

\begin{table*}
\centering
\caption{CIGALE input modules and physical parameter values.}
\label{CIGALE_parameters}
\begin{tabular}{lll} % Usamos 'l' para alinear a la izquierda y que no se disperse
\hline\hline
Module & Parameter & Value \\ \hline
Delayed SFH & 
\begin{tabular}[c]{@{}l@{}}
    $\tau_{\rm{main}}$\\ 
    Age$_{\rm{main}}$ \\ 
    $\tau_{\rm{burst}}$ \\ 
    Age$_{\rm{burst}}$ \\ 
    $f_{\rm{burst}}$ \\ 
    SFR$_{\rm{A}}$
\end{tabular} & 
\begin{tabular}[c]{@{}l@{}}
    7000, 8000, 10000, 12000, 13000\\ 
    9500, 13000\\ 
    5.0, 10.0, 20.0, 50.0, 80.0, 110.0\\ 
    10, 30, 50, 70, 90, 150, 200, 250, 300, 400\\ 
    0.1, 0.2, 0.3, 0.4, 0.5, 0.75\\ 
    1.0
\end{tabular} \\ \hline

SSP & 
\begin{tabular}[c]{@{}l@{}}
    IMF\\ 
    Stellar metallicity
\end{tabular} & 
\begin{tabular}[c]{@{}l@{}}
    \cite{Chabrier2003}\\ 
    0.004, 0.008, 0.02
\end{tabular} \\ \hline

Dust attenuation & 
\begin{tabular}[c]{@{}l@{}}
    $E(B-V)_{\rm{young}}$\\ 
    $E(B-V)_{\rm{old\thinspace factor}}$ \\ 
    UV bump wavelength \\ 
    UV bump width\\ 
    UV bump amplitude\\ 
    Slope delta \\ 
    Filters 
\end{tabular} & 
\begin{tabular}[c]{@{}l@{}} 
     % Corte de línea 1
    0.0625,0.125, 0.25\\   % Corte de línea 2
    0.50\\ 
    217.5\\ 
    35.0\\ 
    0.0\\ 
    0.0\\ 
    FUV
\end{tabular} \\ \hline

%Dust emission & 
%\begin{tabular}[c]{@{}l@{}}
%    $q_{\rm PAH}$ \\ 
%    $U_{\rm min}$\\ 
%    $\alpha$ \\ 
%    $\gamma$ 
%\end{tabular} & 
%\begin{tabular}[c]{@{}l@{}}
%    0.47, 1.12, 1.77\\ 
%    3.5, 4.0, 5.0\\ 
%    1.9, 2.0, 2.1\\ 
%    0.032
%\end{tabular} \\ \hline
\end{tabular}
\tablefoot{$\tau_{\rm{main}}$, Age$_{\rm{main}}$, $\tau_{\rm{burst}}$ and Age$_{\rm{burst}}$ are expressed in $10^{6}$ yr. The SSP, dust attenuation models and dust emission model are described in the cited references. The specific stellar metallicity used for each galaxy is 0.004 for NGC 2976, 0.008 for NGC 4214, NGC 2403, NGC 925, and 0.02 for NGC 3077. {Stellar metallicities ($Z$) were set by converting global gas-phase $12+\log(\text{O/H})$ values to $Z$ and matching them to the closest discrete metallicity grid point available in CIGALE.}}

\end{table*}

\begin{table*}
\centering
\caption{Dust emission parameters used as input for the CIGALE code.}
\label{table:dust_grids}
\begin{tabular}{lllll}
\hline\hline
Galaxy & $q_{\rm PAH}$ & $U_{\rm min}$ & $\alpha$ & $\gamma$ \\ \hline
NGC 4214 & 0.47, 1.12, 1.77 & 3.5, 4.0, 5.0 & 1.9, 2.0, 2.1 & 0.032 \\ \hline
NGC 2976 & 2.50, 3.19, 3.90 & 1.7, 2.0, 2.5 & 1.0, 2.0, 3.0 & \begin{tabular}[c]{@{}l@{}}0.005, 0.01, \\ 0.02\end{tabular} \\ \hline
NGC 3077 & 1.77, 2.50, 3.19 & 1.7, 2.0, 2.5 & 1.0, 2.0, 3.0 & \begin{tabular}[c]{@{}l@{}}0.015, 0.03, \\ 0.06\end{tabular} \\ \hline
NGC 2403 & \begin{tabular}[c]{@{}l@{}}3.19, 3.90, \\ 4.58, 5.26\end{tabular} & \begin{tabular}[c]{@{}l@{}}1.7, 2.0, 2.5 \end{tabular} & \begin{tabular}[c]{@{}l@{}}1.9, 2.0, 2.1, \\ 2.2, 2.3, 2.4\end{tabular} & 0.007 \\ \hline
NGC 925 & 5.95, 6.63, 7.32 & 1.0, 1.2, 1.5 & 2.5, 2.6, 2.7 & 0.008 \\ \hline
\end{tabular}
\tablefoot{Specific parameters for each galaxy in the sample.}
\end{table*}

\end{appendix}

\end{document}